\documentclass[floatfix,prb,aps,showpacs,twocolumn]{revtex4-2}

\usepackage{graphicx,amsmath,amssymb,color}
\usepackage{nicefrac}
\usepackage{multirow, makecell, ragged2e}
\usepackage[titletoc,title]{appendix}
\usepackage{bm}
\usepackage{float,lipsum}
\usepackage{algpseudocode}
\usepackage{algorithm}
\usepackage[final]{changes}

\newcommand{\be}{\begin{equation}}
\newcommand{\ee}{\end{equation}}

\newcommand{\cH}{\mathcal{H}}

\renewcommand{\vec}[1]{\bm{#1}}

\newcommand{\unit}[1]{\vec{\hat{ #1}}}
\DeclareMathOperator{\Tr}{Tr}
\newcommand{\up}{\uparrow}
\newcommand{\down}{\downarrow}

\graphicspath{{./Figures/}}

\begin{document}

\title{\added{Topological superconductivity induced by spin-orbit coupling, perpendicular magnetic field and superlattice potential}}
\author{Jonathan Schirmer$^*$, J. K. Jain, and C. -X. Liu}
\affiliation{Department of Physics, 104 Davey Lab, The Pennsylvania State University, University Park, Pennsylvania 16802}

\begin{abstract}

Topological superconductors support Majorana modes, which are quasiparticles that are their own antiparticles and which obey non-Abelian statistics in which successive exchanges of particles do not always commute. Here we investigate whether a two-dimensional superconductor with ordinary s-wave pairing can be rendered topological by the application of a strong magnetic field. To address this, we obtain the self-consistent solutions to the mean field Bogoliubov-de Gennes equations, which are a large set of nonlinearly coupled equations, for electrons moving on a lattice. We find that the topological ``quantum Hall superconductivity" is facilitated by a combination of spin-orbit coupling, which locks an electron’s spin to its momentum as it moves through a material, and a coupling to an external periodic potential which gives a dispersion to the Landau levels and also distorts the Abrikosov lattice. We find that, for a range of parameters, the Landau levels broadened by the external periodic potential support topological superconductivity, which is typically accompanied by a lattice of ``giant" $h/e$ vortices as opposed to the familiar lattice of $h/2e$ Abrikosov vortices. In the presence of a periodic potential, we find it necessary to use an ansatz for the pairing potential of the form $\Delta(\vec{r})e^{i2\vec{Q}\cdot\vec{r}}$ where $\Delta(\vec{r})$ has a periodicity commensurate with the periodic potential. However, despite this form of the pairing potential, the current in the ground state is zero.  In the region of ordinary superconductivity, we typically find a lattice of dimers of $h/2e$ vortices. Our work suggests a realistic proposal for achieving topological superconductivity, as well as a helical order parameter and unusual Abrikosov lattices.

\end{abstract}
\graphicspath{{./Figures/}}

\maketitle
\section{Introduction}

Majorana modes, which are quasiparticles which may be considered as their own anti-particles, are perhaps the most readily realizable non-Abelian anyons. In topological superconductors, they appear as zero-energy quasiparticles which reside in the cores of Abrikosov vortices, or appear as chiral boundary modes. Their non-Abelian braid statistics, although not yielding all unitary gates, are a major step towards topological quantum computation with anyons  \cite{Stone2006, Oshikawa2007, Nayak2008,Sau2010a,Stanescu2010,Alicea2012,Hung2013,Beenakker2013,Zhou2013,Biswas2013,Sarma2015,Liu2015,Murray2015,Smith2016}. Besides their potential application for quantum computation and being of fundamental scientific interest, achieving Majorana modes in the lab could lead to the physical realization of interesting theoretical models, such as the Sachdev-Ye-Kitaev model \cite{Chowdhury2022}. There are several candidate condensed matter systems which are thought to host Majorana modes, including the $\nu = 5/2$ fractional quantum Hall effect, two dimensional p-wave superconductors, and two dimensional films of $^{3}$He-A superfluid \cite{Bernevig2013}. In addition, there have been proposals to engineer topologically interesting structures which host Majorana modes. In particular, topological p-wave superconductivity (SC) is thought to arise when a spin-orbit coupled electron gas is proximity coupled to an s-wave superconductor, owing to the appearance of a single Fermi surface in certain parameter regimes \cite{Sarma2006,Fu2008,Lutchyn2010,Qi2010,Sau2010,Sau2010b,Alicea2010,Qi2011,BlackSchaffer2011,Mourik2012,Nakosai2012,Stanescu2013,Goertzen2017,Lutchyn2018}. Particularly notable are proposals for quasi-one dimensional systems, e.g. semiconductor wires \cite{Lutchyn2010,Sau2010b,Oreg2010,Mourik2012,NadjPerge2014,Liu2017a,Frolov2020}, magnetic atom chains \cite{Klinovaja2013,Li2014, choy_majorana_2011, nadj-perge_proposal_2013, pientka_topological_2013}, and planar Josephson junctions \cite{Fornieri2019,Pientka2017}, for realizing topological superconductivity (TSC). Some exciting progress has been made on these fronts \cite{Marra2022}. 

In earlier work, Sau {\it et al.} proposed a heterostructure consisting of a spin-orbit coupled two-dimensional electron gas (2DEG) coupled to a ferromagnetic insulator and an s-wave superconductor \cite{Sau2010, Sau2010b}. The ferromagnetic insulator causes a gap in the single-particle spectrum of the 2DEG at zero momentum. This gap, along with spin-orbit coupling (SOC), leads to the appearance of a single Fermi surface for certain values of the chemical potential. When the system is coupled to an s-wave superconductor, effective chiral p-wave TSC is realized in a single band. They studied the properties of the BdG spectrum in the regime where vortices are well-separated. Crucially for these works, it is the SOC, together with the Zeeman-like effect from the magnetic insulator, which gives rise to the topological phase.

A related idea is to apply a magnetic field to a two-dimensional system, utilizing the orbital effect to realize TSC. Indeed the topological nature of the underlying Landau levels (LLs) sets the stage nicely for TSC when a pairing gap is opened \cite{Qi2010,Qi2011}. The combination of SC and quantum Hall (QH) systems, both integer (IQHE) and fractional (FQHE), has been thought to be fertile ground for the realization not only of TSC with the concomitant Majorana particles, but also more exotic topological phases that harbor parafermions and Fibonacci anyons \cite{Lu2010,Clarke2013,Mong2014,alicea2015,Alicea2016,Amet2016,liang2019,Gul2020,shaffer2021}. In order to integrate the orbital effect of the magnetic field and SC, there are, broadly speaking, two approaches. The first, like the above proposal, involves proximity coupling a QH system to a superconductor. It is necessary to take into account the vortices in superconductors exposed to a magnetic field \cite{Zocher2016,Mishmash2019,Jeon2019,Chaudhary2020,pathak2021}.
Zocher and Rosenow \cite{Zocher2016}, Mishmash {\it et al.} \cite{Mishmash2019} and Chaudhary and MacDonald \cite{Chaudhary2020} showed that there is no TSC for particles in the lowest LL (or lowest few LLs)  if the unit cell of the Abrikosov lattice has only one (or in general an odd number of) superconducting flux quanta (that is, the unit cell violates magnetic translation symmetry), as is the case for the usual triangular Abrikosov lattice or even a square Abrikosov lattice.
It was shown that the spectrum of the Bogoliubov-de  Gennes (BdG) equations, which is com­pletely determined by the s-wave pairing potential when the normal state bands are flat, has an even degeneracy in the case of a square or triangular Abrikosov vortex lattice, and therefore any quasiparticle spectral gap closing across a topological phase transition must lead to a  Chern number change by an even integer, thereby precluding  the appearance of TSC {that requires odd integer Chern number}. All of these works considered LLs with SOC. Also, because they considered proximity-induced SC, they did not need to find a self-consistent mean-field state within the 2DEG, since the pairing, and consequently the Abrikosov lattice, are inherited from a nearby bulk superconductor, and hence are a part of the Hamiltonian defining the problem, not of its solution.

The second approach is to apply a strong magnetic field to an intrinsic two-dimensional superconductor, driving it into the Landau level limit. It has been known for some time that BCS mean field theory predicts that superconductivity can be enhanced due to the Landau level structure of systems in a strong magnetic field \cite{Rajagopal1991,tevsanovic1989,tevsanovic1991,PhysRevLett.66.842,Akera1991,Rajagopal1991,rajagopal1991quantum,norman1991landau,Rasolt1992,MacDonald1992,Norman1992,rajagopal1992solutions,macdonald1993quantum,ryan1993manifestations,ryan1993vortex,norman1995,maska2002reentrant,scherpelz2013general,ran2019,Kim2019,chaudhary2021}. Because of the enlargement of the density of states in Landau levels, it has been predicted that the critical temperature in this limit approaches that of the zero-field value \cite{tevsanovic1989,tevsanovic1991}. This has been referred to as quantum Hall superconductivity \cite{chaudhary2021}.  In contrast to the case of proximity coupled SC to LLs, the vortex lattice structure is determined by electrons pairing in LLs within the same sample, and consequently a vortex lattice structure cannot be assumed but must be solved for self-consistently \cite{Akera1991,Norman1992,ryan1993vortex}. In spite of extensive theoretical efforts, convincing experimental demonstrations of these remarkable predictions have been lacking.
In addition, to our knowledge, there has not been any study of the possibility of topological superconductivity in such systems. That is the focus of the present study.

\begin{figure}
\includegraphics[width=.5\textwidth]{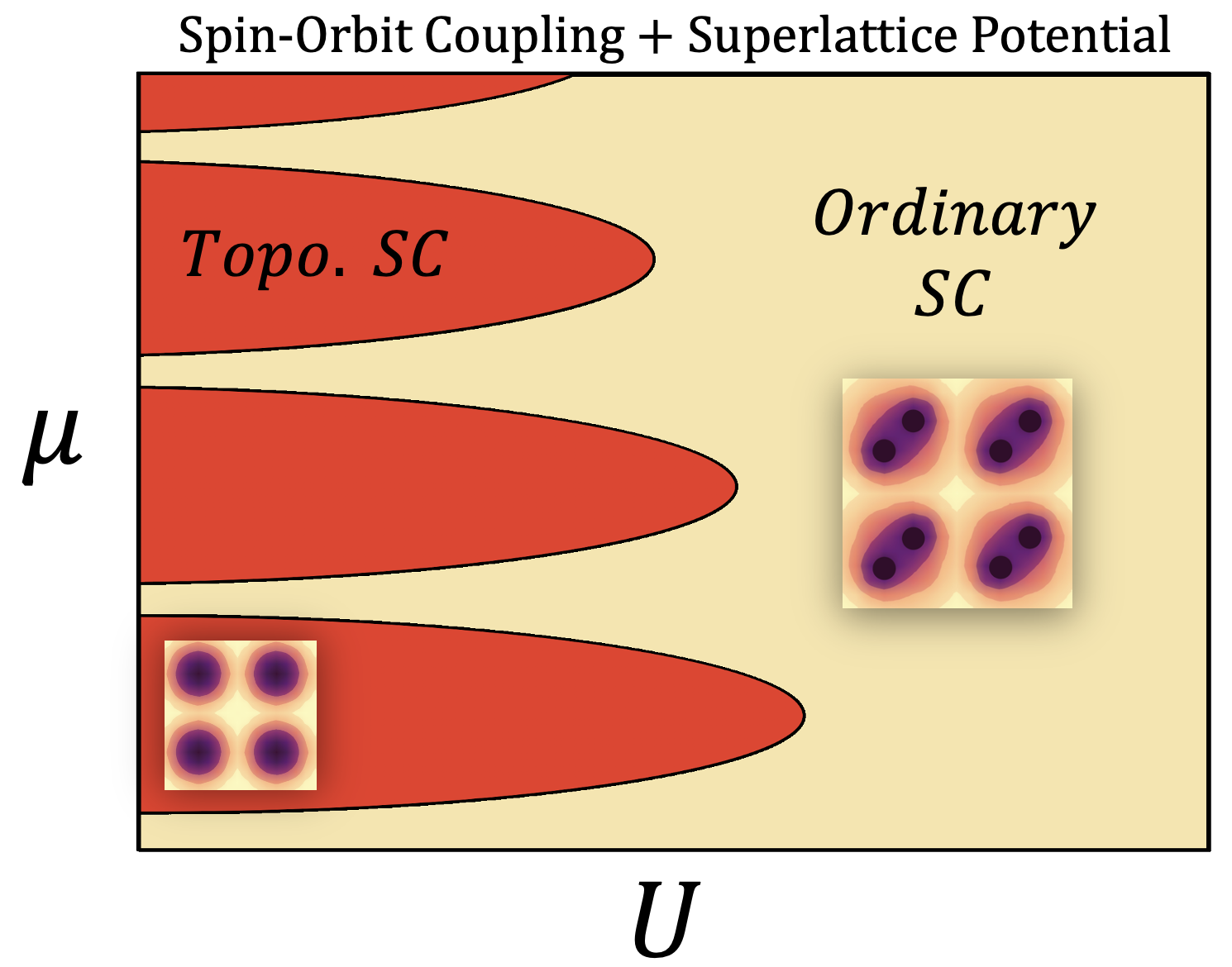}
\caption{A schematic of the numerically determined phase diagram in the presence of both spin-orbit coupling and a superlattice potential as a function of the strength of the onsite attractive interaction $-U$ and the chemical potential $\mu$. The regions of topological superconductivity (TSC), depicted in red, roughly coincide with Landau levels broadened by the superlattice potential. The vortices here typically tend to form a lattice of ``giant" vortices (as shown in the bottom left of the figure), where a giant vortex consists of two superconducting vortices merged into one. In the regions of ordinary superconductivity (shaded yellow), the vortices typically form a lattice of dimers as depicted on the right of the figure. The vortex lattices are square because of our choice of a square superlattice potential.  } \label{schematic-pd}
\end{figure}

In this work, we explore if TSC can arise from pairing between electrons occupying LLs, which may be referred to as ``quantum Hall topological superconductivity." For this purpose, we consider a fully self-consistent mean-field theory of a lattice model under a magnetic field with a phenomenological attractive on-site Hubbard interaction. We also consider Rashba SOC and a periodic superlattice potential. We numerically determine phase diagrams without SOC or superlattice potential, with SOC and without superlattice potential, without SOC and with superlattice potential, and  with SOC and superlattice potential. We find that while SOC is necessary for TSC, the application of a superlattice potential markedly enlarges the regions hosting TSC in the phase diagram. The reason is because the Abrikosov lattice distorted by the superlattice potential has more than one superconducting flux quantum in a unit cell. (As shown below, the TSC often occurs in regions where two Abrikosov vortices merge into a ``giant" vortex.) A schematic phase diagram is shown in Fig. \ref{schematic-pd}.
More detailed phase diagrams are given below.  In order to properly treat systems with a superlattice potential, we find it necessary to consider helical pairing functions, which describe Cooper pairs with non-zero center-of-mass momentum. This ansatz is the same as the one employed by Fulde and Ferrell \cite{fulde_superconductivity_1964}. We also describe vortex lattice structures arising in our model, some of which are quite unexpected. We finally discuss some prospective experimental realizations of this model.

\section{Model}
\subsection{Hamiltonian}
We consider
spin-$\frac{1}{2}$ fermions on a square lattice in a magnetic field with Rashba SOC, a single-particle potential, and onsite attractive interaction. The interacting Hamiltonian is
\be
\cH = \cH_0 + \cH_{\rm SO} + \cH_{\rm I},
\ee
where
\begin{equation} \label{ham}
\begin{split}
\cH_0 =&-\sum_{j,\delta,\sigma}\Big(e^{iA_\delta (\vec{r}_j)}c^\dagger_{j+\delta,\sigma} c_{j,\sigma}+e^{-iA_\delta(\vec{r}_j)}c^{\dagger}_{j,\sigma} c_{j+\delta,\sigma}\Big) \\ &-
 \sum_{j,\sigma} \big(\mu-V(\vec{r}_j) \big) c^\dagger_{j,\sigma} c_{j,\sigma} \\
  \cH_{\rm SO} =& V_{\rm SO} \sum_{j} \bigg( e^{iA_{\unit{x}} (\vec{r}_j)}\Big( c^\dagger_{j+\unit{x},\down}c_{j,\up} -c^\dagger_{j+\unit{x},\up} c_{j,\down} \Big) \\ &+i e^{iA_{\unit{y}} (\vec{r}_j)} \Big(c^\dagger_{j+\unit{y},\down}c_{j,\up} + c^\dagger_{j+\unit{y},\up}c_{j,\down} \Big)  \bigg)  + {\rm h.c.}\\
 \cH_{\rm I} =& -U\sum_{j}c^\dagger_{j,\up}c^\dagger_{j,\down}c_{j,\down}c_{j,\up},
\end{split}
\end{equation}
where $\vec{r}_j$ is the location of site $j$, $V_{\rm SO}$ is the SOC strength, $\mu$ is the chemical potential, $U$ is the interaction strength, $A_\delta (\vec{r}_j)$ are the hopping (Peierls) phases, $\delta=\unit{x},\unit{y}$, $\sigma=\up$, $\down$. The annihilation (creation) operators for fermions of spin $\sigma$ at site $j$ are $c_{j,\sigma}$ ($c^\dagger_{j,\sigma}$). We have set the hopping amplitude to unity. Note that $A_\delta (\vec{r}_j)$ appears both in the hopping term and the SOC term, which is necessary for gauge invariance. The Rashba SOC takes the form of the lattice-discretized version of $H_R=\alpha\hat{z}\cdot(\vec{\sigma\times\vec{\pi}})$ where $\vec{\pi}$ is the kinematic momentum (further discussion of this term can be found in the Supplementary Materials (SM) \cite{Schirmer2022e}). Finally, we include a periodic single-particle potential of the form
\be \label{sp_potential}
V(\vec{r}) = -\frac{V_{\rm sp}}{2} \sum_{\delta=\unit{x},\unit{y}} \cos\big(\vec{\eta}_\delta\cdot \vec{r}\big)
\ee
where $V_{\rm sp}$ is the strength of the periodic superlattice potential and $\eta_{x}$ and $\eta_{y}$ are the wavevectors of the periodic superlattice potential in the x and y directions, respectively.

\subsection{Magnetic unit cell}
A discussion of the translation symmetries of the non-interacting part of the Hamiltonian, i.e. $ \cH_0 + \cH_{\rm SO}$, is in order -- we will discuss the interacting part $\cH_{\rm I}$ below. The magnetic field in our model is chosen so that the magnetic flux through each square plaquette of the lattice is the same rational fraction $\alpha = 1/q$ of the flux quantum $\Phi_0 = h/e$. This rational fraction, and in particular the denominator $q$, determines the translation properties of the non-interacting Hamiltonian by constraining the unit cell of the system -- called the magnetic unit cell (MUC) -- to have a multiple of $q$ sites \cite{harper1955single,azbel1964energy,PhysRev.134.A1602,PhysRev.133.A1038,zak1964magnetic,Hofstadter1976}. For our calculation, we choose $q=64$ and take the MUC to be a square consisting of 
$16 \times 16$ sites. The total flux through this magnetic unit cell is four flux quanta. \added{We will take the pairing potential to have the same periodicity as the MUC. This choice of MUC allows for triangular, square, and dimerized lattice structures for the Abrikosov flux lattice.  (We note that the smallest choice would be an $8\times 8$ MUC, which encloses one flux quantum, but that does not allow a triangular lattice of Abrikosov vortices. We have also examined several cases for other MUC types at the same magnetic field, such as $4 \times 64$ with 4 flux quanta per MUC, and found that in all these cases the energies were higher than those obtained with the $16\times 16$ MUC.)}

The single-particle potential $V(\vec{r})$ is chosen to be commensurate with the MUC, and in particular we take $(\eta_x, \eta_y) = (\frac{2\pi}{8},\frac{2\pi}{8})$. Previous authors \cite{Zocher2016,Mishmash2019,Chaudhary2020} have emphasized the need to break the translation symmetry of the vortex lattice in order to realize TSC in the presence of Abrikosov vortices. Our choice of single-particle potential accomplishes this, since the unit cell for the periodic potential contains one flux quantum $h/e$. In experiments, we expect that the magnetic length can be tuned to the wavelength of the periodic single-particle superlattice potential by tuning the magnetic field.

\subsection{Mean-field theory}
A mean-field factorization of the interacting part of the Hamiltonian $\cH_{\rm I}$ is performed in the pairing channel
\begin{equation}
\begin{split}
\cH_{\rm I}=&- U\sum_{j} c^\dagger_{j,\uparrow}c^\dagger_{j,\downarrow}c_{j,\downarrow}c_{j,\uparrow} \\
\rightarrow &- U\sum_{j} \Big(\langle c_{j,\downarrow}c_{j,\uparrow} \rangle c^\dagger_{j,\uparrow}c^\dagger_{j,\downarrow}+\langle c^\dagger_{j,\uparrow}c^\dagger_{j,\downarrow}\rangle  c_{j,\downarrow}c_{j,\uparrow}\\&-\langle c^\dagger_{j,\uparrow}c^\dagger_{j,\downarrow}\rangle \langle c_{j,\downarrow}c_{j,\uparrow} \rangle \Big)
\end{split}
\end{equation}
In this work, we consider a mean-field ansatz of the form
\be
\cH_{\rm I} \rightarrow \cH_{\Delta} = - \sum_j \Delta_j e^{i2\vec{Q}\cdot\vec{r}_j}c^\dagger_{j,\uparrow}c^\dagger_{j,\downarrow} + {\rm h.c.} + \sum_j \frac{|\Delta_j|^2}{U}
\ee
where the field $\Delta_j$ is assumed to be periodic with the same periodicity as the MUC. The vector $\vec{Q}$, which we will refer to as a boost vector, is treated as a variational parameter; the optimum $\vec{Q}$ is that which leads to a mean-field groundstate with the lowest energy. Note that, due to the phase factor $e^{i2\vec{Q}\cdot\vec{r}_j}$, the pairing potential in this model is {\it not} necessarily periodic. The addition of this phase factor gives the superconducting condensate a finite momentum $2\vec{Q}$ \cite{Yang2000,Kaur2005,Agterberg2007,Agterberg2012,Mironov2017},
although the groundstate with nonzero $\vec{Q}$
does not carry a finite net current. This admits a short proof: since we are to minimize the energy with respect to $\vec{Q}$, we must have $\frac{\delta \mathcal{E}_{\rm MF}}{\delta \vec{Q}} = 0$ where $\mathcal{E}_{\rm MF}$ is the energy. However  $\frac{\delta \mathcal{E}_{\rm MF}}{\delta \vec{Q}}$ is also the net current (see Appendix \ref{total_current}). Thus it vanishes in the ground state.

The self-consistency equations are
\be
\Delta_j e^{i2\vec{Q}\cdot\vec{r}_j} = U\langle \Gamma | c_{j,\downarrow}c_{j,\uparrow} | \Gamma \rangle
\ee
where $| \Gamma \rangle$ is the mean-field groundstate. In order to solve this equation, we first make a boost transformation on the mean-field Hamiltonian. A similar transformation has been employed in studies of helical phases in non-centrosymmetric superconductors \cite{Kaur2005,Agterberg2007,Agterberg2012}. We define boost operators $U_B(\vec{Q})$ which act on the creation operators in position space as
\be
\bar{c}_{i, \sigma} \equiv U^\dagger_B(\vec{Q})c_{i, \sigma}U_B(\vec{Q}) = c_{i, \sigma}e^{i\vec{Q}\cdot\vec{r}_i},
\ee
where the bar over the operators is meant to denote operators in a boosted `frame'. The mean-field Hamiltonian, defined as $\cH_{\rm MF} = \cH_0 + \cH_{\rm SO} + \cH_{\Delta}$, is transformed using the boost operators. The action of the boosts on the terms $\cH_0$ and $\cH_{\rm SO}$ has the effect of shifting the hopping phases by $\vec{A}(\vec{r}_j) \rightarrow \vec{A}(\vec{r}_j) - \vec{Q}$. The boost has the effect on $\cH_{\Delta}$ of cancelling the phase factor $e^{i2\vec{Q}\cdot\vec{r}_j}$. In other words,
\be
\bar{\cH}_{\Delta}= U^\dagger_B(\vec{Q})\cH_{\Delta}U_B(\vec{Q})  =  - \sum_j \Delta_j c^\dagger_{j,\uparrow} c^\dagger_{j,\downarrow} + {\rm h.c.} + \sum_j \frac{|\Delta_j|^2}{U}
\ee
Thus the pairing Hamiltonian in the barred frame has the same periodicity as $\bar{\cH}_{0}$ and $\bar{\cH}_{\rm SO}$, and we can make use of the BdG formalism, using the magnetic Bloch basis, to solve the mean-field problem. The self-consistency equations in the barred frame are
\be \label{selfconeq}
\Delta_j = U \langle \bar{\Gamma} | c_{j,\downarrow}c_{j,\uparrow} |\bar{\Gamma}\rangle
\ee
where $|\bar{\Gamma}\rangle = U^\dagger_B(\vec{Q}) | \Gamma \rangle$ is the mean-field groundstate in the barred frame.

\subsection{Bogoliubov-de Gennes formalism}
The mean-field Hamiltonian in the boosted frame can be expressed in terms of a BdG Hamiltonian by first transforming it into (magnetic) momentum space using the formula $c_{j,\sigma} = \frac{1}{L} \sum_{\vec{k}} e^{i \vec{k} \cdot \vec{R}_j} c_{\tilde{j},\sigma}(\vec{k})$, where $\vec{R}_j$ is the coordinate of the origin of the MUC in which site $j$ resides and $\tilde{j}$ denotes the site {\it within the MUC} of site $j$. In other words $\vec{r}_j = \vec{R}_j +\vec{\tilde{r}}_j$ where $\vec{\tilde{r}}_j$ is the coordinate of $\tilde{j}$ (relative to the origin of the MUC). $L$ is the number of MUCs in both the $x$ and $y$ directions; we consider $L$ large enough so that the results are converged to the thermodynamic limit, where we define convergence according to the following criterion: a solution $\Delta_j^L$ to Eq. \eqref{selfconeq} with system size $L$ is said to be converged to the thermodynamic limit if it is also a solution to Eq. \eqref{selfconeq} with system size $L+2$ to within the tolerance of the iterative algorithm used to solve Eq. \eqref{selfconeq} (see the Appendix \ref{sec5} for more details on the algorithm). Generally, we find that we need $L \geq 32$ for a $16\times 16$ MUC for the pairing potential to be well converged. We also note that the Chern number (discussed below) is well-converged for much smaller system size --  $L\approx 6$ for $16\times 16$ MUC \added{using a $24 \times 24$ momentum space grid for the Chern number calculation (see section \ref{sec:chern} below)}.

The mean-field Hamiltonian may be written as
\be
\begin{split}
\cH_{\rm MF}=\frac{1}{2}\sum_{\vec{k}} \sum_{\alpha, \beta =1}^{4N_{\rm site}}\psi^\dagger_\alpha(\vec{k}) \big(H_{\rm BdG}(\vec{k}) \big)_{\alpha\beta}\psi_\beta(\vec{k})\\+\frac{1}{2}\sum_{\vec{k}}\Tr \big[h(\vec{k}) \big]+\sum_j \frac{|\Delta_j|^2}{U},
\end{split}
\ee
where
\be
\begin{pmatrix}
\psi_{\tilde{j}}(\vec{k}) \\ \psi_{\tilde{j}+N_{\rm site}}(\vec{k})  \\ \psi_{\tilde{j}+2N_{\rm site}}(\vec{k})  \\ \psi_{\tilde{j}+3N_{\rm site}}(\vec{k})
\end{pmatrix}
=
\begin{pmatrix}
c_{\tilde{j},\up}(\vec{k}) \\ c_{\tilde{j},\down}(\vec{k}) \\ c^\dagger_{\tilde{j},\up}(-\vec{k}) \\ c^\dagger_{\tilde{j},\down}(-\vec{k})
\end{pmatrix}
\ee
and

\begin{multline} \label{bdgham}
H_{\rm BdG}(\vec{k}) =\\
=\begin{pmatrix}
h(\vec{k}) & \Sigma(\vec{k}) &0 & \Delta_{\up \down}(\vec{k})  \\
\Sigma^\dagger (\vec{k}) & h(\vec{k}) &\Delta_{\down \up}(\vec{k}) & 0 \\
0 & -\Delta_{\up \down}^*(-\vec{k}) &-h^*(-\vec{k}) & -\Sigma^*(\vec{-k})  \\
-\Delta_{\down \up}^*(-\vec{k}) &0 & -\Sigma^{\rm T} (-\vec{k}) & -h^*(-\vec{k})
\end{pmatrix}
\end{multline}
is the BdG Hamiltonian. $N_{\rm site}$ is the number of sites in the MUC. Each entry in Eq. \eqref{bdgham} represents a matrix: $h(\vec{k}) $ contains all the elements for hoppings (in the barred frame), the chemical potential, and the single-particle potential; $\Sigma(\vec{k})$ contains the SOC elements (in the barred frame), and $\Delta_{\up \down}(\vec{k})$ contains the pairing elements. Fermi statistics and Hermiticity imply $\Delta^{\rm T}_{\down \up}(\vec{k})=-\Delta_{\up \down}(-\vec{k})$. In our model, $\Delta_{\up \up}(\vec{k}) =\Delta_{\down \down}(\vec{k}) =0$ and $\Delta_{\down \up}(\vec{k})=-\Delta_{\up \down}(\vec{k})={\rm diag}(\Delta_{\tilde{j}})$. \added{Note that $H_{\text{BdG}}(\vec{k}) $ is particle-hole symmetric $\mathcal{P}^{-1} H_{\rm BdG}(\vec{k}) \mathcal{P} = -H_{\rm BdG}(-\vec{k})$ with $\mathcal{P}=\mathcal{K}\tau_x$ where $\tau_x$ is a Pauli matrix acting on the particle and hole subspaces and $\mathcal{K}$ is complex conjugation. This symmetry and its implications are further discussed in the SM \cite{Schirmer2022e}}

We define the BdG quasiparticle creation operators as
\be \label{gammas}
\gamma^\dagger_{\beta}(\vec{k})=\sum_{\alpha=1}^{4N_{\rm site}} \psi^\dagger_\alpha(\vec{k}) \mathcal{U}_{\alpha \beta}(\vec{k})
\ee
By assumption, these are eigenoperators, so
\be \label{eq11}
\big[\cH_{\rm MF},\gamma^\dagger_{\beta}(\vec{k}) \big]=E_\beta(\vec{k})\gamma^\dagger_{\beta}(\vec{k}),
\ee
which implies that the columns of $\mathcal{U}_{\alpha \beta}$ must be eigenvectors of the BdG Hamiltonian
\be \label{bdgeq}
\sum_{\alpha=1}^{4N_{\rm site}}\big(H_{\rm BdG}(\vec{k}) \big)_{\delta \alpha}\mathcal{U}_{\alpha \beta}(\vec{k})=E_\beta(\vec{k})\mathcal{U}_{\delta \beta}(\vec{k}).
\ee
We identify \eqref{bdgeq} as the BdG equations. The groundstate $ | \bar{\Gamma} \rangle$ is obtained by filling the vacuum with the negative energy BdG quasiparticle states. The mean-field groundstate energy is given by
\be \label{energy_formula}
\mathcal{E}_{\rm MF}=\frac{1}{2}\sum_{E_\beta(\vec{k})<0}E_\beta(\vec{k}) +\frac{1}{2}\sum_{\vec{k}}\Tr \big[h(\vec{k}) \big]+\sum_j \frac{|\Delta_j|^2}{U}
\ee

\subsection{Self-consistency equations}
To solve the self-consistency equations, we define the Gor'kov Green's function at momentum $\vec{k}$ as
\be \label{densmat}
\begin{split}
\mathcal{G}_{\alpha\beta}(\vec{k}) &= \langle\psi^\dagger_\alpha(\vec{k})\psi_\beta(\vec{k}) \rangle\\&=\sum_{m,n=1}^{4N_{\rm site}}\mathcal{U}^\dagger_{m\alpha}(\vec{k})\langle \gamma_m^\dagger(\vec{k}) \gamma_n(\vec{k})\rangle\mathcal{U}_{\beta n}(\vec{k})\\&=\sum_{m=1}^{4N_{\rm site}}\mathcal{U}^\dagger_{m\alpha}(\vec{k})\mathcal{U}_{\beta m}(\vec{k}) f(E_m(\vec{k}))
\end{split}
\ee
where $f(E)=\frac{1}{\exp(E/k_B T)+1}$ is the Fermi-Dirac distribution function. We consider only $T=0$ in this work. The self-consistency equations can be expressed in terms of $\mathcal{G}$ as
\be 
\begin{split}
\Delta_j&=U\langle \bar{\Gamma} | c_{j,\downarrow}c_{j,\uparrow} |\bar{\Gamma}\rangle\\& = \frac{U}{L^2} \sum_{\vec{k},\vec{k}'} e^{i (\vec{k}+\vec{k}')\cdot \vec{R}_j} \langle  \bar{\Gamma}| c_{\tilde{j},\downarrow}(\vec{k}')c_{\tilde{j},\uparrow}(\vec{k})| \bar{\Gamma} \rangle\\&=\frac{U}{L^2} \sum_{\vec{k}} \mathcal{G}_{\tilde{j}+3N_{\rm site},\tilde{j}}(\vec{k})
\end{split}
\ee
This equation is solved iteratively to achieve self-consistency.

\subsection{Rationale for boosts}
\begin{figure}
\includegraphics[width=.5\textwidth]{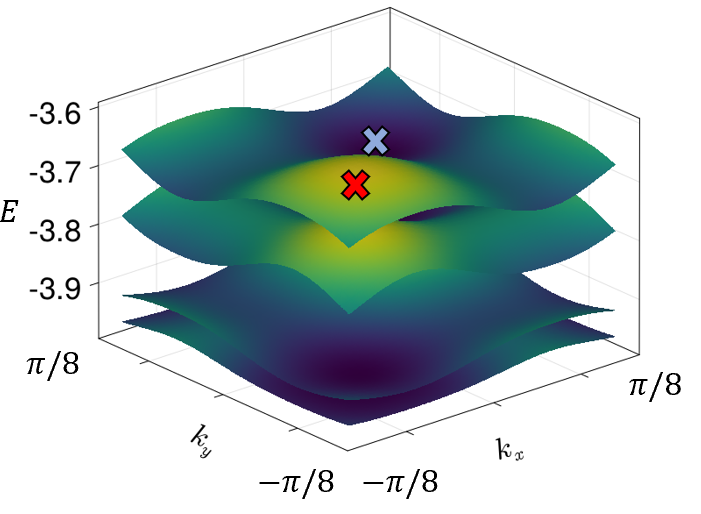}
\caption{The energies of broadened Landau levels plotted in the magnetic Brillouin zone. The single particle potential strength is $V_{\rm sp} = 0.2$ (in units where the hopping amplitude is set to unity).  The flux per plaquette $\alpha=1/64$. The Rashba SOC strength is $V_{\rm SO} = 0.1$. Because of the presence of the single-particle potential, the Landau levels become dispersive, and because of the SOC, the spin degeneracy is lifted. The maximum and minimum of the top band, which like the maxima and minima of other bands occur at finite momentum, are marked with a red and blue X, respectively} \label{spectrum}
\end{figure}

\begin{figure}
\includegraphics[width=.5\textwidth]{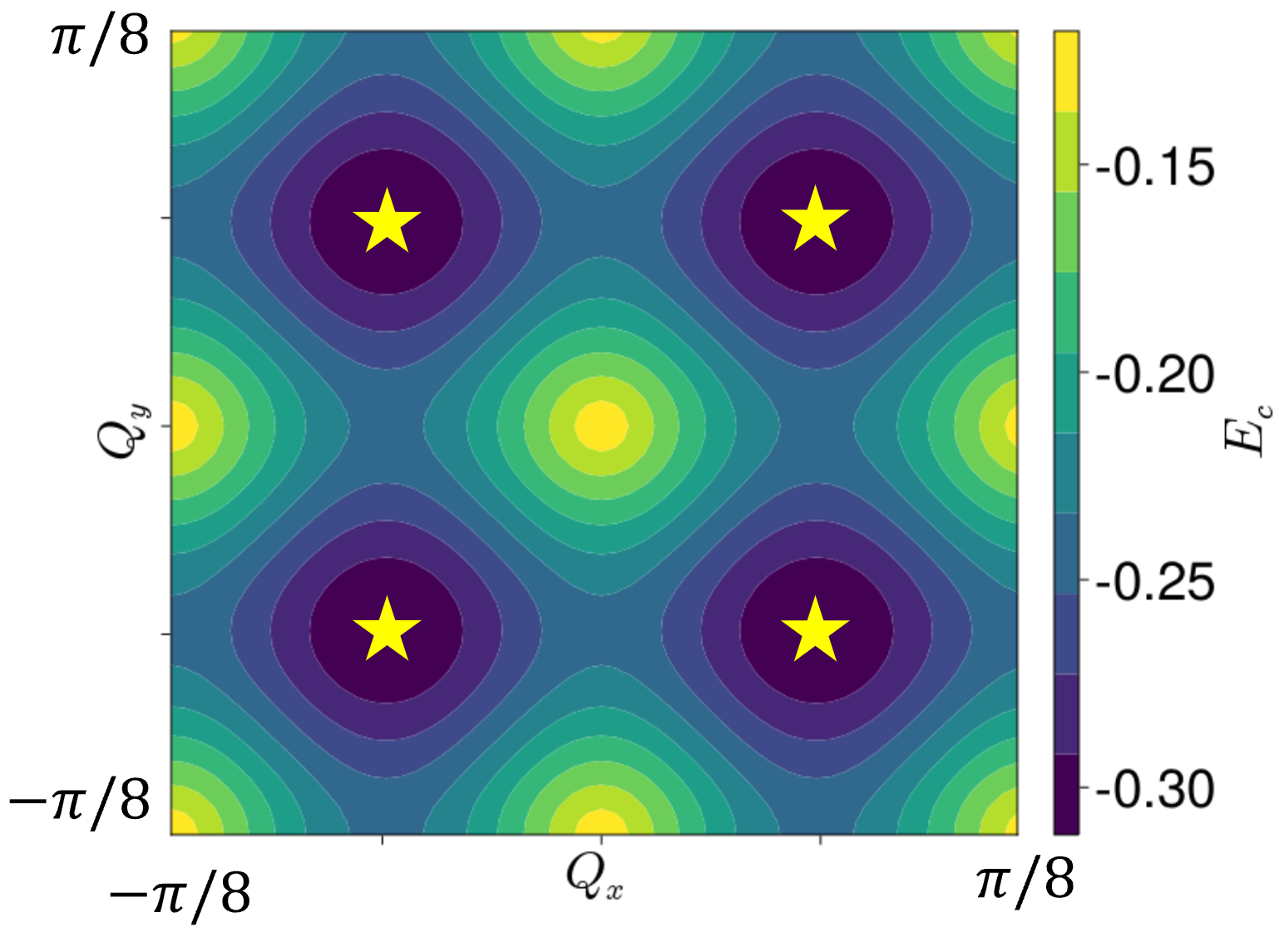}
\caption{Condensation energy as a function of the boost vector $\vec{Q}$ for $V_{\rm sp} = 0.1$, $V_{\rm SO} = 0.1$, $U=3.2$, $\mu = -3.48$ in the magnetic Brillouin zone. The stars indicate the minimum energy states. Note that states differeing by a half reciprocal lattice vector have the same energy.} \label{boost}
\end{figure}

We consider superconductivity arising from Landau levels which are broadened by the presence of the single-particle potential in Eq. \eqref{sp_potential}. A portion of the normal state spectrum (i.e. the energies of $\cH_0 + \cH_{\rm SO}$) is shown in Fig. \ref{spectrum}. Each band is spin-split due to the presence of SOC and, in contrast to a system without a single particle potential, the Landau levels are broadened and acquire a dispersion. The minima and maxima of the bands occur at non-zero momenta and consequently a pairing Hamiltonian describing pairs with finite momentum -- that is, one that pairs particles at $\vec{Q}-\vec{k}$ and $\vec{Q}+\vec{k}$ -- is required to open a superconducting gap at Fermi surfaces which appear when the chemical potential $\mu$ is tuned to within a band.  Fig. \ref{boost} shows the condensation energy as a function of the boost momentum vector $\vec{Q}$. The stars indicate momenta corresponding to self-consistent mean-field states with the lowest energy, and it is at these momenta where a full pairing gap appears. An optimal boost moves the Fermi surface so that it becomes centered at $\vec{k} = \vec{G}/2$, where $\vec{G}$ is a reciprocal lattice vector (more information about the boost transformation can be found in the Appendix \ref{sec1}).

\subsection{Chern number}\label{sec:chern}
We compute the Chern number, an integer-valued bulk topological invariant of the BdG Hamiltonian, to classify the system's topology. The non-Abelian formalism is used to avoid having to keep track of topological transitions of bands far below the Fermi level. We define the Berry connection in terms of its eigenvectors $|u^m(\vec{k})\rangle$:
\begin{equation}
A^{mn}_{\mu}(\vec{k})=i\langle u^m(\vec{k})| \partial_{\mu} |u^n(\vec{k})\rangle,
\end{equation}
where $\partial_{\mu}\equiv\partial/\partial{k_{\mu}}$, $\mu = x, y$ and $m, n$ are band indices. The Berry curvature is defined in terms of $A^{mn}_{\mu}(\vec{k})$
\begin{equation}
 F^{mn}_{\mu\nu}(\vec{k})=\partial_{\mu} A^{mn}_{\nu} - \partial_{\nu} A^{mn}_{\mu} + i \left[{A}_{\mu}, {A}_{\nu} \right]^{mn}
\end{equation}
and the Chern number is given by the integral over states with negative energy of the BdG Hamiltonian
\begin{equation} \label{hall_formula}
C = \frac{1}{2\pi} \int_{\rm MBZ} d^2k  {\rm Tr} \left[ { F}(\vec{k}) \right]_{E(\vec{k})<0}
\end{equation}
We numerically evaluate this integral using the method by Fukui, et. al \cite{fukui2005} which is highly efficient for gapped systems. The Berry curvature is determined on a grid in a discretized MBZ by defining

\begin{equation}
M^{mn}_\lambda(\vec{k_{\alpha}})=\langle u^m (\vec{k_{\alpha}})| u^n (\vec{k_{\alpha}}+\vec{e}_{\lambda}) \rangle. \label{eq19}
\end{equation}
 The points on the grid are labeled by $\vec{k_{\alpha}}$ and the spacing vectors are $\vec{e}_{\lambda}$ where $\lambda=1,2$. In terms of the link variables defined as
\begin{equation}
U_\lambda(\vec{k_\alpha})=\frac{\det M_\lambda(\vec{k_\alpha})}{|\det M_\lambda(\vec{k_\alpha})|},
\end{equation}
 the discrete Berry curvature at each point on the grid is given by
\begin{multline}
\tilde{F}(\vec{k_{\alpha}})= \ln\Big( U_1(\vec{k_{\alpha}})  U_2(\vec{k_{\alpha}}+\vec{e_1})  \\ U_1^{-1}(\vec{k_{\alpha}}+\vec{e_2}) U_2^{-1}(\vec{k_{\alpha}}) \Big)
\end{multline}
The Chern number is then given by
\begin{equation} \label{eq22}
C=\frac{1}{2\pi i}\sum_{\vec{\alpha}}\tilde{F}(\vec{k_{\alpha}}).
\end{equation}

When the Chern number is odd, the system is in the topological superconducting phase hosting non-Abelian Majorana quasiparticles. When the Chern number is even, the system does not host Majorana quasiparticles, and is in the same topological classification as a quantum Hall state. We find both possibilities in our results, but since both cases are, properly speaking, topological, we hereafter distinguish the former class of systems by referring to it as the non-Abelian topological phase.

\section{Results}

\begin{figure*}
\includegraphics[width=\textwidth]{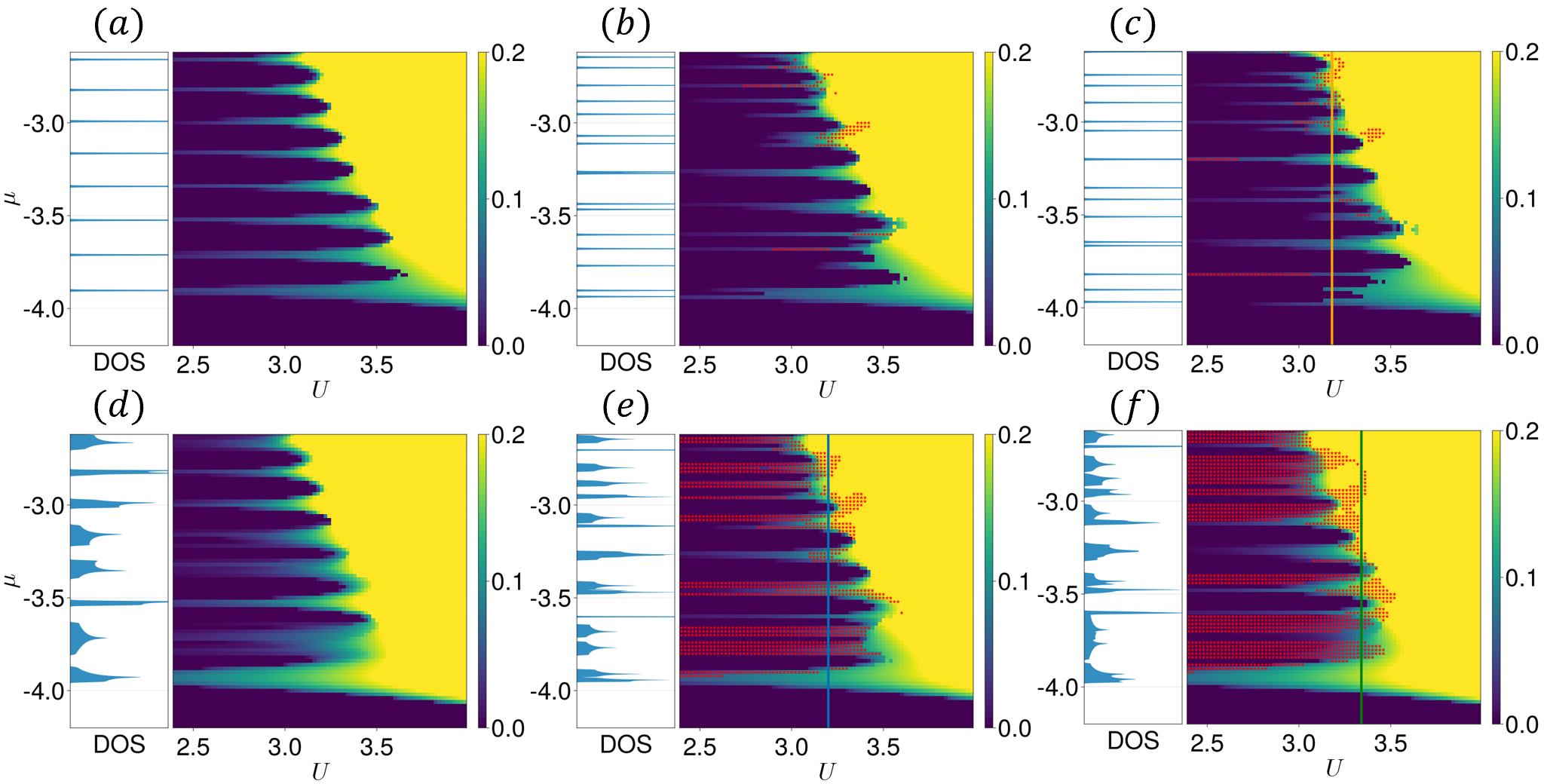}
\caption{The density of states of the system in the normal state on the left and the phase diagram on the right showing $\Delta_{\rm max}$ (colorbar), the maximum value of the modulus of the self-consistent real space pairing potential $|\Delta_j|$ in the ground state. The points where non-Abelian topological superconductivity is realized are denoted with small red stars. The parameters for each phase diagram are (a) $V_{\rm SO}=0$ and $V_{\rm sp} = 0$,  (b) $V_{\rm SO} = 0.1 \approx \hbar \omega_c /2$ and $V_{\rm sp} = 0$, (c) $V_{\rm SO} = 0.15 \approx 3/2 \hbar \omega_c$ and $V_{\rm sp} = 0$, (d) $V_{\rm SO}=0$ and $V_{\rm sp} = 0.2 \approx \hbar \omega_c$,  (e) $V_{\rm SO} = 0.1 \approx \hbar \omega_c /2$ and $V_{\rm sp} = 0.1\approx \hbar \omega_c /2$, (f) $V_{\rm SO} = 0.1 \approx \hbar \omega_c/2$ and $V_{\rm sp} = 0.2\approx \hbar \omega_c $. The self consistency equations for all systems are solved, and the systems with a finite single-particle potential are boosted to an optimum value of $\vec{Q}$. } \label{pd}
\end{figure*}

\begin{figure*}
\includegraphics[width=\textwidth]{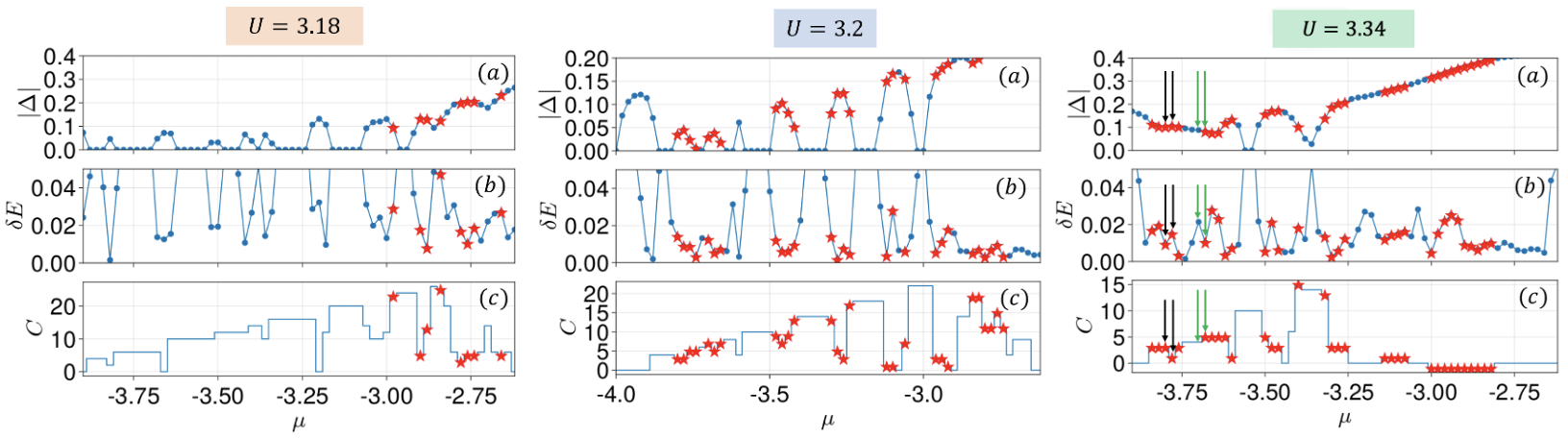}
\caption{(a) The maximum absolute value of the pairing potential; (b) the BdG spectral gap ($\delta E$); and (c) the Chern number (C) for systems with chemical potentials along the orange line in Fig. \ref{pd} (c) at $U=3.18$ (left panel), the blue line in Fig. \ref{pd} (e) at $U=3.2$ (middle panel), and the green line in Fig. \ref{pd} (f) at $U=3.34$ (right panel). The blue dots and red stars indicate systems we have studied. Blue dots correspond to systems with even Chern number and red stars correspond to systems with odd Chern number. The transitions indicated by the black and gree arrows in the right panel are discussed in detail in the SM \cite{Schirmer2022e}.} \label{full-data}
\end{figure*}

\begin{figure*}
\includegraphics[width=.9\textwidth]{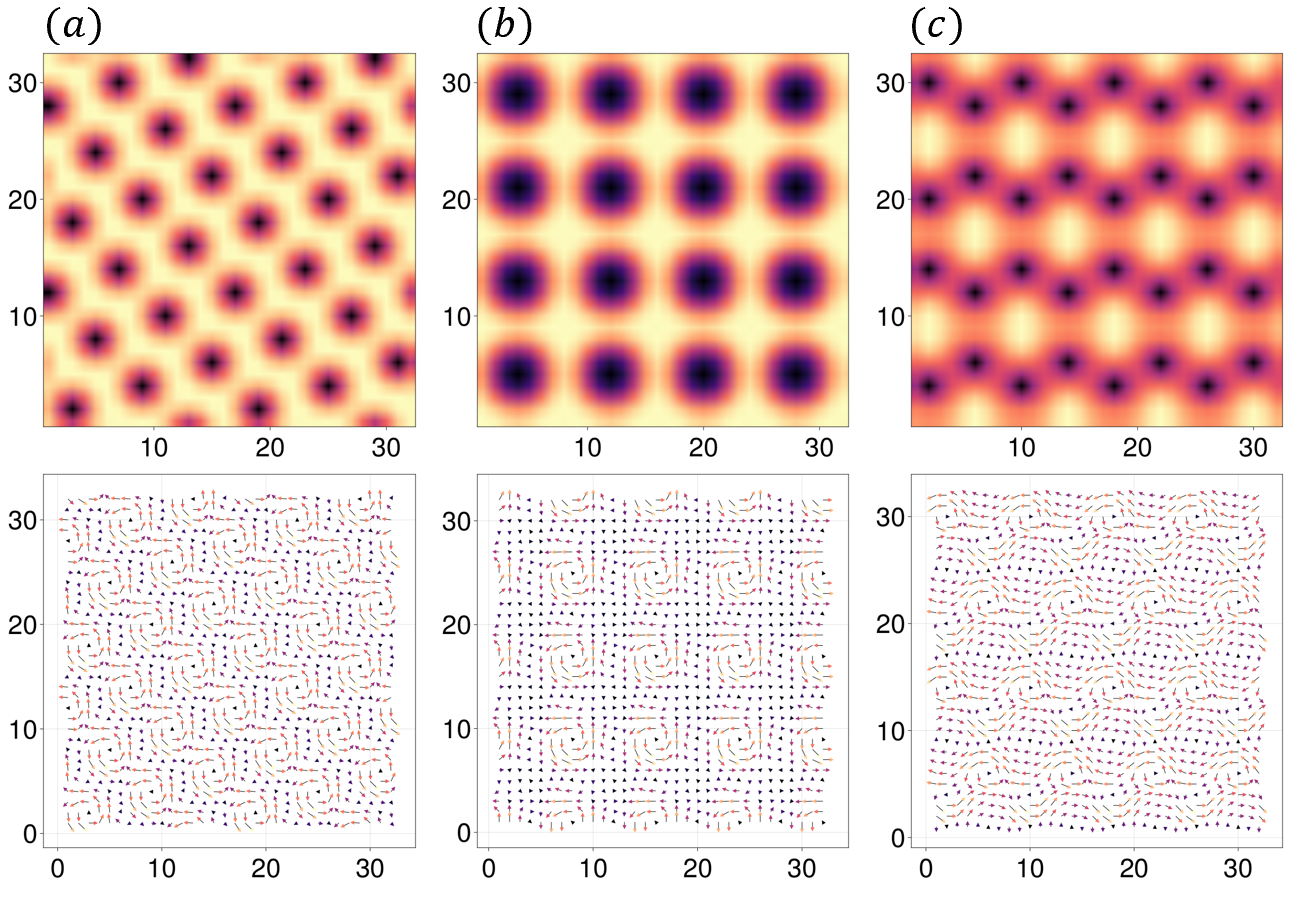}
\caption{Typical vortex lattice structures from the solutions to the self-consistent mean field equations (upper panels) with their current textures (lower panels). (a) The vortex lattice at $V_{\rm sp} = 0$, $V_{\rm SO} = 0.1$, $U=3.52$, $\mu=-3.92$. In the lowest few Landau levels, the self-consistent solutions is a triangular lattice of Abrikosov vortices carrying flux $h/2e$. (b) The vortex lattice at $V_{\rm sp} = 0.2$, $V_{\rm SO} = 0.1$, $U=3.34$, $\mu=-3.66$. The single-particle potential distorts the vortex lattice, drawing the vortices to areas of high potential energy. Typically in the topological region, the system forms a lattice of giant vortices. (c) The vortex lattice at $V_{\rm sp} = 0$, $V_{\rm SO} = 0.1$, $U=3.24$, $\mu=-3.08$. Other kinds of vortex order form at higher chemical potential (Landau level filling factor $\nu\approx 10$). For example, the vortices may form a (distorted) honeycomb lattice. } \label{vortex}
\end{figure*}

We explored phase diagrams for a multitude of values of SOC and periodic potential strength for small Landau level filling factors (up to Landau level index $n=8$), as shown in Fig. \ref{pd}. The self-consistent mean field equations were solved for each point on the phase diagrams, which each consist of 6400 ($80 \times 80$) points. For each point, guesses consisting of random complex numbers at each site within the MUC, as well as uniform real numbers, were used to start the iterative algorithm, and $16 \times 16$ (with flux 4 $h/e$) MUCs were used. The total energy was calculated for each solution (Eq. \ref{energy_formula}) and the solution with the minimum energy was taken to be the groundstate solution.  The color on the phase diagrams denotes $\Delta_{\rm max}$, the maximum value of the modulus of the self-consistent real space pairing potential $|\Delta_j|$ in the ground state. The points marked with red stars are points where the system is in the non-Abelian phase, i.e. the Chern number is an odd integer. The corresponding density of states (DOS) in the normal state is shown to the left of each phase diagram. Let us now consider four different cases:

{\it No SOC or superlattice potential:} Fig. \ref{pd} (a) shows the phase diagram when both the SOC and single-particle potential are set to zero. The single particle spectrum consists of spin-degenerate Landau levels, which have a band width on the order of $10^{-14}$ and their spacing (the cyclotron energy $\hbar \omega_c$) is approximately 0.2. The phase boundary separating non-superconducting and superconducting regions of the phase diagram displays substantial oscillatory behavior due to the Landau level structure. At lower interaction strength, superconductivity is present when the chemical potential is tuned to precisely the energy of a Landau level, but disappears when it is tuned to the gaps between different Landau levels. We hereafter refer to regions of enhanced superconductivity at Landau level energies as Landau level spikes. At larger interaction strength ($U \geq 3.5$), superconductivity is strongly augmented, and oscillations due to the Landau level structure disappear. We will refer to this region as the strong pairing regime. The non-Abelian phase of topological superconductivity does not appear in the system without a periodic potential and SOC. Instead, the Chern numbers for all of the points shown are even integers. This follows from the spin degeneracy of the energy levels (see Appendix \ref{sec9}). In the weak interaction portion of the phase diagram, when the energies are between Landau level spikes, the Chern number is $2\nu$ where $\nu$ is the filling factor, consistent with the system being in a quantum Hall insulating phase. Within the Landau level spikes, the Chern number is an even integer, starting at $2\nu$ when the interaction is weak, and decreases in even increments as the interaction is increased, until the strong pairing regime where the Chern number is zero. It is interesting to note that superconductivity and nontrivial topology coincide on the Landau level spikes, although the topology is Abelian in nature.

 {\it SOC with no superlattice potential:} Figs. \ref{pd} (b) and (c) show the phase diagrams of systems without the single particle potential but with SOC strength $V_{\rm SO} = 0.1$ and  $V_{\rm SO} = 0.15$, respectively. Due to nonzero SOC, the spin degeneracy of the Landau levels is lifted, and the energies are given approximately by
\be
\begin{split}
\epsilon_{0} &= \frac{1}{2}\hbar \omega_c -4 \\
\epsilon_{m,\pm} &= \hbar \omega_c \Bigg( m \pm \sqrt{\frac{\pi}{4}\frac{V_{\rm SO}^2}{\hbar^2 \omega_c^2}m+\frac{1}{4}} \Bigg)-4~~~~ (m>0)
\end{split}
\ee
and accordingly there are Landau level spikes at these energies. Interestingly, we find small regions where the non-Abelian phase of topological superconductivity appears, most markedly at higher filling factors (at or around $\mu = -3$). For weaker interaction strength the Chern number at the non-Abelian points is an odd integer close to twice the filling factor, and decreases as the interaction strength is increased, until the Chern number becomes zero in the strong-pairing regime. The vortex structure at these points will be discussed below. We remark that we do not find non-Abelian TSC arising from superconductivity in the lowest Landau level, which has been the subject of recent studies \cite{Zocher2016,Mishmash2019,Chaudhary2020}.

Plots of the maximum absolute value of the pairing potential, the BdG spectral gap ($\delta E$), and the Chern number (C), for chemical potentials along the orange line in Fig. \ref{pd} (c), at fixed $U = 3.18$, are shown in the left panel of Fig. \ref{full-data}. Points where the system is in the non-Abelian phase are denoted with red stars. Oscillations in Fig. \ref{full-data} (a) (left panel) which correspond to the Landau level spikes. The spectral gaps in Fig. \ref{full-data} (b) (left panel) come in two varieties: superconducting gaps and quantum Hall insulating gaps. The values of the quantum Hall insulating gaps, which occur when $\Delta = 0$, are given by $2{\rm min}(|\mu - \epsilon_{m \pm}|)$, and are much larger than superconducting gaps. When $\Delta \neq 0$, the gaps are superconducting gaps. Generally, the superconducting gaps are small, about an order of magnitude less than the cyclotron energy, in the regime where non-Abelian TSC appears. 
As seen in Fig. \ref{full-data} (c) (left panel), the Chern number exhibits discrete jumps as the chemical potential $\mu$ is tuned. These topological phase transitions come in two varieties: (1) transitions where the change in the Chern number is associated with a closing of the spectral gap, 
and (2) first-order transitions, where the Chern number changes without gap closing; 
these are associated with a change in the vortex lattice structure.
This is discussed in detail with examples in the SM \cite{Schirmer2022e}. 

{\it Superlattice potential with no SOC:} Fig. \ref{pd} (d) shows the phase diagram for the system with periodic potential strength $V_{\rm sp} = 0.2$ but without SOC. The optimum boost vector is used for each point on the phase diagram. Despite the Landau levels becoming significantly broadened, as can be seen in the DOS and in the corresponding smearing of the Landau level spikes, the non-Abelian phase does not appear, and the system transitions from an Abelian TSC or quantum Hall insulating phase at weak interaction to trivial superconductivity at stronger interaction. It comes as no surprise that we do not find the non-Abelian phase when there is a spin degeneracy. This follows from Eqs. (19)-(21) and is established in detail in Appendix \ref{sec9}).

{ \it Both superlattice potential and SOC:} Finally, including both a single-particle potential and SOC leads to a considerable enhancement of non-Abelian TSC in the phase diagram. Fig. \ref{pd} (e) shows the phase diagram for $V_{\rm sp} = 0.1$ and $V_{\rm SO} = 0.1$, and Fig. \ref{pd} (f) shows the phase diagram for $V_{\rm sp} = 0.2$ and $V_{\rm SO} = 0.1$. Non-Abelian TSC, which is marked using small red stars in the phase diagram, generally appears at the boundary between the quantum Hall insulator and superconducting phases, particularly along broadened Landau level spikes. 
Plots of the maximum absolute value of the pairing potential, the BdG spectral gap ($\delta E$), and the Chern number (C), for parameters along the blue line in Fig. \ref{pd} (e) and the green line in Fig. \ref{pd} (f), are shown in the middle panel of Fig. \ref{full-data} and the right panel of Fig. \ref{full-data}, respectively. Points where the system is in the non-Abelian phase are denoted with red stars. Oscillations can be seen in Fig. \ref{full-data} (a) (middle panel), and to a lesser degree in Fig. \ref{full-data} (a) (right panel), which reflect the Landau level spikes. The spectral gaps are about an order of magnitude less than the cyclotron energy, at points where non-Abelian TSC appears. Again, we find topological phase transitions that come in the two varieties mentioned above (gap-closing and first-order).

\subsection{Vortices and Majoranas}
Due to the applied perpendicular magnetic field, the pairing potential experiences orbital frustration and Abrikosov vortices develop at the centers of which the pairing potential is zero. We find a plethora of configurations formed by the Abrikosov vortices, some of which are quite unexpected. As anticipated, Abrikosov vortices form a triangular lattice in the $n=0$ and $n=1$ Landau levels when the SOC and single-particle potential are absent. In addition we found square vortex lattice solutions with approximately $1-2\%$ higher energy than the triangular lattice solutions. With SOC, but without single-particle potential, a triangular lattice of vortices forms when the chemical potential is in the lowest 5 spin-split Landau levels -- an example is shown in Fig. \ref{vortex} (a). In both cases, at higher chemical potential, the vortex structure is far more diverse. Of particular note is the vortex structure at topological points in Figs. \ref{pd} (b) and (c), a typical example of which is shown in Fig. \ref{vortex} (c). Here the vortices form a honeycomb lattice, although not all the bonds are identical. Other interesting structures appear in the non-topological regions in the strong pairing regime at chemical potential above the first few Landau levels. We show these in Fig. \ref{vortex2} of the Appendix. Finally, when a single-particle potential is included, the vortices are distorted at all chemical potentials. In particular, vortices are drawn to locations where the single-particle potential is high. The vortices form dimer pairs, or combine to form giant vortices, consisting of two Abrikosov vortices. Topological points in Figs. \ref{pd} (e) and (f) typically host the latter configuration (Fig. \ref{vortex} (b)).

 As revealed by the finite gap when the system is in the non-Abelian topological phase, we do not find Majorana modes with zero energy in the bulk of the system. Instead the Majorana modes, which are thought to reside at vortex cores, hybridize with their neighbors and open a gap \cite{grosfeld2006electronic,Cheng2009,Cheng2010,Ludwig2011,kraus2011majorana,Lahtinen2012,Laumann2012,Zhou2013,Biswas2013,silaev2013majorana,Murray2015,Liu2015,Ariad2015,Li2018,Mishmash2019}. Nevertheless, because of the non-trivial Chern number and the bulk-boundary correspondence, we expect there to be chiral Majorana edge modes at the boundaries of samples in this phase.

\section{Discussion}
In this work, we investigated the emergence of TSC from topologically flat bands broadened by a superlattice potential. The robust SOC and the superlattice potential are essential components of our model. We comment on the relationship of our findings to potential materials platforms. A variety of systems have been discovered to exhibit superconductivity with strong SOC, including iron-based superconductors, interfaces of insulating oxides, and transition metal dichalcogenides \cite{Bauer2012,Smidman2017}. 
More detailed, material-specific calculations would be required in the future to establish any specific candidate material, which is beyond the scope of this paper.
    
{Our calculations have focused on a specific region of parameter space, which in our model contains a large number of parameters, and we would like to clarify this region to better explain the relationship between our results and any class of candidate materials. In the following discussion, we will express all energies in our calculations in units of the hopping amplitude $t$, which we have set to unity in our model. The cyclotron energy in our calculations is approximately 0.2. 
  We have selected SOC strengths of 0.1 in the phase diagrams shown in figures 3(e) and 3(f) of the main article.
  Our calculations are therefore directly applicable when the SOC strength is less than but close to the cyclotron energy. 
  The cyclotron energy in real materials is determined by the effective mass $m^*$, as well as the strength of the applied magnetic field $B$ through the formula$$\hbar \omega_c = \hbar\frac{e B}{m^*}$$
  As an example, the cyclotron energy for bare electrons in a magnetic field of 20 T is estimated to be about 2 meV. 
  Based on this information, we will now discuss some materials platforms individually. This is by no means intended to be an exhaustive list. A summary of relevant parameters for the following systems is provided in Table \ref{table:1}}

\begin{table*}
\centering
\begin{tabular}{||c c c c c||} 
 \hline
 Material & $V_{\rm SO}$ (meV) &  $\Delta$ (meV)  & $m^*$ ($m_e$) & $\hbar \omega_c$ (meV)\\ [0.5ex] 
 \hline
 \hline
  Iron-based thin films& 5-25 \cite{Borisenko2016} &  5-25 \cite{Borisenko2016,Haindl2021} &10 \cite{Haindl2021}&   .2 \\ 
  LaAlO$_3$/SrTiO$_3$ interfaces & 1-10 \cite{Caviglia2010, Shanavas2014}  &  0.04 \cite{Singh2022} & 3 \cite{Caviglia2010, Mattheiss1972} & .7 \\ TMDs \cite{Xi2015,Xi2016,Ugeda2016,Wang2017, Barrera2018, Xiao2012,Zhu2011,Kosmider2013} & 50-150 (Ising)  & $\sim$ 1 &$\sim$ 0.4 & $\sim$ 6 \\
  
 \hline
\end{tabular}
\caption{{Estimates of key parameters in some select classes of spin-orbit coupled superconductors. $V_{\text{SO}}$ is the SOC strength, determined by the spin-splitting at the Fermi level. $\Delta$ denotes the superconducting gap. $m^*$ is the effective electron mass, expressed in terms of the bare electron mass $m_e$. $\hbar \omega_c$ is the cyclotron energy at a magnetic field of 20 T, computed using the material's effective electron mass.}}
\label{table:1}
\end{table*}

  { \textit{Iron-based superconductors.} The semimetallic iron-pnictides and iron-chalcogenides are a group of high-temperature superconductors that have generated significant enthusiasm in the condensed matter community since their discovery in 2008. These discoveries have ushered in a remarkable era of superconductivity known as the ``iron age." The Rashba SOC in iron-based superconductors is found to be in the range $5-25$ meV \cite{Borisenko2016}.  In addition, the pairing gap is found to be in this same range \cite{Borisenko2016}. However, due to the strong interactions in these materials, the effective mass may be as high as 20 times the bare electron mass \cite{Haindl2021, Baglo2022}. In order to model these materials, it may be preferable to take the SOC to be significantly greater than the cyclotron energy. It is anticipated that increasing the SOC will not diminish TSC, and we have performed additional calculations to support this claim. We show results for Rashba SOC strength of $V_{\text{SO}} = 1.0$ (5 times the cyclotron energy) in Fig. \ref{fig:high-soc}.
  }
  
  {
\textit{Oxide interfaces.} Superconductivity has been reported in the electron gas formed at the interfaces of LaAlO$_3$ and SrTiO$_3$ \cite{Reyren2007}. In addition, tunable Rashba SOC has been explored in this system with Rashba SOC strength in the range $1-10$ meV \cite{Caviglia2010, Shanavas2014}.  The superconducting gap is found to be approximately 40 $\mu$eV \cite{Singh2022} and the effective mass is $m^* = 3m_e$ where $m_e$ is the electron mass \cite{Caviglia2010, Mattheiss1972}. Given these parameters, we are optimistic about the applicability of our results to this material platform.}

{\textit{Transition metal dichalcogenides.} 
  Superconductivity has been reported in layers of transition metal dichalcogenide (TMD) materials, resulting in 2D superconductivity. Examples of materials in this class include MoS$_2$, MoSe$_2$, WS$_2$, WSe$_2$, and NbSe$_2$. Layers of these materials break inversion symmetry within the plane, giving rise to Ising, instead of Rashba, SOC. Ising SOC tends to lock the electron spins perpendicular to the plane in a momentum-dependent way. In TMDs, this manifests as an effective Zeeman field with opposite signs at the K and K' points of the hexagonal Brillouin zone. Because the SOC is not Rashba, our calculations are not readily applicable to these systems. However, Ising superconductivity still gives rise to a momentum-dependent spin-splitting \cite{Hsu2017}, so it may be worthwhile in the future to investigate the possibility of TSC using Ising SOC instead of Rashba SOC, as we did in our calculations. We point out that a crossover from Ising to Rashba SOC has also been recently observed in NbSe$_2$/Bi$_2$Se$_3$ heterostructures \cite{Yi2022}. We hope that our study will stimulate inquiry along this direction.}

  {
In discussing the inclusion of a superlattice potential, the lattice period of the superlattice is an additional relevant parameter. Our calculations have focused on the case where the superlattice period is comparable to the magnetic length $\ell_B$. The magnetic length is determined by the strength of the applied magnetic field $B$ through the formula}
$$
\ell_B = \sqrt{\frac{\hbar}{eB}} \approx \frac{25}{\sqrt{B~\text{in Tesla}}}~\text{nm}
$$

{For reference, $\ell_B \approx $ 6 nm for a magnetic field $B = 20$ T. Below, we list systems where a superlattice potential may be achievable.}

{\textit{Gating periodic patterned dielectric substrates.} The engineering of superlattices has been reported using patterned dielectrics, with strengths up to 50 meV and lattice periods as small as 35 nm \cite{Forsythe2018,Yankowitz2018,Shi2019,Xu2021,Li2021a}. 
In order to make direct connection with our results, let us suppose  that the magnetic field $B \sim 20$ T, so that $\ell_B \approx $ 6 nm. Thus, for our results to directly apply, it may be desirable to reduce the lattice period of the superlattice relative to what has been currently achieved. On physical grounds, however, we expect that TSC is readily achievable with large superlattice periods. This is because the superlattice potential brings about TSC through a distortion of the vortex lattice. This distortion is clearly possible with large superlattice periods. Therefore, we see gating periodic patterned dielectric substrates as a very promising route to inducing TSC using a superlattice potential.}

{\textit{Moiré patterns.} A moiré pattern can be formed by stacking layers of materials with a relative twist, in which case the lattice vectors of the two layers are relatively rotated by an angle $\theta$, or by stacking materials with slightly mismatched lattice periodicities. The moiré pattern leads to a periodic potential with a periodicity determined by $\theta$ with a strength on the order of hundreds of meV \cite{Tilak2022} in bilayer MoS$_2$. 
Exciting recent progress has been reported in the fabrication of Van der Waals superconducting heterostructures wherein moiré patterns have been achieved \cite{zhou_stack_2023}.
Since materials from this class typically host Ising spin-orbit coupling, instead of Rashba spin-orbit coupling, further calculations would be required to make a firm connection between our proposal and these materials.}
Finally, we comment on the connection of the parameters of our ``effective" model to an experimental system. 
Our phase diagrams shows TSC at rather large $U\sim3$, an order of magnitude larger than the cyclotron energy. 
However, the gap $\Delta$ for these values of $U$ is roughly of the same order of magnitude as the cyclotron energy. 
In experiments, it is the pairing gap $\Delta$ that is typically measured rather than the coupling strength $U$ between electrons. 
Therefore, in order to relate our findings to experimental data, the parameter $U$ in our model should be fixed by matching the 
gap $\Delta$ to its experimental value. We have listed some values of $\Delta$ for systems that we discussed in the previous paragraphs in Table \ref{table:1}.

\vspace{5 mm}

Previously, the helical phase of superconductors was discussed in the context of non-centrosymmetric superconductors using Ginzburg-Landau theory \cite{Agterberg2003,Kaur2005,Agterberg2007,Agterberg2012}. Certain terms, known as Lifshitz invariants, can be eliminated from the free energy by performing a helical (or in our language, boost) transformation when inversion symmetry is broken. This is a similar procedure to the one we employ, albeit in a microscopic theory. However, even in the absence of Rashba SOC when inversion symmetry is preserved, helical transformations are necessary in our model because of the presence of a magnetic field and superlattice potential.

 This work has demonstrated the dramatic influence on the vortex lattice structure by a superlattice potential, in the case of $\ell_{\rm B}/a_{\rm M} \approx 1$, where $\ell_{\rm B}$ is the magnetic length and $a_{\rm M}$ is the superlattice vector. Understanding the vortex lattice structure for other values of $\ell_{\rm B}/a_{\rm M}$, particularly irrational values, and the corresponding topological properties in this model, will be left for future work.
 
 Finally, we would like to discuss some issues related to realizing intrinsic quantum Hall superconductivity. It has been over 50 years since quantum Hall superconductivity was first explored \cite{Rajagopal1966}, however, due to the requirement of low densities or very high magnetic fields, it has not been achieved in experiments. In addition, theoretical studies on intrinsic quantum Hall superconductivity, including this one, have relied on mean-field theories which do not allow for all possible instabilities. In a real material, the interaction will be a mixture of attractive and repulsive, and this may lead to competition between other correlated phases including, but not limited to, FQHE, stripe phases, charge density waves, and spin density waves. The question of whether other strongly correlated phases may arise, particularly in the case of flat LLs, is an interesting question beyond mean-field theory, which we leave for the future.

\section*{Acknowledgements}
We thank Liang Fu and Luiz Santos for thought-provoking comments. JS and JKJ were supported in part by the U. S. Department of Energy, Office of Basic Energy Sciences, under Grant no. DE-SC-0005042. We acknowledge Advanced CyberInfrastructure computational resources provided by The Institute for CyberScience at The Pennsylvania State University.

\begin{appendices}

\section{Derivation of the Expression for the Net Current} \label{total_current}
In the main text, we pointed out that the condition that the energy be minimized with respect to $\vec{Q}$ implies that the net current vanishes, since the net current is $\vec{J}_{\text{net}} = \frac{\delta \mathcal{E}_{\rm MF}}{\delta \vec{Q}}$. In this appendix, we derive this expression for the net current.

The current density across a link starting at position $\vec{r}_i$ along the $\mu$-direction ($\mu = x,~y$) is
 \be
 J_{\mu}(\vec{r}_i)=-\frac{\partial \cH_{\rm MF}}{\partial A_\mu(\vec{r}_i)}  \ee
 where $A_\mu(\vec{r}_i)$ is the hopping phase across the same link and $\cH_{\rm MF}$ is the mean-field Hamiltonian. The net current in the $\mu$-direction is the sum over all of the sites of $J_{\mu}(\vec{r}_i)$
 \be
 (\vec{J}_{\text{net}} )_{\mu} = - \sum_i \frac{\partial \cH_{\rm MF}}{\partial A_\mu(\vec{r}_i)}
 \ee
 As we have pointed out in the main text, shifting $\vec{Q}$ is equivalent to shifting all of the hopping phases by a position-independent amount $\vec{A}(\vec{r}_j) \rightarrow \vec{A}(\vec{r}_j) - \vec{Q}$. Under a small shift $\delta \vec{Q}$, the change in $\cH_{\rm MF}$, to first order in $\delta \vec{Q}$, is given by
 \begin{equation}
 \begin{split}
 &\cH_{\rm MF} \Big( \{ A_\mu(\vec{r}_i) - \delta Q_\mu \} \Big) \\ = &\cH_{\rm MF} \Big( \{ A_\mu(\vec{r}_i)\} \Big) - \sum_{\mu = x,y}\delta Q_\mu \sum_{i} \frac{\partial \cH_{\rm MF}}{\partial A_\mu(\vec{r}_i)} \\ 
 = &\cH_{\rm MF} \Big( \{ A_\mu(\vec{r}_i)\} \Big) + \delta \vec{Q} \cdot  \vec{J}_{\text{net}} \end{split}
 \end{equation}
 Therefore
 \be
  (\vec{J}_{\text{net}} )_{\mu} = \frac{\delta \cH_{\rm MF}}{\delta Q_\mu}
 \ee
 Taking the groundstate expectation value (and using the Feynman-Hellmann theorem), we find $\vec{J}_{\text{net}} = \frac{\delta \mathcal{E}_{\rm MF}}{\delta \vec{Q}}$.

\section{Boosts}\label{sec1}
 \begin{figure}
      \centering
      \includegraphics[width=\linewidth]{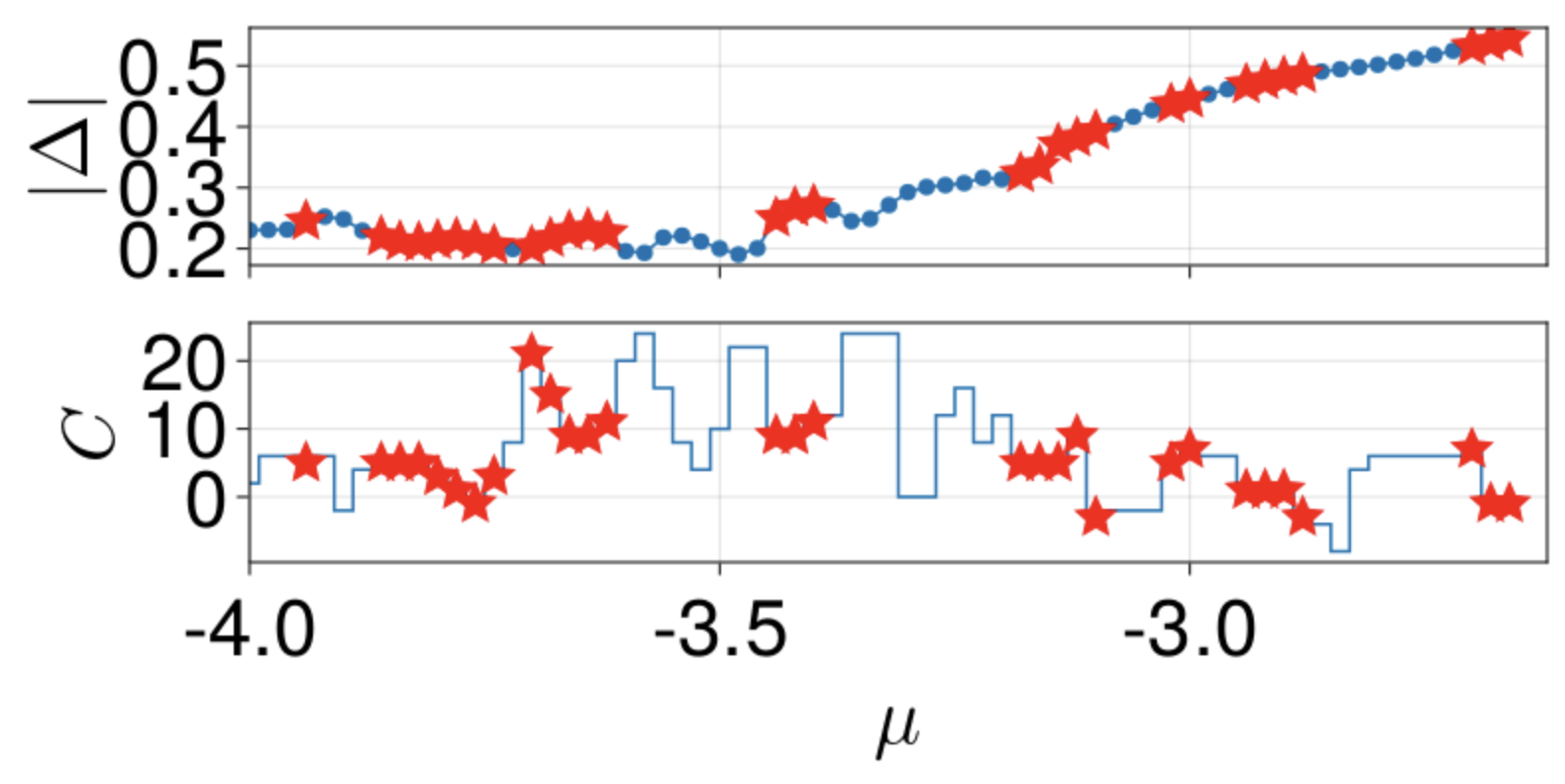}
      \caption{{Topological superconductivity still appears (denoted with red stars) when the spin-orbit coupling strength is high. In this figure, the spin-orbit coupling strength is $V_{\text{SO}} = 1.0$, the superlattice potential strength is $V_{\text{sp}} = 0.2$ and the interaction strength $U = 3.98$. The Zeeman energy is zero.}}
      \label{fig:high-soc}
  \end{figure}
Non-relativistic physics is invariant under Galilean boosts, whereas relativistic physics is invariant under Lorentz boosts. In classical mechanics, the transformation rule for a Galilean boost by velocity $\vec{v}$ is
\be
\begin{split}
\vec{\bar{x}} &= \vec{x}+\vec{v}t \\
\bar{t} &= t
\end{split}
\ee
This increases the velocity of every particle in the system by $\vec{v}$.

Continuing to the quantum mechanical case, suppose we have $N$ identical fermions with mass $m$ and momenta $\vec{k}_1 \neq  \vec{k}_2 \neq \cdots \neq \vec{k}_N$. We will neglect spin for the moment since Galilean boosts do not affect the spins of the particles. The action of a Galilean boost is given by an operator $U_B(\vec{v})$ which acts on a first-quantized momentum eigenstate as follows
\be
U_B(\vec{v}) |\vec{k}_1, \dots, \vec{k}_N \rangle = |\vec{k}_1 + m\vec{v}, \dots, \vec{k}_N + m\vec{v} \rangle
\ee
To deduce the operator form of $U_B(\vec{v})$, we determine its action on position space eigenkets. In fact:
\be
\begin{split}
&U_B(\vec{v})  |\vec{x}_1, \dots, \vec{x}_N \rangle \\= &\frac{1}{L^N} \sum_{\{ \vec{k} \}} \exp{\Big( -i\sum_i \vec{k}_i\cdot \vec{x}_i\Big)}U_B(\vec{v}) |\vec{k}_1, \dots, \vec{k}_N \rangle \\
= &\frac{1}{L^N} \sum_{\{ \vec{k} \}} \exp{\Big( -i\sum_i \vec{k}_i\cdot \vec{x}_i\Big)}|\vec{k}_1 + m\vec{v}, \dots, \vec{k}_N + m\vec{v} \rangle \\
= &\frac{1}{L^N} \sum_{\{ \vec{k} \}} \exp{\Big( -i\sum_i (\vec{k}_i - m \vec{v})\cdot \vec{x}_i\Big)}|\vec{k}_1, \dots, \vec{k}_N \rangle \\
= &\exp{\Big( i m \vec{v}\cdot \sum_i  \vec{x}_i\Big)}|\vec{x}_1, \dots, \vec{x}_N \rangle,
\end{split}
\ee
so we see that the position kets are eigenstates of the boost operator. This also allows us to deduce that the operator form of $U_B(\vec{v})$ is
\be
U_B(\vec{v}) =  \exp{\Big( i m \vec{v}\cdot \vec{X} \Big)} = \exp{\Big( i M \vec{v}\cdot \vec{R}_{\rm CM} \Big)}
\ee
where $\vec{X} = \sum_i  \vec{x}_i$ is the sum of the positions of all the particles, $M = \sum_i m_i = N m$ is the total mass of the system, and $\vec{R}_{\rm CM}$ is the center of mass coordinate of the system. We see that $U_B(\vec{v})$ is unitary. Using the form $U_B(\vec{v}) =  \exp{\Big( i m \vec{v}\cdot \vec{X} \Big)} $, we can write $U_B(\vec{v})$ in second quantized representation as
\be
U_B(\vec{v}) =  \exp{\Bigg(i m \vec{v}\cdot \int d^2\vec{r} ~ \vec{r} c_{\vec{r}}^\dagger c_{\vec{r}} \Bigg)}
\ee

For notational consistency with the main text, we will set $m\vec{v}=\vec{Q}$ and define $U_B(\vec{v})  \equiv U_B(\vec{Q})$. Now consider the action of $U_B(\vec{Q})$ on the field operators. We have (we may use the Baker-Campbell-Hausdorff formula to show this)
\be \label{eq42}
\bar{c}_{\vec{r}} := U^\dagger_B(\vec{Q})c_{\vec{r}}U_B(\vec{Q}) = c_{\vec{r}}e^{i\vec{Q}\cdot\vec{r}}
\ee
where barred operators are operators for the ``boosted frame''. As an example, let us apply this boost to a system of free spinless fermions described by the Hamiltonian
\be \label{eq43}
\cH = \sum_{\vec{k}}\big(\epsilon(\vec{k}) - \mu\big)c^\dagger_{\vec{k}}c_{\vec{k}}
\ee
where $c_{\vec{k}} = \frac{1}{L}\sum_{\vec{r}}e^{-i\vec{k}\cdot\vec{r}}c_{\vec{r}}$. Note that by Eq. \eqref{eq42}, we have 
\be
U^\dagger_B(\vec{Q})c_{\vec{k}}U_B(\vec{Q}) = \bar{c}_{\vec{k}} =  c_{\vec{k}-\vec{Q}}
\ee
Suppose we view the system in Eq. \eqref{eq43} in a frame boosted by $\vec{Q}$, then the Hamiltonian $\bar{\cH}$ in this frame is
\begin{equation}
\begin{split}
\bar{\cH }=   U^\dagger_B(\vec{Q}) \cH U_B(\vec{Q}) =  \sum_{\vec{k}}\big(\epsilon(\vec{k}) - \mu\big)\bar{c}^\dagger_{\vec{k}}\bar{c}_{\vec{k}} \\ = \sum_{\vec{k}}\big(\epsilon(\vec{k}) - \mu\big)c^\dagger_{\vec{k}-\vec{Q}}c_{\vec{k}-\vec{Q}} \\= \sum_{\vec{k}}\big(\epsilon(\vec{k}+\vec{Q}) - \mu\big)c^\dagger_{\vec{k}}c_{\vec{k}}
\end{split}
\end{equation}
It will be useful to see how the ground states in the lab and boosted frames are related. In the lab frame, the ground state is
\be
|\Phi_0 \rangle  =\bigg( \prod_{\substack{\vec{k}~{\rm s.t.} \\  \epsilon(\vec{k})<\mu}} c^\dagger_{\vec{k}} \bigg) |0\rangle
\ee
whereas in the boosted frame
\be
\begin{split}
|\bar{\Phi}_0 \rangle  =\bigg( \prod_{\substack{\vec{k}~{\rm s.t.} \\  \epsilon(\vec{k})<\mu}} \bar{c}^\dagger_{\vec{k}} \bigg) |0\rangle  = \bigg( \prod_{\substack{\vec{k}~{\rm s.t.} \\  \epsilon(\vec{k})<\mu}} c^\dagger_{\vec{k}-\vec{Q}} \bigg) |0\rangle \\ = \bigg( \prod_{\substack{\vec{k}~{\rm s.t.} \\  \epsilon(\vec{k}+\vec{Q})<\mu}} c^\dagger_{\vec{k}} \bigg) |0\rangle
\end{split}
\ee
These are related as
\be
|\Phi_0 \rangle = U_B(\vec{Q}) |\bar{\Phi}_0 \rangle 
\ee
This relation holds generally and can be used to find the groundstate in the lab frame once the ground state in the boosted frame is known.

\begin{figure}
\includegraphics[width=.45\textwidth]{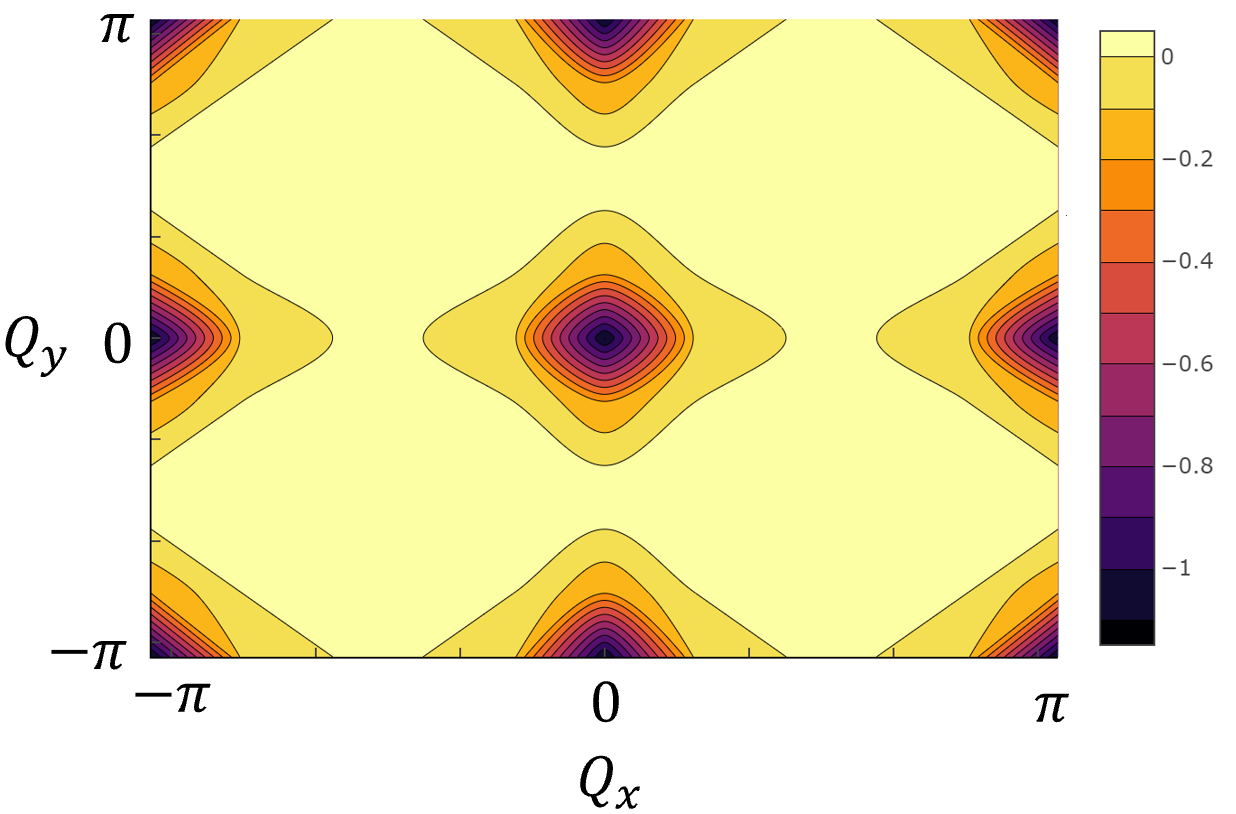}
\caption{The total energy, measured relative to the normal state (i.e. the condensation energy), as a function of the boost vector $\vec{Q}$. The hopping amplitude in the square lattice model has been set to unity. The minima occur when the Fermi surface is situated such that $\vec{k}_{+} + \vec{k}_{-} = \vec{G}$ where $\vec{k}_{\pm}$ are momenta on opposite sides of the Fermi surface and $\vec{G}$ is a reciprocal lattice vector. See Fig. \ref{fig2}.} \label{fig1}
\end{figure}

\begin{figure}
\includegraphics[width=.5\textwidth]{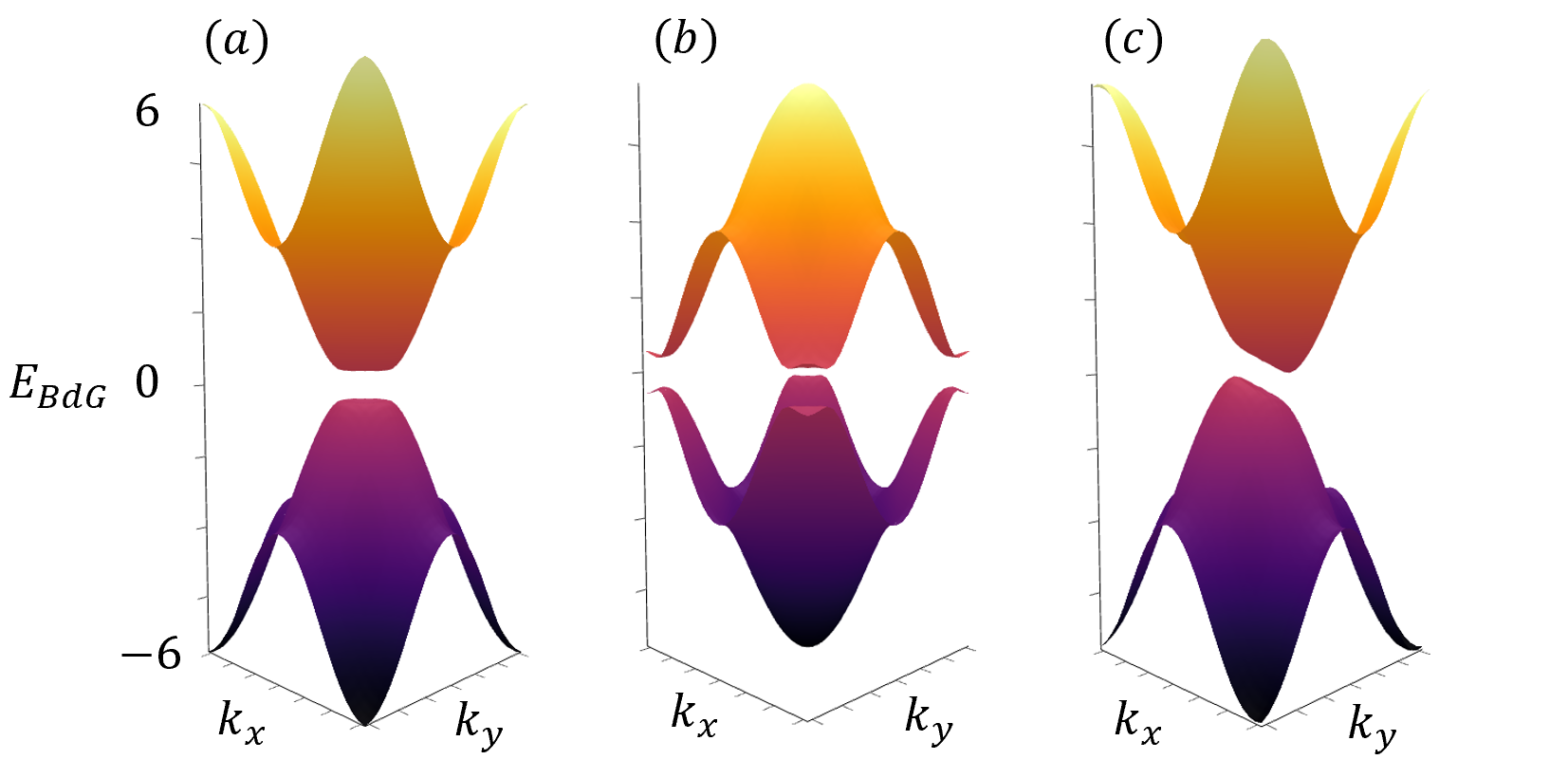}
\caption{ The Bogoliubov-de Gennes spectrum for boost vectors (a) at the minimum energy point $\vec{Q} = 0$, where it can be seen that a robust pairing gap opens; (b) at the minimum energy point $\vec{Q} = (\pi, \pi)$, where it can be seen that a robust pairing gap also opens;  and (c) slightly away from $\vec{Q} = 0$. In general, away from the minimum energy $\vec{Q}$, one generally finds small indirect gaps, or indirect gap closings in the spectrum.  } \label{fig2}
\end{figure}

As a simple example to elucidate the procedure in the main text, we may take a lattice system, in zero magnetic field, described by the Hamiltonian
\be
\cH = \sum_{\vec{k}, \sigma}\big(\epsilon(\vec{k}) - \mu\big)c^\dagger_{\vec{k} \sigma}c_{\vec{k} \sigma} - U\sum_{j} c^\dagger_{j,\uparrow}c^\dagger_{j,\downarrow}c_{j,\downarrow}c_{j,\uparrow}
\ee
where $\epsilon(\vec{k}) = -2 \big( \cos{k_x} + \cos{k_y} \big)$. We perform a mean-field decoupling with a helical ansatz, as discussed in the main article, to determine the self-consistent superconducting groundstate. The condensation energies are shown in Fig. \ref{fig1}, which is used to determined the ground state, and the accompanying Bogoliubov-de Gennes spectra for several boost vectors $\vec{Q}$ are shown in Fig. \ref{fig2}, where it can be seen how it changes for different choices of $\vec{Q}$.
 
\begin{figure*}
\includegraphics[width=.7\linewidth]{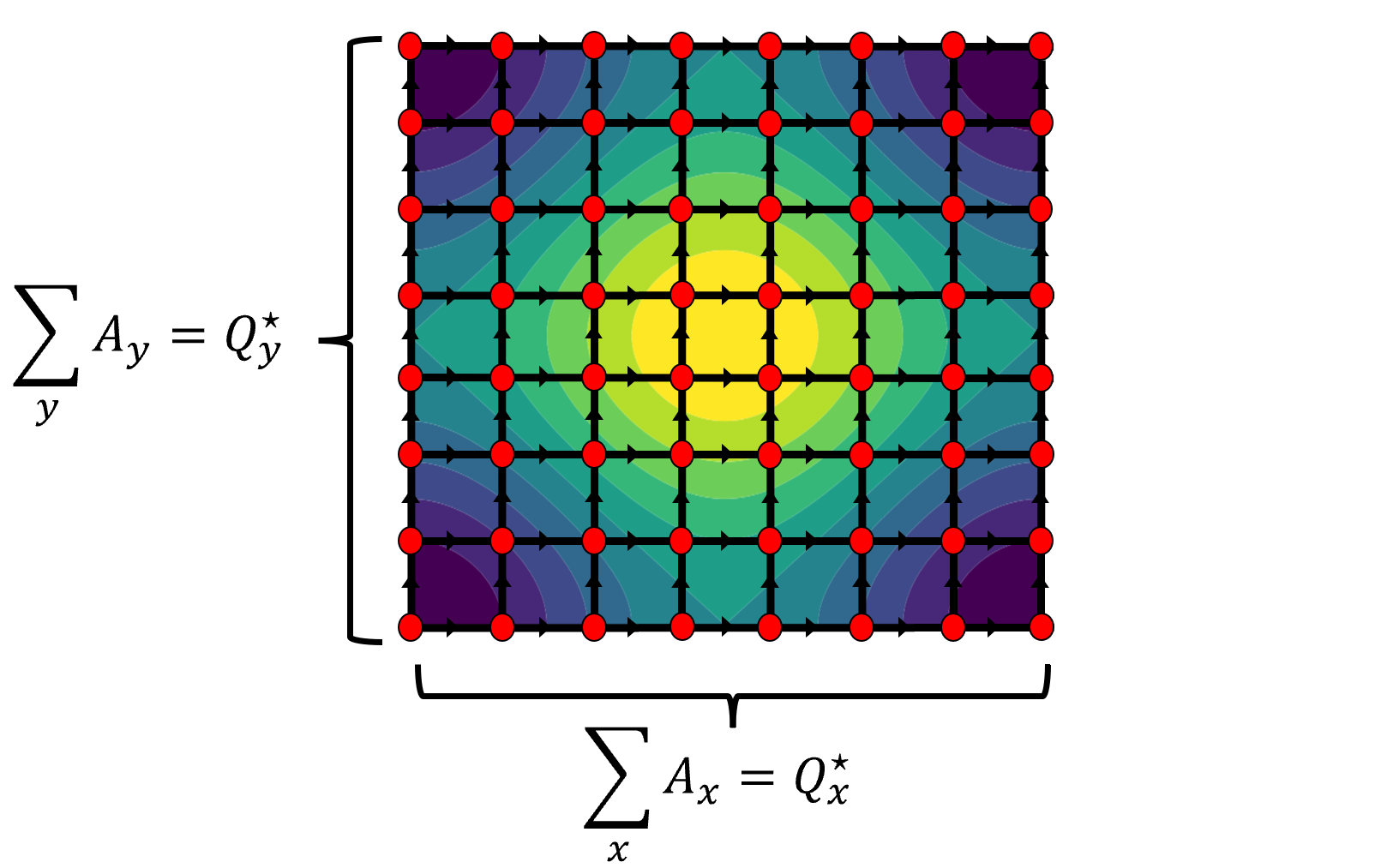}
\caption{The unit cell of the superlattice potential used in the main text. The color plot underneath the lattice shows the contours of the potential with the maximum denoted in bright yellow and the minimum in deep blue. For a given choice of gauge $\vec{A}$, an optimum boost vector (one that leads to the lowest mean-field energy) is given by the sum of the hopping phases along edges connecting the minima of the superlattice potential. } \label{bestQ}
\end{figure*}

\begin{figure*}
\includegraphics[width=.9\textwidth]{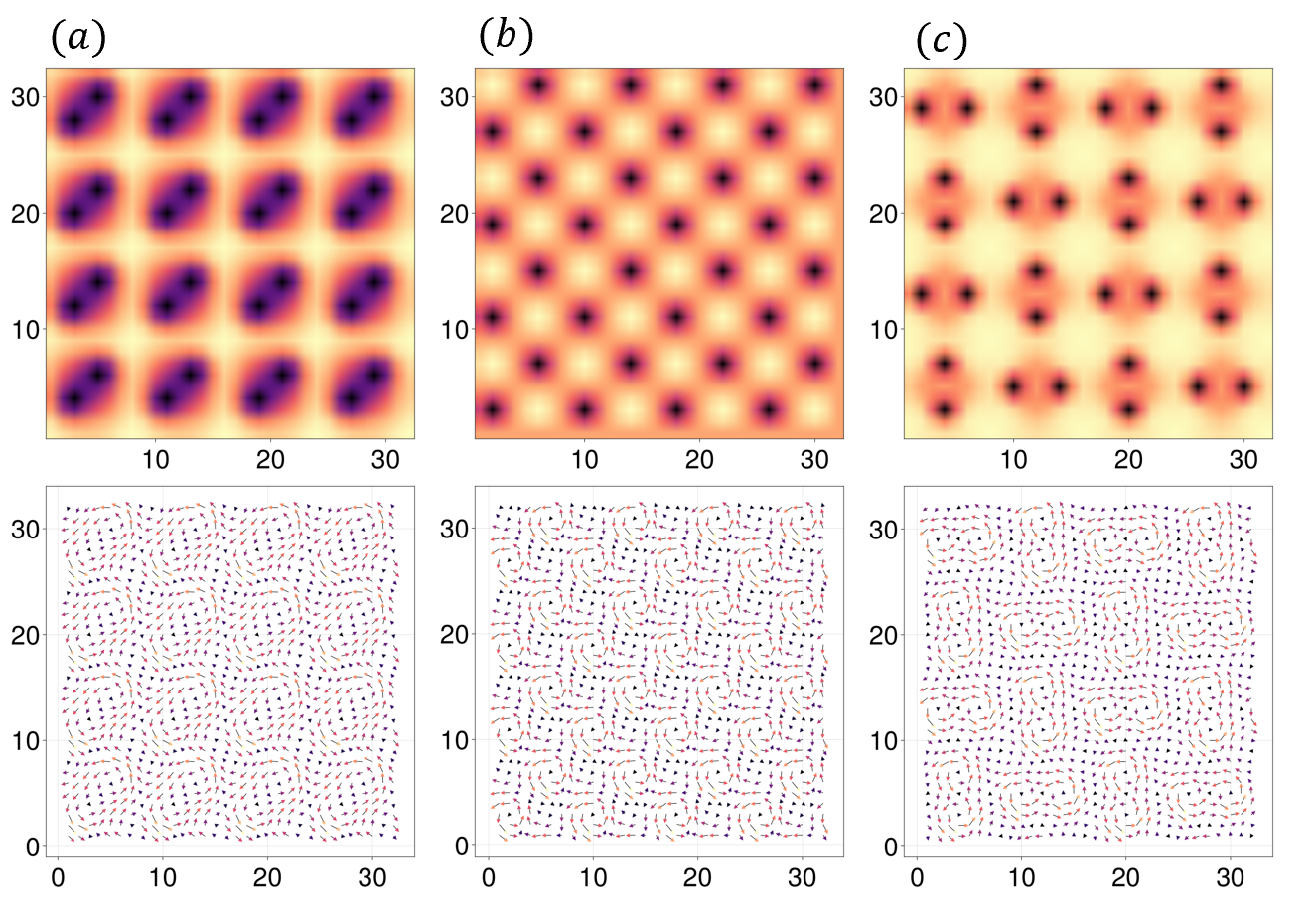}
\caption{ Additional unexpected vortex lattice structures (above) from the solutions to the self-consistent mean field equations (in the main text) with their current textures (below). (a) Dimer vortex configuration at $V_{\rm sp} = 0.2$, $V_{\rm SO} = 0.1$, $U=3.48$, $\mu=-3.66$.  Dimer configurations of vortices can form due to the influence of the superlattice potential. We find that the appearance of dimerized vortices signals the exit from the non-Abelian topological phase. (b-c) Unexpected vortex configurations may appear even in the absence of a superlattice potential particularly at higher Landau level filling (or chemical potential). Shown here are (b) a stabilized square lattice configuration at $V_{\rm sp} = 0$, $V_{\rm SO} = 0.1$, $U=3.2$, $\mu=-2.92$ and (c) a ``double dimer'' at $V_{\rm sp} = 0$, $V_{\rm SO} = 0.1$, $U=3.54$, $\mu=-3.22$. } \label{vortex2}
\end{figure*}

\begin{figure}
\includegraphics[width=\linewidth]{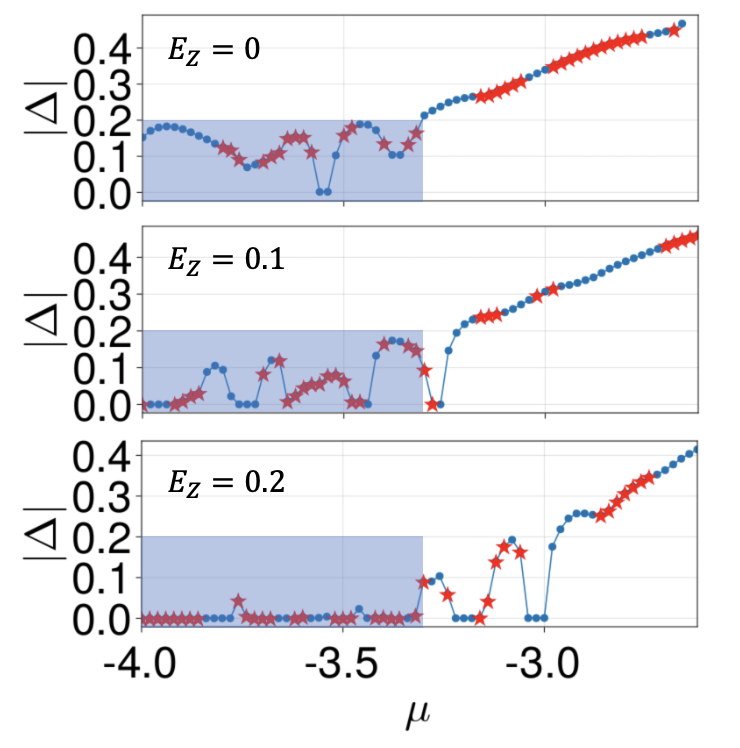}
\caption{{The pairing potential is reduced based on the Pauli limit. When the value of $E_Z$ reaches 0.2, all the points with pairing potential less than or equal to approximately 0.2 in the absence of Zeeman field ($\mu \approx -4$ to $\mu \approx -3.3$, highlighted with a blue rectangle) are significantly reduced. However, we still find topological superconductivity (denoted with red stars). Here we have set the attractive interaction $U=3.38$, the superlattice potential strength $V_{\text{sp}}=0.2$, the spin-orbit coupling strength $V_{\text{SO}} = 0.1$.}}\label{fig:zeeman}
\end{figure}

\begin{figure}
\includegraphics[width=\linewidth]{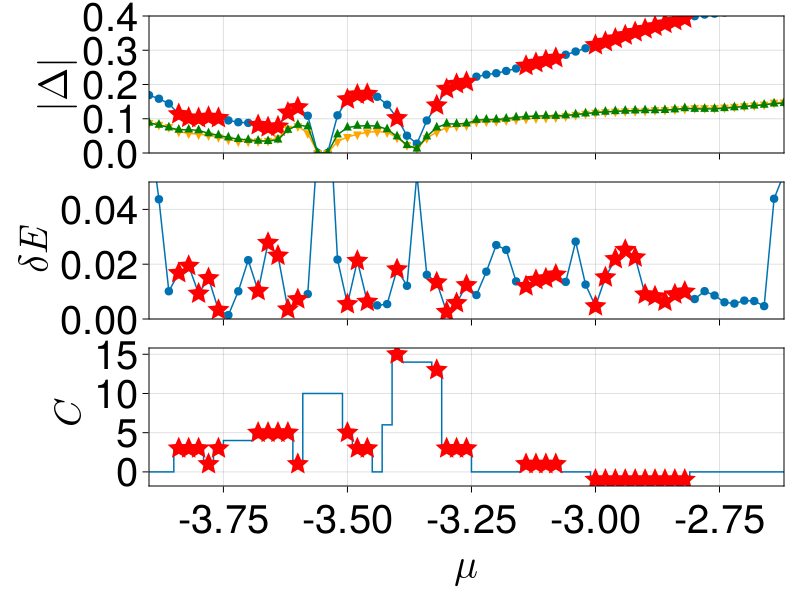}
\caption{{Top: The blue dots and red stars are the maximum absolute value of the pairing potential. Blue dots correspond to systems with even Chern number and red stars correspond to systems with odd Chern number. The green triangles are $10U\text{max}\left(|\Psi_{j\downarrow+}|\right)$ and the orange triangles are $10U\text{max}\left(|\Psi_{j\downarrow+}|\right)$. Middle: the BdG spectral gap ($\delta E$). Bottom: the Chern number (C).  The parameters are $U=3.34$, $V_{\text{sp}} = 0.2$, and $V_{\text{SO}} = 0.1$ (the same as in Fig. 6 of the main article).}}\label{fig:pwave}
\end{figure}

\section{Algorithm}\label{sec5}
The following is an algorithm to solve the mean-field problem self-consistently. Its inputs are the dimensions of the MUC, which contains $\Phi=n\Phi_0$ flux where $n \in \mathbb{Z}$ ($\Phi_0=h/e$ is the flux quantum), the interaction strength $U$, the chemical potential $\mu$, the spin-orbit coupling strength $V_{\rm SO}$, the superlattice potential strength $V_{\rm sp}$, and the initial guess for the $s$-wave pairing potential $\Delta_j^{(0)}$. Its outputs are the Gor'kov Green's function $\mathcal{G}(\vec{k})$ and the mean-field groundstate energy $\mathcal{E}_{\rm MF}$ of the system.
\begin{algorithm}[H]
\caption{Iterative algorithm for solving the mean-field equations.}\label{alg}
\begin{algorithmic}
\Require{$M$, $N$, $n$, $U$, $\mu$, $V_{\rm SO}$, $V_{\rm sp}$, $\Delta_j^{(0)}$}
\Ensure{$\Delta_j$, $\mathcal{G}(\vec{k})~\forall \vec{k}\in$ MBZ, $\mathcal{E}_{\rm MF}$ }
\State{Build the diagonal blocks of $H_{\rm BdG}(\vec{k})$ using $h(\vec{k})$, $\Sigma(\vec{k})$ $\forall \vec{k} \in \rm HBZ$};
 \State{$\delta \Delta = \infty$};
 \State{$l=0$}
 \While{$\delta \Delta>\tau$} \Comment{The tolerance for differences in $\Delta$ is $\tau=10^{-5}$.}
 \State{Update/build $\Delta_{\down \up}$ using $\Delta^{(l)}_j$};
 \For{$\vec{k} \in$ HBZ}
 \State{Update off-diagonal blocks of $H_{\rm BdG}(\vec{k})$ using $\Delta_{\down \up}$};
  \State Compute $\mathcal{G}(\vec{k})$ by diagonalization of  $H_{\rm BdG}(\vec{k})$, using \eqref{densmat} and particle-hole symmetry for $\mathcal{G}(-\vec{k})$;
	\State{Do a running sum to compute $\Delta^{(l+1)}_j$};
	\EndFor
		\State{Compute pairing potential difference $\delta \Delta=$ max$_j\Big(\big|\Delta^{(l+1)}_j -\Delta^{(l)}_j\big|\Big)$};
	\State{$l \gets l+1$};
 \EndWhile
 \State{Compute $\mathcal{E}_{\rm MF}$}
 
\end{algorithmic}
\end{algorithm}

\section{How to determine an optimum boost vector analytically} \label{sec8}
For a particular gauge, the sum of the hopping phases along edges linking the minima of the superlattice potential determines an optimum boost vector (one that results in the lowest mean-field energy). See Fig. \ref{bestQ}. Such a boost leads to zero accumulated phase for particles with $\vec{k}=0$ traversing the unit cell of the potential in both the $x$ and $y$ directions, thereby respecting the $C_ 4$ symmetry of the superlattice potential and resulting in the minimum energy.

\section{The reason there is no TSC when ${\bm V_{\rm SO} = 0}$} \label{sec9}

We do not find topological superconductivity without spin-orbit coupling. This follows from the spin degeneracy in the spectrum of our BdG Hamiltonian in the absence of spin-orbit coupling. Without spin-orbit coupling, the mean field Hamiltonian reads
\begin{widetext}
\begin{equation}
\cH_{\rm MF}=-\sum_{j,\delta,\sigma}\Big(e^{iA_\delta (\vec{r}_j)}c^\dagger_{j+\delta,\sigma} c_{j,\sigma}+e^{-iA_\delta(\vec{r}_j)}c^{\dagger}_{j,\sigma} c_{j+\delta,\sigma}\Big) - 
 \sum_{j,\sigma} \big(\mu-V(\vec{r}_j) \big) c^\dagger_{j,\sigma} c_{j\sigma}
 -\sum_{j} \Big(\Delta_j c^\dagger_{j,\uparrow}c^\dagger_{j,\downarrow}+\Delta_j^*  c_{j,\downarrow}c_{j,\uparrow}-\frac{|\Delta_j|^2}{U} \Big)
\end{equation}
Consider the spin-flip operator $\hat{F}$ defined as $\hat{F} c_{j,\sigma} \hat{F}^\dagger = c_{j,-\sigma}$. Clearly the hopping and chemical potential terms are invariant under the operation of $F$. The action of $F$ on the pairing term is
\be
\hat{F} \cH_{\Delta} \hat{F}^\dagger =- \sum_{j} \Big(\Delta_j c^\dagger_{j,\downarrow}c^\dagger_{j,\uparrow}+\Delta_j^*  c_{j,\uparrow}c_{j,\downarrow}-\frac{|\Delta_j|^2}{U} \Big) = -\sum_{j} \Big(-\Delta_j c^\dagger_{j,\uparrow}c^\dagger_{j,\downarrow}-\Delta_j^*  c_{j,\downarrow}c_{j,\uparrow}-\frac{|\Delta_j|^2}{U} \Big)
\ee
\end{widetext}
So $\cH_\Delta$ is, strictly speaking, not invariant under $\hat{F}$. It is, however, invariant up to a global gauge transformation which renders $\Delta_j \rightarrow - \Delta_j$. Thus the mean-field Hamiltonian has a symmetry $\hat{S} = e^{-i  \hat{N} \pi/2} \hat{F}$ where $\hat{N} = \sum_{j,\sigma}c^\dagger_{j,\sigma} c_{j\sigma}$. Under this symmetry, the eigenvectors of the BdG Hamiltonian generally transform as
\be
\begin{pmatrix} u_\uparrow(\vec{k}) \\  u_\downarrow(\vec{k}) \\  v_\uparrow(\vec{k}) \\  v_\downarrow(\vec{k}) \end{pmatrix} \rightarrow i \begin{pmatrix} u_\downarrow(\vec{k}) \\  u_\uparrow(\vec{k}) \\  -v_\downarrow(\vec{k}) \\  -v_\uparrow(\vec{k}) \end{pmatrix}
\ee
A spin-up state has the form $\begin{pmatrix} u(\vec{k}) &  0 &  0 &  v(\vec{k}) \end{pmatrix}^{\rm T}$ and so the corresponding spin-down state is $i\begin{pmatrix} 0 &  u(\vec{k}) &  -v(\vec{k}) &  0 \end{pmatrix}^{\rm T}$. Constructing the Berry connection matrix out of these two bands, we see that
\begin{equation}
A^{\down \down}_{\mu}(\vec{k})= i \Big[u^*(\vec{k})\partial_{\mu}u(\vec{k})+ v^*(\vec{k})\partial_{\mu}v(\vec{k}) \Big]
\end{equation}
and we find the same result for $A^{\up \up}_{\mu}(\vec{k})$. The Berry curvatures for these bands are thus the same $ F^{\uparrow \uparrow}_{\mu\nu}(\vec{k})= F^{\downarrow \downarrow}_{\mu\nu}(\vec{k})$. The  Chern number for these bands is given by

\be
\begin{split}
C &= \frac{1}{2\pi} \int_{\rm MBZ} d^2k  \Big[F^{\uparrow \uparrow}_{\rm xy}(\vec{k}) + F^{\downarrow \downarrow}_{\rm xy}(\vec{k}) \Big] \\&= 2\frac{1}{2\pi} \int_{\rm MBZ} d^2k  \Big[F^{\uparrow \uparrow}_{\rm xy}(\vec{k}) \Big]\in 2 \mathbb{Z}
\end{split}
\ee
Therefore the sum of the Chern numbers from all the filled bands is an even integer, thus excluding the non-Abelian topological phase.

\section{{Role of Zeeman field}} \label{zeeman_sec}
{We find that upon including the Zeeman effect into the calculation, superconductivity is suppressed due to the Pauli limit} 
\begin{equation}
E_Z^{P} \approx  |\Delta|
\end{equation}
{where $E_Z^{P}$ is the Pauli limited Zeeman energy and $\Delta$ is the pairing potential. However, TSC survives. This is shown in Fig. \ref{fig:zeeman} and described in detail in the caption.} 

\section{{Emergence of $p$-wave superconductivity}}\label{sec::pwave}
{
In this section, We comment on the emergence of $p$-wave component of superconductivity in our model, which can arise from  $s$-wave superconductivity and Rashba spin-orbit coupling due to singlet-triplet mixing, as inversion symmetry is broken by Rashba spin-orbit coupling \cite{Bauer2012}. We explicitly show this by calculating $p$-wave components of the order parameter. 
}

{
Let us define}
\begin{equation}
\Psi_{j\sigma\delta} = \langle c_{j,\sigma}c_{j+\delta,\sigma} \rangle
\end{equation}
{The expectation values can be determined from the Gor'kov Green's function (see the SM \cite{Schirmer2022e}) It is convenient to transform these variables to the $p_x + ip_y$ component $\Psi_{j\sigma+}$ and $p_x - ip_y$ component $\Psi_{j\sigma-}$.}

\begin{equation}
\begin{split}
\Psi_{j\sigma+} = \frac{1}{2} \left[ \Psi_{j\sigma \hat{x}}  - i\Psi_{j\sigma\hat{y}}  \right]\\
\Psi_{j\sigma-} = \frac{1}{2} \left[ \Psi_{j\sigma \hat{x}}  + i\Psi_{j\sigma\hat{y}}  \right]
\end{split}
\end{equation}
{Out of the components $\Psi_{j\uparrow+}$, $\Psi_{j\downarrow+}$, $\Psi_{j\uparrow-}$, and $\Psi_{j\downarrow-}$, we find that $\Psi_{j\uparrow-}$ and $\Psi_{j\downarrow+}$ are dominantly the largest. In Fig. \ref{fig:pwave}, we show these components, together with the $s$-wave component, for parameters $U=3.34$, $V_{\text{sp}} = 0.2$, and $V_{\text{SO}} = 0.1$ (the same as in Fig. 6 of the main article). The green curve in the upper panel is $10U\text{max}\left(|\Psi_{j\downarrow+}|\right)$ and the orange curve in the upper panel is $10U\text{max}\left(|\Psi_{j\downarrow+}|\right)$. The curve with blue dots and red stars is $\text{max}\left(|\Delta_j|\right) = U \text{max}\left(|\langle c_{j,\downarrow}c_{j,\uparrow}\rangle| \right)$. It is thus clear that the $s$-wave component is much larger than the $p$-wave component by roughly a factor of 60. We also see that the emergence of a $p$-wave component accompanies the appearance of \textit{superconductivity}, not just \textit{topological} superconductivity. This is to be expected because superconductors with broken parity and time-reversal symmetries typically exhibit singlet-triplet mixing. 
}

\end{appendices}

\clearpage
\bibliography{my}

\begin{thebibliography}{129}%
\makeatletter
\providecommand \@ifxundefined [1]{%
 \@ifx{#1\undefined}
}%
\providecommand \@ifnum [1]{%
 \ifnum #1\expandafter \@firstoftwo
 \else \expandafter \@secondoftwo
 \fi
}%
\providecommand \@ifx [1]{%
 \ifx #1\expandafter \@firstoftwo
 \else \expandafter \@secondoftwo
 \fi
}%
\providecommand \natexlab [1]{#1}%
\providecommand \enquote  [1]{``#1''}%
\providecommand \bibnamefont  [1]{#1}%
\providecommand \bibfnamefont [1]{#1}%
\providecommand \citenamefont [1]{#1}%
\providecommand \href@noop [0]{\@secondoftwo}%
\providecommand \href [0]{\begingroup \@sanitize@url \@href}%
\providecommand \@href[1]{\@@startlink{#1}\@@href}%
\providecommand \@@href[1]{\endgroup#1\@@endlink}%
\providecommand \@sanitize@url [0]{\catcode `\\12\catcode `\$12\catcode
  `\&12\catcode `\#12\catcode `\^12\catcode `\_12\catcode `\%12\relax}%
\providecommand \@@startlink[1]{}%
\providecommand \@@endlink[0]{}%
\providecommand \url  [0]{\begingroup\@sanitize@url \@url }%
\providecommand \@url [1]{\endgroup\@href {#1}{\urlprefix }}%
\providecommand \urlprefix  [0]{URL }%
\providecommand \Eprint [0]{\href }%
\providecommand \doibase [0]{http://dx.doi.org/}%
\providecommand \selectlanguage [0]{\@gobble}%
\providecommand \bibinfo  [0]{\@secondoftwo}%
\providecommand \bibfield  [0]{\@secondoftwo}%
\providecommand \translation [1]{[#1]}%
\providecommand \BibitemOpen [0]{}%
\providecommand \bibitemStop [0]{}%
\providecommand \bibitemNoStop [0]{.\EOS\space}%
\providecommand \EOS [0]{\spacefactor3000\relax}%
\providecommand \BibitemShut  [1]{\csname bibitem#1\endcsname}%
\let\auto@bib@innerbib\@empty
%</preamble>
\bibitem [{\citenamefont {Stone}\ and\ \citenamefont
  {Chung}(2006)}]{Stone2006}%
  \BibitemOpen
  \bibfield  {author} {\bibinfo {author} {\bibfnamefont {M.}~\bibnamefont
  {Stone}}\ and\ \bibinfo {author} {\bibfnamefont {S.-B.}\ \bibnamefont
  {Chung}},\ }\href@noop {} {\bibfield  {journal} {\bibinfo  {journal}
  {Physical Review B}\ }\textbf {\bibinfo {volume} {73}},\ \bibinfo {pages}
  {014505} (\bibinfo {year} {2006})}\BibitemShut {NoStop}%
\bibitem [{\citenamefont {Oshikawa}\ \emph {et~al.}(2007)\citenamefont
  {Oshikawa}, \citenamefont {Kim}, \citenamefont {Shtengel}, \citenamefont
  {Nayak},\ and\ \citenamefont {Tewari}}]{Oshikawa2007}%
  \BibitemOpen
  \bibfield  {author} {\bibinfo {author} {\bibfnamefont {M.}~\bibnamefont
  {Oshikawa}}, \bibinfo {author} {\bibfnamefont {Y.~B.}\ \bibnamefont {Kim}},
  \bibinfo {author} {\bibfnamefont {K.}~\bibnamefont {Shtengel}}, \bibinfo
  {author} {\bibfnamefont {C.}~\bibnamefont {Nayak}}, \ and\ \bibinfo {author}
  {\bibfnamefont {S.}~\bibnamefont {Tewari}},\ }\href {\doibase
  https://doi.org/10.1016/j.aop.2006.08.001} {\bibfield  {journal} {\bibinfo
  {journal} {Annals of Physics}\ }\textbf {\bibinfo {volume} {322}},\ \bibinfo
  {pages} {1477 } (\bibinfo {year} {2007})}\BibitemShut {NoStop}%
\bibitem [{\citenamefont {Nayak}\ \emph {et~al.}(2008)\citenamefont {Nayak},
  \citenamefont {Simon}, \citenamefont {Stern}, \citenamefont {Freedman},\ and\
  \citenamefont {Das~Sarma}}]{Nayak2008}%
  \BibitemOpen
  \bibfield  {author} {\bibinfo {author} {\bibfnamefont {C.}~\bibnamefont
  {Nayak}}, \bibinfo {author} {\bibfnamefont {S.~H.}\ \bibnamefont {Simon}},
  \bibinfo {author} {\bibfnamefont {A.}~\bibnamefont {Stern}}, \bibinfo
  {author} {\bibfnamefont {M.}~\bibnamefont {Freedman}}, \ and\ \bibinfo
  {author} {\bibfnamefont {S.}~\bibnamefont {Das~Sarma}},\ }\href {\doibase
  10.1103/RevModPhys.80.1083} {\bibfield  {journal} {\bibinfo  {journal}
  {Reviews of Modern Physics}\ }\textbf {\bibinfo {volume} {80}},\ \bibinfo
  {pages} {1083} (\bibinfo {year} {2008})}\BibitemShut {NoStop}%
\bibitem [{\citenamefont {Sau}\ \emph {et~al.}(2010{\natexlab{a}})\citenamefont
  {Sau}, \citenamefont {Lutchyn}, \citenamefont {Tewari},\ and\ \citenamefont
  {Sarma}}]{Sau2010a}%
  \BibitemOpen
  \bibfield  {author} {\bibinfo {author} {\bibfnamefont {J.~D.}\ \bibnamefont
  {Sau}}, \bibinfo {author} {\bibfnamefont {R.~M.}\ \bibnamefont {Lutchyn}},
  \bibinfo {author} {\bibfnamefont {S.}~\bibnamefont {Tewari}}, \ and\ \bibinfo
  {author} {\bibfnamefont {S.~D.}\ \bibnamefont {Sarma}},\ }\href@noop {}
  {\bibfield  {journal} {\bibinfo  {journal} {Physical Review Letters}\
  }\textbf {\bibinfo {volume} {104}},\ \bibinfo {pages} {040502} (\bibinfo
  {year} {2010}{\natexlab{a}})}\BibitemShut {NoStop}%
\bibitem [{\citenamefont {Stanescu}\ \emph {et~al.}(2010)\citenamefont
  {Stanescu}, \citenamefont {Sau}, \citenamefont {Lutchyn},\ and\ \citenamefont
  {Sarma}}]{Stanescu2010}%
  \BibitemOpen
  \bibfield  {author} {\bibinfo {author} {\bibfnamefont {T.~D.}\ \bibnamefont
  {Stanescu}}, \bibinfo {author} {\bibfnamefont {J.~D.}\ \bibnamefont {Sau}},
  \bibinfo {author} {\bibfnamefont {R.~M.}\ \bibnamefont {Lutchyn}}, \ and\
  \bibinfo {author} {\bibfnamefont {S.~D.}\ \bibnamefont {Sarma}},\ }\href@noop
  {} {\bibfield  {journal} {\bibinfo  {journal} {Physical Review B}\ }\textbf
  {\bibinfo {volume} {81}},\ \bibinfo {pages} {241310} (\bibinfo {year}
  {2010})}\BibitemShut {NoStop}%
\bibitem [{\citenamefont {Alicea}(2012)}]{Alicea2012}%
  \BibitemOpen
  \bibfield  {author} {\bibinfo {author} {\bibfnamefont {J.}~\bibnamefont
  {Alicea}},\ }\href@noop {} {\bibfield  {journal} {\bibinfo  {journal}
  {Reports on Progress in Physics}\ }\textbf {\bibinfo {volume} {75}},\
  \bibinfo {pages} {076501} (\bibinfo {year} {2012})}\BibitemShut {NoStop}%
\bibitem [{\citenamefont {Hung}\ \emph {et~al.}(2013)\citenamefont {Hung},
  \citenamefont {Ghaemi}, \citenamefont {Hughes},\ and\ \citenamefont
  {Gilbert}}]{Hung2013}%
  \BibitemOpen
  \bibfield  {author} {\bibinfo {author} {\bibfnamefont {H.-H.}\ \bibnamefont
  {Hung}}, \bibinfo {author} {\bibfnamefont {P.}~\bibnamefont {Ghaemi}},
  \bibinfo {author} {\bibfnamefont {T.~L.}\ \bibnamefont {Hughes}}, \ and\
  \bibinfo {author} {\bibfnamefont {M.~J.}\ \bibnamefont {Gilbert}},\
  }\href@noop {} {\bibfield  {journal} {\bibinfo  {journal} {Physical Review
  B}\ }\textbf {\bibinfo {volume} {87}},\ \bibinfo {pages} {035401} (\bibinfo
  {year} {2013})}\BibitemShut {NoStop}%
\bibitem [{\citenamefont {Beenakker}(2013)}]{Beenakker2013}%
  \BibitemOpen
  \bibfield  {author} {\bibinfo {author} {\bibfnamefont {C.}~\bibnamefont
  {Beenakker}},\ }\href {\doibase 10.1146/annurev-conmatphys-030212-184337}
  {\bibfield  {journal} {\bibinfo  {journal} {Annual Review of Condensed Matter
  Physics}\ }\textbf {\bibinfo {volume} {4}},\ \bibinfo {pages} {113} (\bibinfo
  {year} {2013})}\BibitemShut {NoStop}%
\bibitem [{\citenamefont {Zhou}\ \emph {et~al.}(2013)\citenamefont {Zhou},
  \citenamefont {Wu}, \citenamefont {Li}, \citenamefont {He},\ and\
  \citenamefont {Kou}}]{Zhou2013}%
  \BibitemOpen
  \bibfield  {author} {\bibinfo {author} {\bibfnamefont {J.}~\bibnamefont
  {Zhou}}, \bibinfo {author} {\bibfnamefont {Y.-J.}\ \bibnamefont {Wu}},
  \bibinfo {author} {\bibfnamefont {R.-W.}\ \bibnamefont {Li}}, \bibinfo
  {author} {\bibfnamefont {J.}~\bibnamefont {He}}, \ and\ \bibinfo {author}
  {\bibfnamefont {S.-P.}\ \bibnamefont {Kou}},\ }\href@noop {} {\bibfield
  {journal} {\bibinfo  {journal} {Europhysics Letters}\ }\textbf {\bibinfo
  {volume} {102}},\ \bibinfo {pages} {47005} (\bibinfo {year}
  {2013})}\BibitemShut {NoStop}%
\bibitem [{\citenamefont {Biswas}(2013)}]{Biswas2013}%
  \BibitemOpen
  \bibfield  {author} {\bibinfo {author} {\bibfnamefont {R.~R.}\ \bibnamefont
  {Biswas}},\ }\href@noop {} {\bibfield  {journal} {\bibinfo  {journal}
  {Physical Review Letters}\ }\textbf {\bibinfo {volume} {111}},\ \bibinfo
  {pages} {136401} (\bibinfo {year} {2013})}\BibitemShut {NoStop}%
\bibitem [{\citenamefont {Sarma}\ \emph {et~al.}(2015)\citenamefont {Sarma},
  \citenamefont {Freedman},\ and\ \citenamefont {Nayak}}]{Sarma2015}%
  \BibitemOpen
  \bibfield  {author} {\bibinfo {author} {\bibfnamefont {S.~D.}\ \bibnamefont
  {Sarma}}, \bibinfo {author} {\bibfnamefont {M.}~\bibnamefont {Freedman}}, \
  and\ \bibinfo {author} {\bibfnamefont {C.}~\bibnamefont {Nayak}},\
  }\href@noop {} {\bibfield  {journal} {\bibinfo  {journal} {npj Quantum
  Information}\ }\textbf {\bibinfo {volume} {1}},\ \bibinfo {pages} {1}
  (\bibinfo {year} {2015})}\BibitemShut {NoStop}%
\bibitem [{\citenamefont {Liu}\ and\ \citenamefont {Franz}(2015)}]{Liu2015}%
  \BibitemOpen
  \bibfield  {author} {\bibinfo {author} {\bibfnamefont {T.}~\bibnamefont
  {Liu}}\ and\ \bibinfo {author} {\bibfnamefont {M.}~\bibnamefont {Franz}},\
  }\href {\doibase 10.1103/PhysRevB.92.134519} {\bibfield  {journal} {\bibinfo
  {journal} {Physical Review B}\ }\textbf {\bibinfo {volume} {92}},\ \bibinfo
  {pages} {134519} (\bibinfo {year} {2015})}\BibitemShut {NoStop}%
\bibitem [{\citenamefont {Murray}\ and\ \citenamefont
  {Vafek}(2015)}]{Murray2015}%
  \BibitemOpen
  \bibfield  {author} {\bibinfo {author} {\bibfnamefont {J.~M.}\ \bibnamefont
  {Murray}}\ and\ \bibinfo {author} {\bibfnamefont {O.}~\bibnamefont {Vafek}},\
  }\href@noop {} {\bibfield  {journal} {\bibinfo  {journal} {Physical Review
  B}\ }\textbf {\bibinfo {volume} {92}},\ \bibinfo {pages} {134520} (\bibinfo
  {year} {2015})}\BibitemShut {NoStop}%
\bibitem [{\citenamefont {Smith}\ \emph {et~al.}(2016)\citenamefont {Smith},
  \citenamefont {Tanaka},\ and\ \citenamefont {Nagai}}]{Smith2016}%
  \BibitemOpen
  \bibfield  {author} {\bibinfo {author} {\bibfnamefont {E.~D.}\ \bibnamefont
  {Smith}}, \bibinfo {author} {\bibfnamefont {K.}~\bibnamefont {Tanaka}}, \
  and\ \bibinfo {author} {\bibfnamefont {Y.}~\bibnamefont {Nagai}},\
  }\href@noop {} {\bibfield  {journal} {\bibinfo  {journal} {Physical Review
  B}\ }\textbf {\bibinfo {volume} {94}},\ \bibinfo {pages} {064515} (\bibinfo
  {year} {2016})}\BibitemShut {NoStop}%
\bibitem [{\citenamefont {Chowdhury}\ \emph {et~al.}(2022)\citenamefont
  {Chowdhury}, \citenamefont {Georges}, \citenamefont {Parcollet},\ and\
  \citenamefont {Sachdev}}]{Chowdhury2022}%
  \BibitemOpen
  \bibfield  {author} {\bibinfo {author} {\bibfnamefont {D.}~\bibnamefont
  {Chowdhury}}, \bibinfo {author} {\bibfnamefont {A.}~\bibnamefont {Georges}},
  \bibinfo {author} {\bibfnamefont {O.}~\bibnamefont {Parcollet}}, \ and\
  \bibinfo {author} {\bibfnamefont {S.}~\bibnamefont {Sachdev}},\ }\href@noop
  {} {\bibfield  {journal} {\bibinfo  {journal} {Reviews of Modern Physics}\
  }\textbf {\bibinfo {volume} {94}},\ \bibinfo {pages} {035004} (\bibinfo
  {year} {2022})}\BibitemShut {NoStop}%
\bibitem [{\citenamefont {Bernevig}(2013)}]{Bernevig2013}%
  \BibitemOpen
  \bibfield  {author} {\bibinfo {author} {\bibfnamefont {B.~A.}\ \bibnamefont
  {Bernevig}},\ }\href@noop {} {\emph {\bibinfo {title} {Topological insulators
  and topological superconductors}}}\ (\bibinfo  {publisher} {Princeton
  University Press},\ \bibinfo {year} {2013})\BibitemShut {NoStop}%
\bibitem [{\citenamefont {Sarma}\ \emph {et~al.}(2006)\citenamefont {Sarma},
  \citenamefont {Nayak},\ and\ \citenamefont {Tewari}}]{Sarma2006}%
  \BibitemOpen
  \bibfield  {author} {\bibinfo {author} {\bibfnamefont {S.~D.}\ \bibnamefont
  {Sarma}}, \bibinfo {author} {\bibfnamefont {C.}~\bibnamefont {Nayak}}, \ and\
  \bibinfo {author} {\bibfnamefont {S.}~\bibnamefont {Tewari}},\ }\href@noop {}
  {\bibfield  {journal} {\bibinfo  {journal} {Physical Review B}\ }\textbf
  {\bibinfo {volume} {73}},\ \bibinfo {pages} {220502} (\bibinfo {year}
  {2006})}\BibitemShut {NoStop}%
\bibitem [{\citenamefont {Fu}\ and\ \citenamefont {Kane}(2008)}]{Fu2008}%
  \BibitemOpen
  \bibfield  {author} {\bibinfo {author} {\bibfnamefont {L.}~\bibnamefont
  {Fu}}\ and\ \bibinfo {author} {\bibfnamefont {C.~L.}\ \bibnamefont {Kane}},\
  }\href {\doibase 10.1103/PhysRevLett.100.096407} {\bibfield  {journal}
  {\bibinfo  {journal} {Physical Review Letters}\ }\textbf {\bibinfo {volume}
  {100}},\ \bibinfo {pages} {096407} (\bibinfo {year} {2008})}\BibitemShut
  {NoStop}%
\bibitem [{\citenamefont {Lutchyn}\ \emph {et~al.}(2010)\citenamefont
  {Lutchyn}, \citenamefont {Sau},\ and\ \citenamefont
  {Das~Sarma}}]{Lutchyn2010}%
  \BibitemOpen
  \bibfield  {author} {\bibinfo {author} {\bibfnamefont {R.~M.}\ \bibnamefont
  {Lutchyn}}, \bibinfo {author} {\bibfnamefont {J.~D.}\ \bibnamefont {Sau}}, \
  and\ \bibinfo {author} {\bibfnamefont {S.}~\bibnamefont {Das~Sarma}},\ }\href
  {\doibase 10.1103/PhysRevLett.105.077001} {\bibfield  {journal} {\bibinfo
  {journal} {Physical Review Letters}\ }\textbf {\bibinfo {volume} {105}},\
  \bibinfo {pages} {077001} (\bibinfo {year} {2010})}\BibitemShut {NoStop}%
\bibitem [{\citenamefont {Qi}\ \emph {et~al.}(2010)\citenamefont {Qi},
  \citenamefont {Hughes},\ and\ \citenamefont {Zhang}}]{Qi2010}%
  \BibitemOpen
  \bibfield  {author} {\bibinfo {author} {\bibfnamefont {X.-L.}\ \bibnamefont
  {Qi}}, \bibinfo {author} {\bibfnamefont {T.~L.}\ \bibnamefont {Hughes}}, \
  and\ \bibinfo {author} {\bibfnamefont {S.-C.}\ \bibnamefont {Zhang}},\
  }\href@noop {} {\bibfield  {journal} {\bibinfo  {journal} {Physical Review
  B}\ }\textbf {\bibinfo {volume} {82}},\ \bibinfo {pages} {184516} (\bibinfo
  {year} {2010})}\BibitemShut {NoStop}%
\bibitem [{\citenamefont {Sau}\ \emph {et~al.}(2010{\natexlab{b}})\citenamefont
  {Sau}, \citenamefont {Lutchyn}, \citenamefont {Tewari},\ and\ \citenamefont
  {Sarma}}]{Sau2010}%
  \BibitemOpen
  \bibfield  {author} {\bibinfo {author} {\bibfnamefont {J.~D.}\ \bibnamefont
  {Sau}}, \bibinfo {author} {\bibfnamefont {R.~M.}\ \bibnamefont {Lutchyn}},
  \bibinfo {author} {\bibfnamefont {S.}~\bibnamefont {Tewari}}, \ and\ \bibinfo
  {author} {\bibfnamefont {S.~D.}\ \bibnamefont {Sarma}},\ }\href@noop {}
  {\bibfield  {journal} {\bibinfo  {journal} {Physical Review Letters}\
  }\textbf {\bibinfo {volume} {104}},\ \bibinfo {pages} {040502} (\bibinfo
  {year} {2010}{\natexlab{b}})}\BibitemShut {NoStop}%
\bibitem [{\citenamefont {Sau}\ \emph {et~al.}(2010{\natexlab{c}})\citenamefont
  {Sau}, \citenamefont {Tewari}, \citenamefont {Lutchyn}, \citenamefont
  {Stanescu},\ and\ \citenamefont {Sarma}}]{Sau2010b}%
  \BibitemOpen
  \bibfield  {author} {\bibinfo {author} {\bibfnamefont {J.~D.}\ \bibnamefont
  {Sau}}, \bibinfo {author} {\bibfnamefont {S.}~\bibnamefont {Tewari}},
  \bibinfo {author} {\bibfnamefont {R.~M.}\ \bibnamefont {Lutchyn}}, \bibinfo
  {author} {\bibfnamefont {T.~D.}\ \bibnamefont {Stanescu}}, \ and\ \bibinfo
  {author} {\bibfnamefont {S.~D.}\ \bibnamefont {Sarma}},\ }\href@noop {}
  {\bibfield  {journal} {\bibinfo  {journal} {Physical Review B}\ }\textbf
  {\bibinfo {volume} {82}},\ \bibinfo {pages} {214509} (\bibinfo {year}
  {2010}{\natexlab{c}})}\BibitemShut {NoStop}%
\bibitem [{\citenamefont {Alicea}(2010)}]{Alicea2010}%
  \BibitemOpen
  \bibfield  {author} {\bibinfo {author} {\bibfnamefont {J.}~\bibnamefont
  {Alicea}},\ }\href@noop {} {\bibfield  {journal} {\bibinfo  {journal}
  {Physical Review B}\ }\textbf {\bibinfo {volume} {81}},\ \bibinfo {pages}
  {125318} (\bibinfo {year} {2010})}\BibitemShut {NoStop}%
\bibitem [{\citenamefont {Qi}\ and\ \citenamefont {Zhang}(2011)}]{Qi2011}%
  \BibitemOpen
  \bibfield  {author} {\bibinfo {author} {\bibfnamefont {X.-L.}\ \bibnamefont
  {Qi}}\ and\ \bibinfo {author} {\bibfnamefont {S.-C.}\ \bibnamefont {Zhang}},\
  }\href@noop {} {\bibfield  {journal} {\bibinfo  {journal} {Reviews of Modern
  Physics}\ }\textbf {\bibinfo {volume} {83}},\ \bibinfo {pages} {1057}
  (\bibinfo {year} {2011})}\BibitemShut {NoStop}%
\bibitem [{\citenamefont {Black-Schaffer}(2011)}]{BlackSchaffer2011}%
  \BibitemOpen
  \bibfield  {author} {\bibinfo {author} {\bibfnamefont {A.~M.}\ \bibnamefont
  {Black-Schaffer}},\ }\href@noop {} {\bibfield  {journal} {\bibinfo  {journal}
  {Physical Review B}\ }\textbf {\bibinfo {volume} {83}},\ \bibinfo {pages}
  {060504} (\bibinfo {year} {2011})}\BibitemShut {NoStop}%
\bibitem [{\citenamefont {Mourik}\ \emph {et~al.}(2012)\citenamefont {Mourik},
  \citenamefont {Zuo}, \citenamefont {Frolov}, \citenamefont {Plissard},
  \citenamefont {Bakkers},\ and\ \citenamefont {Kouwenhoven}}]{Mourik2012}%
  \BibitemOpen
  \bibfield  {author} {\bibinfo {author} {\bibfnamefont {V.}~\bibnamefont
  {Mourik}}, \bibinfo {author} {\bibfnamefont {K.}~\bibnamefont {Zuo}},
  \bibinfo {author} {\bibfnamefont {S.~M.}\ \bibnamefont {Frolov}}, \bibinfo
  {author} {\bibfnamefont {S.}~\bibnamefont {Plissard}}, \bibinfo {author}
  {\bibfnamefont {E.~P.}\ \bibnamefont {Bakkers}}, \ and\ \bibinfo {author}
  {\bibfnamefont {L.~P.}\ \bibnamefont {Kouwenhoven}},\ }\href@noop {}
  {\bibfield  {journal} {\bibinfo  {journal} {Science}\ }\textbf {\bibinfo
  {volume} {336}},\ \bibinfo {pages} {1003} (\bibinfo {year}
  {2012})}\BibitemShut {NoStop}%
\bibitem [{\citenamefont {Nakosai}\ \emph {et~al.}(2012)\citenamefont
  {Nakosai}, \citenamefont {Tanaka},\ and\ \citenamefont
  {Nagaosa}}]{Nakosai2012}%
  \BibitemOpen
  \bibfield  {author} {\bibinfo {author} {\bibfnamefont {S.}~\bibnamefont
  {Nakosai}}, \bibinfo {author} {\bibfnamefont {Y.}~\bibnamefont {Tanaka}}, \
  and\ \bibinfo {author} {\bibfnamefont {N.}~\bibnamefont {Nagaosa}},\
  }\href@noop {} {\bibfield  {journal} {\bibinfo  {journal} {Physical Review
  Letters}\ }\textbf {\bibinfo {volume} {108}},\ \bibinfo {pages} {147003}
  (\bibinfo {year} {2012})}\BibitemShut {NoStop}%
\bibitem [{\citenamefont {Stanescu}\ and\ \citenamefont
  {Tewari}(2013)}]{Stanescu2013}%
  \BibitemOpen
  \bibfield  {author} {\bibinfo {author} {\bibfnamefont {T.~D.}\ \bibnamefont
  {Stanescu}}\ and\ \bibinfo {author} {\bibfnamefont {S.}~\bibnamefont
  {Tewari}},\ }\href {\doibase 10.1088/0953-8984/25/23/233201} {\bibfield
  {journal} {\bibinfo  {journal} {Journal of Physics: Condensed Matter}\
  }\textbf {\bibinfo {volume} {25}},\ \bibinfo {pages} {233201} (\bibinfo
  {year} {2013})}\BibitemShut {NoStop}%
\bibitem [{\citenamefont {Goertzen}\ \emph {et~al.}(2017)\citenamefont
  {Goertzen}, \citenamefont {Tanaka},\ and\ \citenamefont
  {Nagai}}]{Goertzen2017}%
  \BibitemOpen
  \bibfield  {author} {\bibinfo {author} {\bibfnamefont {S.}~\bibnamefont
  {Goertzen}}, \bibinfo {author} {\bibfnamefont {K.}~\bibnamefont {Tanaka}}, \
  and\ \bibinfo {author} {\bibfnamefont {Y.}~\bibnamefont {Nagai}},\
  }\href@noop {} {\bibfield  {journal} {\bibinfo  {journal} {Physical Review
  B}\ }\textbf {\bibinfo {volume} {95}},\ \bibinfo {pages} {064509} (\bibinfo
  {year} {2017})}\BibitemShut {NoStop}%
\bibitem [{\citenamefont {Lutchyn}\ \emph {et~al.}(2018)\citenamefont
  {Lutchyn}, \citenamefont {Bakkers}, \citenamefont {Kouwenhoven},
  \citenamefont {Krogstrup}, \citenamefont {Marcus},\ and\ \citenamefont
  {Oreg}}]{Lutchyn2018}%
  \BibitemOpen
  \bibfield  {author} {\bibinfo {author} {\bibfnamefont {R.~M.}\ \bibnamefont
  {Lutchyn}}, \bibinfo {author} {\bibfnamefont {E.~P.}\ \bibnamefont
  {Bakkers}}, \bibinfo {author} {\bibfnamefont {L.~P.}\ \bibnamefont
  {Kouwenhoven}}, \bibinfo {author} {\bibfnamefont {P.}~\bibnamefont
  {Krogstrup}}, \bibinfo {author} {\bibfnamefont {C.~M.}\ \bibnamefont
  {Marcus}}, \ and\ \bibinfo {author} {\bibfnamefont {Y.}~\bibnamefont
  {Oreg}},\ }\href@noop {} {\bibfield  {journal} {\bibinfo  {journal} {Nature
  Reviews Materials}\ }\textbf {\bibinfo {volume} {3}},\ \bibinfo {pages} {52}
  (\bibinfo {year} {2018})}\BibitemShut {NoStop}%
\bibitem [{\citenamefont {Oreg}\ \emph {et~al.}(2010)\citenamefont {Oreg},
  \citenamefont {Refael},\ and\ \citenamefont {Von~Oppen}}]{Oreg2010}%
  \BibitemOpen
  \bibfield  {author} {\bibinfo {author} {\bibfnamefont {Y.}~\bibnamefont
  {Oreg}}, \bibinfo {author} {\bibfnamefont {G.}~\bibnamefont {Refael}}, \ and\
  \bibinfo {author} {\bibfnamefont {F.}~\bibnamefont {Von~Oppen}},\ }\href@noop
  {} {\bibfield  {journal} {\bibinfo  {journal} {Physical review letters}\
  }\textbf {\bibinfo {volume} {105}},\ \bibinfo {pages} {177002} (\bibinfo
  {year} {2010})}\BibitemShut {NoStop}%
\bibitem [{\citenamefont {Nadj-Perge}\ \emph {et~al.}(2014)\citenamefont
  {Nadj-Perge}, \citenamefont {Drozdov}, \citenamefont {Li}, \citenamefont
  {Chen}, \citenamefont {Jeon}, \citenamefont {Seo}, \citenamefont {MacDonald},
  \citenamefont {Bernevig},\ and\ \citenamefont {Yazdani}}]{NadjPerge2014}%
  \BibitemOpen
  \bibfield  {author} {\bibinfo {author} {\bibfnamefont {S.}~\bibnamefont
  {Nadj-Perge}}, \bibinfo {author} {\bibfnamefont {I.~K.}\ \bibnamefont
  {Drozdov}}, \bibinfo {author} {\bibfnamefont {J.}~\bibnamefont {Li}},
  \bibinfo {author} {\bibfnamefont {H.}~\bibnamefont {Chen}}, \bibinfo {author}
  {\bibfnamefont {S.}~\bibnamefont {Jeon}}, \bibinfo {author} {\bibfnamefont
  {J.}~\bibnamefont {Seo}}, \bibinfo {author} {\bibfnamefont {A.~H.}\
  \bibnamefont {MacDonald}}, \bibinfo {author} {\bibfnamefont {B.~A.}\
  \bibnamefont {Bernevig}}, \ and\ \bibinfo {author} {\bibfnamefont
  {A.}~\bibnamefont {Yazdani}},\ }\href@noop {} {\bibfield  {journal} {\bibinfo
   {journal} {Science}\ }\textbf {\bibinfo {volume} {346}},\ \bibinfo {pages}
  {602} (\bibinfo {year} {2014})}\BibitemShut {NoStop}%
\bibitem [{\citenamefont {Liu}\ \emph {et~al.}(2017)\citenamefont {Liu},
  \citenamefont {Sau}, \citenamefont {Stanescu},\ and\ \citenamefont
  {Sarma}}]{Liu2017a}%
  \BibitemOpen
  \bibfield  {author} {\bibinfo {author} {\bibfnamefont {C.-X.}\ \bibnamefont
  {Liu}}, \bibinfo {author} {\bibfnamefont {J.~D.}\ \bibnamefont {Sau}},
  \bibinfo {author} {\bibfnamefont {T.~D.}\ \bibnamefont {Stanescu}}, \ and\
  \bibinfo {author} {\bibfnamefont {S.~D.}\ \bibnamefont {Sarma}},\ }\href@noop
  {} {\bibfield  {journal} {\bibinfo  {journal} {Physical Review B}\ }\textbf
  {\bibinfo {volume} {96}},\ \bibinfo {pages} {075161} (\bibinfo {year}
  {2017})}\BibitemShut {NoStop}%
\bibitem [{\citenamefont {Frolov}\ \emph {et~al.}(2020)\citenamefont {Frolov},
  \citenamefont {Manfra},\ and\ \citenamefont {Sau}}]{Frolov2020}%
  \BibitemOpen
  \bibfield  {author} {\bibinfo {author} {\bibfnamefont {S.}~\bibnamefont
  {Frolov}}, \bibinfo {author} {\bibfnamefont {M.}~\bibnamefont {Manfra}}, \
  and\ \bibinfo {author} {\bibfnamefont {J.}~\bibnamefont {Sau}},\ }\href@noop
  {} {\bibfield  {journal} {\bibinfo  {journal} {Nature Physics}\ }\textbf
  {\bibinfo {volume} {16}},\ \bibinfo {pages} {718} (\bibinfo {year}
  {2020})}\BibitemShut {NoStop}%
\bibitem [{\citenamefont {Klinovaja}\ \emph {et~al.}(2013)\citenamefont
  {Klinovaja}, \citenamefont {Stano}, \citenamefont {Yazdani},\ and\
  \citenamefont {Loss}}]{Klinovaja2013}%
  \BibitemOpen
  \bibfield  {author} {\bibinfo {author} {\bibfnamefont {J.}~\bibnamefont
  {Klinovaja}}, \bibinfo {author} {\bibfnamefont {P.}~\bibnamefont {Stano}},
  \bibinfo {author} {\bibfnamefont {A.}~\bibnamefont {Yazdani}}, \ and\
  \bibinfo {author} {\bibfnamefont {D.}~\bibnamefont {Loss}},\ }\href@noop {}
  {\bibfield  {journal} {\bibinfo  {journal} {Physical review letters}\
  }\textbf {\bibinfo {volume} {111}},\ \bibinfo {pages} {186805} (\bibinfo
  {year} {2013})}\BibitemShut {NoStop}%
\bibitem [{\citenamefont {Li}\ \emph {et~al.}(2014)\citenamefont {Li},
  \citenamefont {Chen}, \citenamefont {Drozdov}, \citenamefont {Yazdani},
  \citenamefont {Bernevig},\ and\ \citenamefont {MacDonald}}]{Li2014}%
  \BibitemOpen
  \bibfield  {author} {\bibinfo {author} {\bibfnamefont {J.}~\bibnamefont
  {Li}}, \bibinfo {author} {\bibfnamefont {H.}~\bibnamefont {Chen}}, \bibinfo
  {author} {\bibfnamefont {I.~K.}\ \bibnamefont {Drozdov}}, \bibinfo {author}
  {\bibfnamefont {A.}~\bibnamefont {Yazdani}}, \bibinfo {author} {\bibfnamefont
  {B.~A.}\ \bibnamefont {Bernevig}}, \ and\ \bibinfo {author} {\bibfnamefont
  {A.}~\bibnamefont {MacDonald}},\ }\href@noop {} {\bibfield  {journal}
  {\bibinfo  {journal} {Physical Review B}\ }\textbf {\bibinfo {volume} {90}},\
  \bibinfo {pages} {235433} (\bibinfo {year} {2014})}\BibitemShut {NoStop}%
\bibitem [{\citenamefont {Choy}\ \emph {et~al.}(2011)\citenamefont {Choy},
  \citenamefont {Edge}, \citenamefont {Akhmerov},\ and\ \citenamefont
  {Beenakker}}]{choy_majorana_2011}%
  \BibitemOpen
  \bibfield  {author} {\bibinfo {author} {\bibfnamefont {T.-P.}\ \bibnamefont
  {Choy}}, \bibinfo {author} {\bibfnamefont {J.~M.}\ \bibnamefont {Edge}},
  \bibinfo {author} {\bibfnamefont {A.~R.}\ \bibnamefont {Akhmerov}}, \ and\
  \bibinfo {author} {\bibfnamefont {C.~W.~J.}\ \bibnamefont {Beenakker}},\
  }\href {\doibase 10.1103/PhysRevB.84.195442} {\bibfield  {journal} {\bibinfo
  {journal} {Physical Review B}\ }\textbf {\bibinfo {volume} {84}},\ \bibinfo
  {pages} {195442} (\bibinfo {year} {2011})}\BibitemShut {NoStop}%
\bibitem [{\citenamefont {Nadj-Perge}\ \emph {et~al.}(2013)\citenamefont
  {Nadj-Perge}, \citenamefont {Drozdov}, \citenamefont {Bernevig},\ and\
  \citenamefont {Yazdani}}]{nadj-perge_proposal_2013}%
  \BibitemOpen
  \bibfield  {author} {\bibinfo {author} {\bibfnamefont {S.}~\bibnamefont
  {Nadj-Perge}}, \bibinfo {author} {\bibfnamefont {I.~K.}\ \bibnamefont
  {Drozdov}}, \bibinfo {author} {\bibfnamefont {B.~A.}\ \bibnamefont
  {Bernevig}}, \ and\ \bibinfo {author} {\bibfnamefont {A.}~\bibnamefont
  {Yazdani}},\ }\href {\doibase 10.1103/PhysRevB.88.020407} {\bibfield
  {journal} {\bibinfo  {journal} {Physical Review B}\ }\textbf {\bibinfo
  {volume} {88}},\ \bibinfo {pages} {020407} (\bibinfo {year}
  {2013})}\BibitemShut {NoStop}%
\bibitem [{\citenamefont {Pientka}\ \emph {et~al.}(2013)\citenamefont
  {Pientka}, \citenamefont {Glazman},\ and\ \citenamefont
  {Von~Oppen}}]{pientka_topological_2013}%
  \BibitemOpen
  \bibfield  {author} {\bibinfo {author} {\bibfnamefont {F.}~\bibnamefont
  {Pientka}}, \bibinfo {author} {\bibfnamefont {L.~I.}\ \bibnamefont
  {Glazman}}, \ and\ \bibinfo {author} {\bibfnamefont {F.}~\bibnamefont
  {Von~Oppen}},\ }\href {\doibase 10.1103/PhysRevB.88.155420} {\bibfield
  {journal} {\bibinfo  {journal} {Physical Review B}\ }\textbf {\bibinfo
  {volume} {88}},\ \bibinfo {pages} {155420} (\bibinfo {year}
  {2013})}\BibitemShut {NoStop}%
\bibitem [{\citenamefont {Fornieri}\ \emph {et~al.}(2019)\citenamefont
  {Fornieri}, \citenamefont {Whiticar}, \citenamefont {Setiawan}, \citenamefont
  {Portol{\'e}s}, \citenamefont {Drachmann}, \citenamefont {Keselman},
  \citenamefont {Gronin}, \citenamefont {Thomas}, \citenamefont {Wang},
  \citenamefont {Kallaher} \emph {et~al.}}]{Fornieri2019}%
  \BibitemOpen
  \bibfield  {author} {\bibinfo {author} {\bibfnamefont {A.}~\bibnamefont
  {Fornieri}}, \bibinfo {author} {\bibfnamefont {A.~M.}\ \bibnamefont
  {Whiticar}}, \bibinfo {author} {\bibfnamefont {F.}~\bibnamefont {Setiawan}},
  \bibinfo {author} {\bibfnamefont {E.}~\bibnamefont {Portol{\'e}s}}, \bibinfo
  {author} {\bibfnamefont {A.~C.}\ \bibnamefont {Drachmann}}, \bibinfo {author}
  {\bibfnamefont {A.}~\bibnamefont {Keselman}}, \bibinfo {author}
  {\bibfnamefont {S.}~\bibnamefont {Gronin}}, \bibinfo {author} {\bibfnamefont
  {C.}~\bibnamefont {Thomas}}, \bibinfo {author} {\bibfnamefont
  {T.}~\bibnamefont {Wang}}, \bibinfo {author} {\bibfnamefont {R.}~\bibnamefont
  {Kallaher}},  \emph {et~al.},\ }\href@noop {} {\bibfield  {journal} {\bibinfo
   {journal} {Nature}\ }\textbf {\bibinfo {volume} {569}},\ \bibinfo {pages}
  {89} (\bibinfo {year} {2019})}\BibitemShut {NoStop}%
\bibitem [{\citenamefont {Pientka}\ \emph {et~al.}(2017)\citenamefont
  {Pientka}, \citenamefont {Keselman}, \citenamefont {Berg}, \citenamefont
  {Yacoby}, \citenamefont {Stern},\ and\ \citenamefont
  {Halperin}}]{Pientka2017}%
  \BibitemOpen
  \bibfield  {author} {\bibinfo {author} {\bibfnamefont {F.}~\bibnamefont
  {Pientka}}, \bibinfo {author} {\bibfnamefont {A.}~\bibnamefont {Keselman}},
  \bibinfo {author} {\bibfnamefont {E.}~\bibnamefont {Berg}}, \bibinfo {author}
  {\bibfnamefont {A.}~\bibnamefont {Yacoby}}, \bibinfo {author} {\bibfnamefont
  {A.}~\bibnamefont {Stern}}, \ and\ \bibinfo {author} {\bibfnamefont {B.~I.}\
  \bibnamefont {Halperin}},\ }\href@noop {} {\bibfield  {journal} {\bibinfo
  {journal} {Physical Review X}\ }\textbf {\bibinfo {volume} {7}},\ \bibinfo
  {pages} {021032} (\bibinfo {year} {2017})}\BibitemShut {NoStop}%
\bibitem [{\citenamefont {Marra}(2022)}]{Marra2022}%
  \BibitemOpen
  \bibfield  {author} {\bibinfo {author} {\bibfnamefont {P.}~\bibnamefont
  {Marra}},\ }\href@noop {} {\bibfield  {journal} {\bibinfo  {journal} {Journal
  of Applied Physics}\ }\textbf {\bibinfo {volume} {132}},\ \bibinfo {pages}
  {231101} (\bibinfo {year} {2022})}\BibitemShut {NoStop}%
\bibitem [{\citenamefont {Lu}\ \emph {et~al.}(2010)\citenamefont {Lu},
  \citenamefont {Sarma},\ and\ \citenamefont {Park}}]{Lu2010}%
  \BibitemOpen
  \bibfield  {author} {\bibinfo {author} {\bibfnamefont {H.}~\bibnamefont
  {Lu}}, \bibinfo {author} {\bibfnamefont {S.~D.}\ \bibnamefont {Sarma}}, \
  and\ \bibinfo {author} {\bibfnamefont {K.}~\bibnamefont {Park}},\ }\href@noop
  {} {\bibfield  {journal} {\bibinfo  {journal} {Physical Review B}\ }\textbf
  {\bibinfo {volume} {82}},\ \bibinfo {pages} {201303} (\bibinfo {year}
  {2010})}\BibitemShut {NoStop}%
\bibitem [{\citenamefont {Clarke}\ \emph {et~al.}(2013)\citenamefont {Clarke},
  \citenamefont {Alicea},\ and\ \citenamefont {Shtengel}}]{Clarke2013}%
  \BibitemOpen
  \bibfield  {author} {\bibinfo {author} {\bibfnamefont {D.~J.}\ \bibnamefont
  {Clarke}}, \bibinfo {author} {\bibfnamefont {J.}~\bibnamefont {Alicea}}, \
  and\ \bibinfo {author} {\bibfnamefont {K.}~\bibnamefont {Shtengel}},\
  }\href@noop {} {\bibfield  {journal} {\bibinfo  {journal} {Nature
  communications}\ }\textbf {\bibinfo {volume} {4}},\ \bibinfo {pages} {1}
  (\bibinfo {year} {2013})}\BibitemShut {NoStop}%
\bibitem [{\citenamefont {Mong}\ \emph {et~al.}(2014)\citenamefont {Mong},
  \citenamefont {Clarke}, \citenamefont {Alicea}, \citenamefont {Lindner},
  \citenamefont {Fendley}, \citenamefont {Nayak}, \citenamefont {Oreg},
  \citenamefont {Stern}, \citenamefont {Berg}, \citenamefont {Shtengel} \emph
  {et~al.}}]{Mong2014}%
  \BibitemOpen
  \bibfield  {author} {\bibinfo {author} {\bibfnamefont {R.~S.}\ \bibnamefont
  {Mong}}, \bibinfo {author} {\bibfnamefont {D.~J.}\ \bibnamefont {Clarke}},
  \bibinfo {author} {\bibfnamefont {J.}~\bibnamefont {Alicea}}, \bibinfo
  {author} {\bibfnamefont {N.~H.}\ \bibnamefont {Lindner}}, \bibinfo {author}
  {\bibfnamefont {P.}~\bibnamefont {Fendley}}, \bibinfo {author} {\bibfnamefont
  {C.}~\bibnamefont {Nayak}}, \bibinfo {author} {\bibfnamefont
  {Y.}~\bibnamefont {Oreg}}, \bibinfo {author} {\bibfnamefont {A.}~\bibnamefont
  {Stern}}, \bibinfo {author} {\bibfnamefont {E.}~\bibnamefont {Berg}},
  \bibinfo {author} {\bibfnamefont {K.}~\bibnamefont {Shtengel}},  \emph
  {et~al.},\ }\href@noop {} {\bibfield  {journal} {\bibinfo  {journal}
  {Physical Review X}\ }\textbf {\bibinfo {volume} {4}},\ \bibinfo {pages}
  {011036} (\bibinfo {year} {2014})}\BibitemShut {NoStop}%
\bibitem [{\citenamefont {Alicea}\ and\ \citenamefont
  {Stern}(2015)}]{alicea2015}%
  \BibitemOpen
  \bibfield  {author} {\bibinfo {author} {\bibfnamefont {J.}~\bibnamefont
  {Alicea}}\ and\ \bibinfo {author} {\bibfnamefont {A.}~\bibnamefont {Stern}},\
  }\href {\doibase 10.1088/0031-8949/2015/t164/014006} {\bibfield  {journal}
  {\bibinfo  {journal} {Physica Scripta}\ }\textbf {\bibinfo {volume} {T164}},\
  \bibinfo {pages} {014006} (\bibinfo {year} {2015})}\BibitemShut {NoStop}%
\bibitem [{\citenamefont {Alicea}\ and\ \citenamefont
  {Fendley}(2016)}]{Alicea2016}%
  \BibitemOpen
  \bibfield  {author} {\bibinfo {author} {\bibfnamefont {J.}~\bibnamefont
  {Alicea}}\ and\ \bibinfo {author} {\bibfnamefont {P.}~\bibnamefont
  {Fendley}},\ }\href@noop {} {\bibfield  {journal} {\bibinfo  {journal}
  {Annual Review of Condensed Matter Physics}\ }\textbf {\bibinfo {volume}
  {7}},\ \bibinfo {pages} {119} (\bibinfo {year} {2016})}\BibitemShut {NoStop}%
\bibitem [{\citenamefont {Amet}\ \emph {et~al.}(2016)\citenamefont {Amet},
  \citenamefont {Ke}, \citenamefont {Borzenets}, \citenamefont {Wang},
  \citenamefont {Watanabe}, \citenamefont {Taniguchi}, \citenamefont {Deacon},
  \citenamefont {Yamamoto}, \citenamefont {Bomze}, \citenamefont {Tarucha},\
  and\ \citenamefont {Finkelstein}}]{Amet2016}%
  \BibitemOpen
  \bibfield  {author} {\bibinfo {author} {\bibfnamefont {F.}~\bibnamefont
  {Amet}}, \bibinfo {author} {\bibfnamefont {C.~T.}\ \bibnamefont {Ke}},
  \bibinfo {author} {\bibfnamefont {I.~V.}\ \bibnamefont {Borzenets}}, \bibinfo
  {author} {\bibfnamefont {J.}~\bibnamefont {Wang}}, \bibinfo {author}
  {\bibfnamefont {K.}~\bibnamefont {Watanabe}}, \bibinfo {author}
  {\bibfnamefont {T.}~\bibnamefont {Taniguchi}}, \bibinfo {author}
  {\bibfnamefont {R.~S.}\ \bibnamefont {Deacon}}, \bibinfo {author}
  {\bibfnamefont {M.}~\bibnamefont {Yamamoto}}, \bibinfo {author}
  {\bibfnamefont {Y.}~\bibnamefont {Bomze}}, \bibinfo {author} {\bibfnamefont
  {S.}~\bibnamefont {Tarucha}}, \ and\ \bibinfo {author} {\bibfnamefont
  {G.}~\bibnamefont {Finkelstein}},\ }\href {\doibase 10.1126/science.aad6203}
  {\bibfield  {journal} {\bibinfo  {journal} {Science}\ }\textbf {\bibinfo
  {volume} {352}},\ \bibinfo {pages} {966} (\bibinfo {year}
  {2016})}\BibitemShut {NoStop}%
\bibitem [{\citenamefont {Liang}\ \emph {et~al.}(2019)\citenamefont {Liang},
  \citenamefont {Simion},\ and\ \citenamefont {Lyanda-Geller}}]{liang2019}%
  \BibitemOpen
  \bibfield  {author} {\bibinfo {author} {\bibfnamefont {J.}~\bibnamefont
  {Liang}}, \bibinfo {author} {\bibfnamefont {G.}~\bibnamefont {Simion}}, \
  and\ \bibinfo {author} {\bibfnamefont {Y.}~\bibnamefont {Lyanda-Geller}},\
  }\href@noop {} {\bibfield  {journal} {\bibinfo  {journal} {Physical Review
  B}\ }\textbf {\bibinfo {volume} {100}},\ \bibinfo {pages} {075155} (\bibinfo
  {year} {2019})}\BibitemShut {NoStop}%
\bibitem [{\citenamefont {G{\"u}l}\ \emph {et~al.}(2022)\citenamefont
  {G{\"u}l}, \citenamefont {Ronen}, \citenamefont {Lee}, \citenamefont
  {Shapourian}, \citenamefont {Zauberman}, \citenamefont {Lee}, \citenamefont
  {Watanabe}, \citenamefont {Taniguchi}, \citenamefont {Vishwanath},
  \citenamefont {Yacoby} \emph {et~al.}}]{Gul2020}%
  \BibitemOpen
  \bibfield  {author} {\bibinfo {author} {\bibfnamefont {{\"O}.}~\bibnamefont
  {G{\"u}l}}, \bibinfo {author} {\bibfnamefont {Y.}~\bibnamefont {Ronen}},
  \bibinfo {author} {\bibfnamefont {S.~Y.}\ \bibnamefont {Lee}}, \bibinfo
  {author} {\bibfnamefont {H.}~\bibnamefont {Shapourian}}, \bibinfo {author}
  {\bibfnamefont {J.}~\bibnamefont {Zauberman}}, \bibinfo {author}
  {\bibfnamefont {Y.~H.}\ \bibnamefont {Lee}}, \bibinfo {author} {\bibfnamefont
  {K.}~\bibnamefont {Watanabe}}, \bibinfo {author} {\bibfnamefont
  {T.}~\bibnamefont {Taniguchi}}, \bibinfo {author} {\bibfnamefont
  {A.}~\bibnamefont {Vishwanath}}, \bibinfo {author} {\bibfnamefont
  {A.}~\bibnamefont {Yacoby}},  \emph {et~al.},\ }\href@noop {} {\bibfield
  {journal} {\bibinfo  {journal} {Physical Review X}\ }\textbf {\bibinfo
  {volume} {12}},\ \bibinfo {pages} {021057} (\bibinfo {year}
  {2022})}\BibitemShut {NoStop}%
\bibitem [{\citenamefont {Shaffer}\ \emph {et~al.}(2021)\citenamefont
  {Shaffer}, \citenamefont {Wang},\ and\ \citenamefont {Santos}}]{shaffer2021}%
  \BibitemOpen
  \bibfield  {author} {\bibinfo {author} {\bibfnamefont {D.}~\bibnamefont
  {Shaffer}}, \bibinfo {author} {\bibfnamefont {J.}~\bibnamefont {Wang}}, \
  and\ \bibinfo {author} {\bibfnamefont {L.~H.}\ \bibnamefont {Santos}},\
  }\href@noop {} {\bibfield  {journal} {\bibinfo  {journal} {Physical Review
  B}\ }\textbf {\bibinfo {volume} {104}},\ \bibinfo {pages} {184501} (\bibinfo
  {year} {2021})}\BibitemShut {NoStop}%
\bibitem [{\citenamefont {Zocher}\ and\ \citenamefont
  {Rosenow}(2016)}]{Zocher2016}%
  \BibitemOpen
  \bibfield  {author} {\bibinfo {author} {\bibfnamefont {B.}~\bibnamefont
  {Zocher}}\ and\ \bibinfo {author} {\bibfnamefont {B.}~\bibnamefont
  {Rosenow}},\ }\href@noop {} {\bibfield  {journal} {\bibinfo  {journal}
  {Physical Review B}\ }\textbf {\bibinfo {volume} {93}},\ \bibinfo {pages}
  {214504} (\bibinfo {year} {2016})}\BibitemShut {NoStop}%
\bibitem [{\citenamefont {Mishmash}\ \emph {et~al.}(2019)\citenamefont
  {Mishmash}, \citenamefont {Yazdani},\ and\ \citenamefont
  {Zaletel}}]{Mishmash2019}%
  \BibitemOpen
  \bibfield  {author} {\bibinfo {author} {\bibfnamefont {R.~V.}\ \bibnamefont
  {Mishmash}}, \bibinfo {author} {\bibfnamefont {A.}~\bibnamefont {Yazdani}}, \
  and\ \bibinfo {author} {\bibfnamefont {M.~P.}\ \bibnamefont {Zaletel}},\
  }\href@noop {} {\bibfield  {journal} {\bibinfo  {journal} {Physical Review
  B}\ }\textbf {\bibinfo {volume} {99}},\ \bibinfo {pages} {115427} (\bibinfo
  {year} {2019})}\BibitemShut {NoStop}%
\bibitem [{\citenamefont {Jeon}\ \emph {et~al.}(2019)\citenamefont {Jeon},
  \citenamefont {Jain},\ and\ \citenamefont {Liu}}]{Jeon2019}%
  \BibitemOpen
  \bibfield  {author} {\bibinfo {author} {\bibfnamefont {G.~S.}\ \bibnamefont
  {Jeon}}, \bibinfo {author} {\bibfnamefont {J.}~\bibnamefont {Jain}}, \ and\
  \bibinfo {author} {\bibfnamefont {C.-X.}\ \bibnamefont {Liu}},\ }\href@noop
  {} {\bibfield  {journal} {\bibinfo  {journal} {Physical Review B}\ }\textbf
  {\bibinfo {volume} {99}},\ \bibinfo {pages} {094509} (\bibinfo {year}
  {2019})}\BibitemShut {NoStop}%
\bibitem [{\citenamefont {Chaudhary}\ and\ \citenamefont
  {MacDonald}(2020)}]{Chaudhary2020}%
  \BibitemOpen
  \bibfield  {author} {\bibinfo {author} {\bibfnamefont {G.}~\bibnamefont
  {Chaudhary}}\ and\ \bibinfo {author} {\bibfnamefont {A.~H.}\ \bibnamefont
  {MacDonald}},\ }\href@noop {} {\bibfield  {journal} {\bibinfo  {journal}
  {Physical Review B}\ }\textbf {\bibinfo {volume} {101}},\ \bibinfo {pages}
  {024516} (\bibinfo {year} {2020})}\BibitemShut {NoStop}%
\bibitem [{\citenamefont {Pathak}\ \emph {et~al.}(2021)\citenamefont {Pathak},
  \citenamefont {Plugge},\ and\ \citenamefont {Franz}}]{pathak2021}%
  \BibitemOpen
  \bibfield  {author} {\bibinfo {author} {\bibfnamefont {V.}~\bibnamefont
  {Pathak}}, \bibinfo {author} {\bibfnamefont {S.}~\bibnamefont {Plugge}}, \
  and\ \bibinfo {author} {\bibfnamefont {M.}~\bibnamefont {Franz}},\
  }\href@noop {} {\bibfield  {journal} {\bibinfo  {journal} {Annals of
  Physics}\ ,\ \bibinfo {pages} {168431}} (\bibinfo {year} {2021})}\BibitemShut
  {NoStop}%
\bibitem [{\citenamefont {Rajagopal}\ and\ \citenamefont
  {Vasudevan}(1991)}]{Rajagopal1991}%
  \BibitemOpen
  \bibfield  {author} {\bibinfo {author} {\bibfnamefont {A.}~\bibnamefont
  {Rajagopal}}\ and\ \bibinfo {author} {\bibfnamefont {R.}~\bibnamefont
  {Vasudevan}},\ }\href@noop {} {\bibfield  {journal} {\bibinfo  {journal}
  {Physical Review B}\ }\textbf {\bibinfo {volume} {44}},\ \bibinfo {pages}
  {2807} (\bibinfo {year} {1991})}\BibitemShut {NoStop}%
\bibitem [{\citenamefont {Tesanovic}\ \emph {et~al.}(1989)\citenamefont
  {Tesanovic}, \citenamefont {Rasolt},\ and\ \citenamefont
  {Xing}}]{tevsanovic1989}%
  \BibitemOpen
  \bibfield  {author} {\bibinfo {author} {\bibfnamefont {Z.}~\bibnamefont
  {Tesanovic}}, \bibinfo {author} {\bibfnamefont {M.}~\bibnamefont {Rasolt}}, \
  and\ \bibinfo {author} {\bibfnamefont {L.}~\bibnamefont {Xing}},\ }\href@noop
  {} {\bibfield  {journal} {\bibinfo  {journal} {Physical Review Letters}\
  }\textbf {\bibinfo {volume} {63}},\ \bibinfo {pages} {2425} (\bibinfo {year}
  {1989})}\BibitemShut {NoStop}%
\bibitem [{\citenamefont {Tesanovic}\ \emph {et~al.}(1991)\citenamefont
  {Tesanovic}, \citenamefont {Rasolt},\ and\ \citenamefont
  {Xing}}]{tevsanovic1991}%
  \BibitemOpen
  \bibfield  {author} {\bibinfo {author} {\bibfnamefont {Z.}~\bibnamefont
  {Tesanovic}}, \bibinfo {author} {\bibfnamefont {M.}~\bibnamefont {Rasolt}}, \
  and\ \bibinfo {author} {\bibfnamefont {L.}~\bibnamefont {Xing}},\ }\href@noop
  {} {\bibfield  {journal} {\bibinfo  {journal} {Physical Review B}\ }\textbf
  {\bibinfo {volume} {43}},\ \bibinfo {pages} {288} (\bibinfo {year}
  {1991})}\BibitemShut {NoStop}%
\bibitem [{\citenamefont {Norman}(1991)}]{PhysRevLett.66.842}%
  \BibitemOpen
  \bibfield  {author} {\bibinfo {author} {\bibfnamefont {M.~R.}\ \bibnamefont
  {Norman}},\ }\href {\doibase 10.1103/PhysRevLett.66.842} {\bibfield
  {journal} {\bibinfo  {journal} {Physical Review Letters}\ }\textbf {\bibinfo
  {volume} {66}},\ \bibinfo {pages} {842} (\bibinfo {year} {1991})}\BibitemShut
  {NoStop}%
\bibitem [{\citenamefont {Akera}\ \emph {et~al.}(1991)\citenamefont {Akera},
  \citenamefont {MacDonald}, \citenamefont {Girvin},\ and\ \citenamefont
  {Norman}}]{Akera1991}%
  \BibitemOpen
  \bibfield  {author} {\bibinfo {author} {\bibfnamefont {H.}~\bibnamefont
  {Akera}}, \bibinfo {author} {\bibfnamefont {A.}~\bibnamefont {MacDonald}},
  \bibinfo {author} {\bibfnamefont {S.}~\bibnamefont {Girvin}}, \ and\ \bibinfo
  {author} {\bibfnamefont {M.}~\bibnamefont {Norman}},\ }\href@noop {}
  {\bibfield  {journal} {\bibinfo  {journal} {Physical review letters}\
  }\textbf {\bibinfo {volume} {67}},\ \bibinfo {pages} {2375} (\bibinfo {year}
  {1991})}\BibitemShut {NoStop}%
\bibitem [{\citenamefont {Rajagopal}\ and\ \citenamefont
  {Ryan}(1991)}]{rajagopal1991quantum}%
  \BibitemOpen
  \bibfield  {author} {\bibinfo {author} {\bibfnamefont {A.}~\bibnamefont
  {Rajagopal}}\ and\ \bibinfo {author} {\bibfnamefont {J.}~\bibnamefont
  {Ryan}},\ }\href@noop {} {\bibfield  {journal} {\bibinfo  {journal} {Physical
  Review B}\ }\textbf {\bibinfo {volume} {44}},\ \bibinfo {pages} {10280}
  (\bibinfo {year} {1991})}\BibitemShut {NoStop}%
\bibitem [{\citenamefont {Norman}\ \emph {et~al.}(1991)\citenamefont {Norman},
  \citenamefont {Akera},\ and\ \citenamefont {MacDonald}}]{norman1991landau}%
  \BibitemOpen
  \bibfield  {author} {\bibinfo {author} {\bibfnamefont {M.}~\bibnamefont
  {Norman}}, \bibinfo {author} {\bibfnamefont {H.}~\bibnamefont {Akera}}, \
  and\ \bibinfo {author} {\bibfnamefont {A.}~\bibnamefont {MacDonald}},\
  }\href@noop {} {\emph {\bibinfo {title} {Landau quantization and
  superconductivity at high magnetic fields}}},\ \bibinfo {type} {Tech. Rep.}\
  (\bibinfo {year} {1991})\BibitemShut {NoStop}%
\bibitem [{\citenamefont {Rasolt}\ and\ \citenamefont
  {Tesanovic}(1992)}]{Rasolt1992}%
  \BibitemOpen
  \bibfield  {author} {\bibinfo {author} {\bibfnamefont {M.}~\bibnamefont
  {Rasolt}}\ and\ \bibinfo {author} {\bibfnamefont {Z.}~\bibnamefont
  {Tesanovic}},\ }\href@noop {} {\bibfield  {journal} {\bibinfo  {journal}
  {Reviews of Modern Physics}\ }\textbf {\bibinfo {volume} {64}},\ \bibinfo
  {pages} {709} (\bibinfo {year} {1992})}\BibitemShut {NoStop}%
\bibitem [{\citenamefont {MacDonald}\ \emph {et~al.}(1992)\citenamefont
  {MacDonald}, \citenamefont {Akera},\ and\ \citenamefont
  {Norman}}]{MacDonald1992}%
  \BibitemOpen
  \bibfield  {author} {\bibinfo {author} {\bibfnamefont {A.}~\bibnamefont
  {MacDonald}}, \bibinfo {author} {\bibfnamefont {H.}~\bibnamefont {Akera}}, \
  and\ \bibinfo {author} {\bibfnamefont {M.}~\bibnamefont {Norman}},\
  }\href@noop {} {\bibfield  {journal} {\bibinfo  {journal} {Physical Review
  B}\ }\textbf {\bibinfo {volume} {45}},\ \bibinfo {pages} {10147} (\bibinfo
  {year} {1992})}\BibitemShut {NoStop}%
\bibitem [{\citenamefont {Norman}\ \emph {et~al.}(1992)\citenamefont {Norman},
  \citenamefont {Akera},\ and\ \citenamefont {MacDonald}}]{Norman1992}%
  \BibitemOpen
  \bibfield  {author} {\bibinfo {author} {\bibfnamefont {M.}~\bibnamefont
  {Norman}}, \bibinfo {author} {\bibfnamefont {H.}~\bibnamefont {Akera}}, \
  and\ \bibinfo {author} {\bibfnamefont {A.}~\bibnamefont {MacDonald}},\
  }\href@noop {} {\bibfield  {journal} {\bibinfo  {journal} {Physica C:
  Superconductivity}\ }\textbf {\bibinfo {volume} {196}},\ \bibinfo {pages}
  {43} (\bibinfo {year} {1992})}\BibitemShut {NoStop}%
\bibitem [{\citenamefont {Rajagopal}(1992)}]{rajagopal1992solutions}%
  \BibitemOpen
  \bibfield  {author} {\bibinfo {author} {\bibfnamefont {A.}~\bibnamefont
  {Rajagopal}},\ }\href@noop {} {\bibfield  {journal} {\bibinfo  {journal}
  {Physical Review B}\ }\textbf {\bibinfo {volume} {46}},\ \bibinfo {pages}
  {1224} (\bibinfo {year} {1992})}\BibitemShut {NoStop}%
\bibitem [{\citenamefont {MacDonald}\ \emph {et~al.}(1993)\citenamefont
  {MacDonald}, \citenamefont {Akera},\ and\ \citenamefont
  {Norman}}]{macdonald1993quantum}%
  \BibitemOpen
  \bibfield  {author} {\bibinfo {author} {\bibfnamefont {A.}~\bibnamefont
  {MacDonald}}, \bibinfo {author} {\bibfnamefont {H.}~\bibnamefont {Akera}}, \
  and\ \bibinfo {author} {\bibfnamefont {M.}~\bibnamefont {Norman}},\
  }\href@noop {} {\bibfield  {journal} {\bibinfo  {journal} {Australian journal
  of physics}\ }\textbf {\bibinfo {volume} {46}},\ \bibinfo {pages} {333}
  (\bibinfo {year} {1993})}\BibitemShut {NoStop}%
\bibitem [{\citenamefont {Ryan}\ and\ \citenamefont
  {Rajagopal}(1993{\natexlab{a}})}]{ryan1993manifestations}%
  \BibitemOpen
  \bibfield  {author} {\bibinfo {author} {\bibfnamefont {J.}~\bibnamefont
  {Ryan}}\ and\ \bibinfo {author} {\bibfnamefont {A.}~\bibnamefont
  {Rajagopal}},\ }\href@noop {} {\bibfield  {journal} {\bibinfo  {journal}
  {Journal of Physics and Chemistry of Solids}\ }\textbf {\bibinfo {volume}
  {54}},\ \bibinfo {pages} {1281} (\bibinfo {year}
  {1993}{\natexlab{a}})}\BibitemShut {NoStop}%
\bibitem [{\citenamefont {Ryan}\ and\ \citenamefont
  {Rajagopal}(1993{\natexlab{b}})}]{ryan1993vortex}%
  \BibitemOpen
  \bibfield  {author} {\bibinfo {author} {\bibfnamefont {J.}~\bibnamefont
  {Ryan}}\ and\ \bibinfo {author} {\bibfnamefont {A.}~\bibnamefont
  {Rajagopal}},\ }\href@noop {} {\bibfield  {journal} {\bibinfo  {journal}
  {Physical Review B}\ }\textbf {\bibinfo {volume} {47}},\ \bibinfo {pages}
  {8843} (\bibinfo {year} {1993}{\natexlab{b}})}\BibitemShut {NoStop}%
\bibitem [{\citenamefont {Norman}\ \emph {et~al.}(1995)\citenamefont {Norman},
  \citenamefont {MacDonald},\ and\ \citenamefont {Akera}}]{norman1995}%
  \BibitemOpen
  \bibfield  {author} {\bibinfo {author} {\bibfnamefont {M.}~\bibnamefont
  {Norman}}, \bibinfo {author} {\bibfnamefont {A.}~\bibnamefont {MacDonald}}, \
  and\ \bibinfo {author} {\bibfnamefont {H.}~\bibnamefont {Akera}},\
  }\href@noop {} {\bibfield  {journal} {\bibinfo  {journal} {Physical Review
  B}\ }\textbf {\bibinfo {volume} {51}},\ \bibinfo {pages} {5927} (\bibinfo
  {year} {1995})}\BibitemShut {NoStop}%
\bibitem [{\citenamefont {Ma{\'s}ka}(2002)}]{maska2002reentrant}%
  \BibitemOpen
  \bibfield  {author} {\bibinfo {author} {\bibfnamefont {M.~M.}\ \bibnamefont
  {Ma{\'s}ka}},\ }\href@noop {} {\bibfield  {journal} {\bibinfo  {journal}
  {Physical Review B}\ }\textbf {\bibinfo {volume} {66}},\ \bibinfo {pages}
  {054533} (\bibinfo {year} {2002})}\BibitemShut {NoStop}%
\bibitem [{\citenamefont {Scherpelz}\ \emph {et~al.}(2013)\citenamefont
  {Scherpelz}, \citenamefont {Wulin}, \citenamefont {{\v{S}}op{\'\i}k},
  \citenamefont {Levin},\ and\ \citenamefont
  {Rajagopal}}]{scherpelz2013general}%
  \BibitemOpen
  \bibfield  {author} {\bibinfo {author} {\bibfnamefont {P.}~\bibnamefont
  {Scherpelz}}, \bibinfo {author} {\bibfnamefont {D.}~\bibnamefont {Wulin}},
  \bibinfo {author} {\bibfnamefont {B.}~\bibnamefont {{\v{S}}op{\'\i}k}},
  \bibinfo {author} {\bibfnamefont {K.}~\bibnamefont {Levin}}, \ and\ \bibinfo
  {author} {\bibfnamefont {A.}~\bibnamefont {Rajagopal}},\ }\href@noop {}
  {\bibfield  {journal} {\bibinfo  {journal} {Physical Review B}\ }\textbf
  {\bibinfo {volume} {87}},\ \bibinfo {pages} {024516} (\bibinfo {year}
  {2013})}\BibitemShut {NoStop}%
\bibitem [{\citenamefont {Ran}\ \emph {et~al.}(2019)\citenamefont {Ran},
  \citenamefont {Eckberg}, \citenamefont {Ding}, \citenamefont {Furukawa},
  \citenamefont {Metz}, \citenamefont {Saha}, \citenamefont {Liu},
  \citenamefont {Zic}, \citenamefont {Kim}, \citenamefont {Paglione} \emph
  {et~al.}}]{ran2019}%
  \BibitemOpen
  \bibfield  {author} {\bibinfo {author} {\bibfnamefont {S.}~\bibnamefont
  {Ran}}, \bibinfo {author} {\bibfnamefont {C.}~\bibnamefont {Eckberg}},
  \bibinfo {author} {\bibfnamefont {Q.}~\bibnamefont {Ding}}, \bibinfo {author}
  {\bibfnamefont {Y.}~\bibnamefont {Furukawa}}, \bibinfo {author}
  {\bibfnamefont {T.}~\bibnamefont {Metz}}, \bibinfo {author} {\bibfnamefont
  {S.}~\bibnamefont {Saha}}, \bibinfo {author} {\bibfnamefont {I.}~\bibnamefont
  {Liu}}, \bibinfo {author} {\bibfnamefont {M.}~\bibnamefont {Zic}}, \bibinfo
  {author} {\bibfnamefont {H.}~\bibnamefont {Kim}}, \bibinfo {author}
  {\bibfnamefont {J.}~\bibnamefont {Paglione}},  \emph {et~al.},\ }\href@noop
  {} {\bibfield  {journal} {\bibinfo  {journal} {Science}\ }\textbf {\bibinfo
  {volume} {6454}},\ \bibinfo {pages} {365} (\bibinfo {year}
  {2019})}\BibitemShut {NoStop}%
\bibitem [{\citenamefont {Kim}\ \emph {et~al.}(2019)\citenamefont {Kim},
  \citenamefont {Chien},\ and\ \citenamefont {Lin}}]{Kim2019}%
  \BibitemOpen
  \bibfield  {author} {\bibinfo {author} {\bibfnamefont {T.}~\bibnamefont
  {Kim}}, \bibinfo {author} {\bibfnamefont {C.-C.}\ \bibnamefont {Chien}}, \
  and\ \bibinfo {author} {\bibfnamefont {S.-Z.}\ \bibnamefont {Lin}},\
  }\href@noop {} {\bibfield  {journal} {\bibinfo  {journal} {Physical Review
  B}\ }\textbf {\bibinfo {volume} {99}},\ \bibinfo {pages} {054509} (\bibinfo
  {year} {2019})}\BibitemShut {NoStop}%
\bibitem [{\citenamefont {Chaudhary}\ \emph {et~al.}(2021)\citenamefont
  {Chaudhary}, \citenamefont {MacDonald},\ and\ \citenamefont
  {Norman}}]{chaudhary2021}%
  \BibitemOpen
  \bibfield  {author} {\bibinfo {author} {\bibfnamefont {G.}~\bibnamefont
  {Chaudhary}}, \bibinfo {author} {\bibfnamefont {A.}~\bibnamefont
  {MacDonald}}, \ and\ \bibinfo {author} {\bibfnamefont {M.}~\bibnamefont
  {Norman}},\ }\href@noop {} {\bibfield  {journal} {\bibinfo  {journal}
  {Physical Review Research}\ }\textbf {\bibinfo {volume} {3}},\ \bibinfo
  {pages} {033260} (\bibinfo {year} {2021})}\BibitemShut {NoStop}%
\bibitem [{\citenamefont {Fulde}\ and\ \citenamefont
  {Ferrell}(1964)}]{fulde_superconductivity_1964}%
  \BibitemOpen
  \bibfield  {author} {\bibinfo {author} {\bibfnamefont {P.}~\bibnamefont
  {Fulde}}\ and\ \bibinfo {author} {\bibfnamefont {R.~A.}\ \bibnamefont
  {Ferrell}},\ }\href {\doibase 10.1103/PhysRev.135.A550} {\bibfield  {journal}
  {\bibinfo  {journal} {Physical Review}\ }\textbf {\bibinfo {volume} {135}},\
  \bibinfo {pages} {A550} (\bibinfo {year} {1964})}\BibitemShut {NoStop}%
\bibitem [{\citenamefont {Schirmer}\ \emph {et~al.}(2022)\citenamefont
  {Schirmer}, \citenamefont {Jain},\ and\ \citenamefont {Liu}}]{Schirmer2022e}%
  \BibitemOpen
  \bibfield  {author} {\bibinfo {author} {\bibfnamefont {J.}~\bibnamefont
  {Schirmer}}, \bibinfo {author} {\bibfnamefont {J.~K.}\ \bibnamefont {Jain}},
  \ and\ \bibinfo {author} {\bibfnamefont {C.~X.}\ \bibnamefont {Liu}},\
  }\href@noop {} {\enquote {\bibinfo {title} {Supplementary materials:
  Topological superconductivity induced by spin-orbit coupling, perpendicular
  magnetic field and superlattice potential},}\ } (\bibinfo {year}
  {2022})\BibitemShut {NoStop}%
\bibitem [{\citenamefont {Harper}(1955)}]{harper1955single}%
  \BibitemOpen
  \bibfield  {author} {\bibinfo {author} {\bibfnamefont {P.~G.}\ \bibnamefont
  {Harper}},\ }\href@noop {} {\bibfield  {journal} {\bibinfo  {journal}
  {Proceedings of the Physical Society. Section A}\ }\textbf {\bibinfo {volume}
  {68}},\ \bibinfo {pages} {874} (\bibinfo {year} {1955})}\BibitemShut
  {NoStop}%
\bibitem [{\citenamefont {Azbel}(1964)}]{azbel1964energy}%
  \BibitemOpen
  \bibfield  {author} {\bibinfo {author} {\bibfnamefont {M.~Y.}\ \bibnamefont
  {Azbel}},\ }\href@noop {} {\bibfield  {journal} {\bibinfo  {journal} {Sov
  Phys JETP}\ }\textbf {\bibinfo {volume} {19}},\ \bibinfo {pages} {634}
  (\bibinfo {year} {1964})}\BibitemShut {NoStop}%
\bibitem [{\citenamefont {Zak}(1964{\natexlab{a}})}]{PhysRev.134.A1602}%
  \BibitemOpen
  \bibfield  {author} {\bibinfo {author} {\bibfnamefont {J.}~\bibnamefont
  {Zak}},\ }\href {\doibase 10.1103/PhysRev.134.A1602} {\bibfield  {journal}
  {\bibinfo  {journal} {Physical Review}\ }\textbf {\bibinfo {volume} {134}},\
  \bibinfo {pages} {A1602} (\bibinfo {year} {1964}{\natexlab{a}})}\BibitemShut
  {NoStop}%
\bibitem [{\citenamefont {Brown}(1964)}]{PhysRev.133.A1038}%
  \BibitemOpen
  \bibfield  {author} {\bibinfo {author} {\bibfnamefont {E.}~\bibnamefont
  {Brown}},\ }\href {\doibase 10.1103/PhysRev.133.A1038} {\bibfield  {journal}
  {\bibinfo  {journal} {Physical Review}\ }\textbf {\bibinfo {volume} {133}},\
  \bibinfo {pages} {A1038} (\bibinfo {year} {1964})}\BibitemShut {NoStop}%
\bibitem [{\citenamefont {Zak}(1964{\natexlab{b}})}]{zak1964magnetic}%
  \BibitemOpen
  \bibfield  {author} {\bibinfo {author} {\bibfnamefont {J.}~\bibnamefont
  {Zak}},\ }\href@noop {} {\bibfield  {journal} {\bibinfo  {journal} {Physical
  Review}\ }\textbf {\bibinfo {volume} {134}},\ \bibinfo {pages} {A1607}
  (\bibinfo {year} {1964}{\natexlab{b}})}\BibitemShut {NoStop}%
\bibitem [{\citenamefont {Hofstadter}(1976)}]{Hofstadter1976}%
  \BibitemOpen
  \bibfield  {author} {\bibinfo {author} {\bibfnamefont {D.~R.}\ \bibnamefont
  {Hofstadter}},\ }\href@noop {} {\bibfield  {journal} {\bibinfo  {journal}
  {Physical Review B}\ }\textbf {\bibinfo {volume} {14}},\ \bibinfo {pages}
  {2239} (\bibinfo {year} {1976})}\BibitemShut {NoStop}%
\bibitem [{\citenamefont {Yang}\ and\ \citenamefont
  {Agterberg}(2000)}]{Yang2000}%
  \BibitemOpen
  \bibfield  {author} {\bibinfo {author} {\bibfnamefont {K.}~\bibnamefont
  {Yang}}\ and\ \bibinfo {author} {\bibfnamefont {D.~F.}\ \bibnamefont
  {Agterberg}},\ }\href@noop {} {\bibfield  {journal} {\bibinfo  {journal}
  {Physical Review Letters}\ }\textbf {\bibinfo {volume} {84}},\ \bibinfo
  {pages} {4970} (\bibinfo {year} {2000})}\BibitemShut {NoStop}%
\bibitem [{\citenamefont {Kaur}\ \emph {et~al.}(2005)\citenamefont {Kaur},
  \citenamefont {Agterberg},\ and\ \citenamefont {Sigrist}}]{Kaur2005}%
  \BibitemOpen
  \bibfield  {author} {\bibinfo {author} {\bibfnamefont {R.}~\bibnamefont
  {Kaur}}, \bibinfo {author} {\bibfnamefont {D.}~\bibnamefont {Agterberg}}, \
  and\ \bibinfo {author} {\bibfnamefont {M.}~\bibnamefont {Sigrist}},\
  }\href@noop {} {\bibfield  {journal} {\bibinfo  {journal} {Physical Review
  Letters}\ }\textbf {\bibinfo {volume} {94}},\ \bibinfo {pages} {137002}
  (\bibinfo {year} {2005})}\BibitemShut {NoStop}%
\bibitem [{\citenamefont {Agterberg}\ and\ \citenamefont
  {Kaur}(2007)}]{Agterberg2007}%
  \BibitemOpen
  \bibfield  {author} {\bibinfo {author} {\bibfnamefont {D.~F.}\ \bibnamefont
  {Agterberg}}\ and\ \bibinfo {author} {\bibfnamefont {R.~P.}\ \bibnamefont
  {Kaur}},\ }\href@noop {} {\bibfield  {journal} {\bibinfo  {journal} {Physical
  Review B}\ }\textbf {\bibinfo {volume} {75}},\ \bibinfo {pages} {064511}
  (\bibinfo {year} {2007})}\BibitemShut {NoStop}%
\bibitem [{\citenamefont {Agterberg}(2012)}]{Agterberg2012}%
  \BibitemOpen
  \bibfield  {author} {\bibinfo {author} {\bibfnamefont {D.~F.}\ \bibnamefont
  {Agterberg}},\ }in\ \href@noop {} {\emph {\bibinfo {booktitle}
  {Non-Centrosymmetric Superconductors}}}\ (\bibinfo  {publisher} {Springer},\
  \bibinfo {year} {2012})\ pp.\ \bibinfo {pages} {155--170}\BibitemShut
  {NoStop}%
\bibitem [{\citenamefont {Mironov}\ and\ \citenamefont
  {Buzdin}(2017)}]{Mironov2017}%
  \BibitemOpen
  \bibfield  {author} {\bibinfo {author} {\bibfnamefont {S.}~\bibnamefont
  {Mironov}}\ and\ \bibinfo {author} {\bibfnamefont {A.}~\bibnamefont
  {Buzdin}},\ }\href@noop {} {\bibfield  {journal} {\bibinfo  {journal}
  {Physical Review Letters}\ }\textbf {\bibinfo {volume} {118}},\ \bibinfo
  {pages} {077001} (\bibinfo {year} {2017})}\BibitemShut {NoStop}%
\bibitem [{\citenamefont {Fukui}\ \emph {et~al.}(2005)\citenamefont {Fukui},
  \citenamefont {Hatsugai},\ and\ \citenamefont {Suzuki}}]{fukui2005}%
  \BibitemOpen
  \bibfield  {author} {\bibinfo {author} {\bibfnamefont {T.}~\bibnamefont
  {Fukui}}, \bibinfo {author} {\bibfnamefont {Y.}~\bibnamefont {Hatsugai}}, \
  and\ \bibinfo {author} {\bibfnamefont {H.}~\bibnamefont {Suzuki}},\
  }\href@noop {} {\bibfield  {journal} {\bibinfo  {journal} {Journal of the
  Physical Society of Japan}\ }\textbf {\bibinfo {volume} {74}},\ \bibinfo
  {pages} {1674} (\bibinfo {year} {2005})}\BibitemShut {NoStop}%
\bibitem [{\citenamefont {Grosfeld}\ and\ \citenamefont
  {Stern}(2006)}]{grosfeld2006electronic}%
  \BibitemOpen
  \bibfield  {author} {\bibinfo {author} {\bibfnamefont {E.}~\bibnamefont
  {Grosfeld}}\ and\ \bibinfo {author} {\bibfnamefont {A.}~\bibnamefont
  {Stern}},\ }\href@noop {} {\bibfield  {journal} {\bibinfo  {journal}
  {Physical Review B}\ }\textbf {\bibinfo {volume} {73}},\ \bibinfo {pages}
  {201303} (\bibinfo {year} {2006})}\BibitemShut {NoStop}%
\bibitem [{\citenamefont {Cheng}\ \emph {et~al.}(2009)\citenamefont {Cheng},
  \citenamefont {Lutchyn}, \citenamefont {Galitski},\ and\ \citenamefont
  {Sarma}}]{Cheng2009}%
  \BibitemOpen
  \bibfield  {author} {\bibinfo {author} {\bibfnamefont {M.}~\bibnamefont
  {Cheng}}, \bibinfo {author} {\bibfnamefont {R.~M.}\ \bibnamefont {Lutchyn}},
  \bibinfo {author} {\bibfnamefont {V.}~\bibnamefont {Galitski}}, \ and\
  \bibinfo {author} {\bibfnamefont {S.~D.}\ \bibnamefont {Sarma}},\ }\href@noop
  {} {\bibfield  {journal} {\bibinfo  {journal} {Physical Review Letters}\
  }\textbf {\bibinfo {volume} {103}},\ \bibinfo {pages} {107001} (\bibinfo
  {year} {2009})}\BibitemShut {NoStop}%
\bibitem [{\citenamefont {Cheng}\ \emph {et~al.}(2010)\citenamefont {Cheng},
  \citenamefont {Lutchyn}, \citenamefont {Galitski},\ and\ \citenamefont
  {Das~Sarma}}]{Cheng2010}%
  \BibitemOpen
  \bibfield  {author} {\bibinfo {author} {\bibfnamefont {M.}~\bibnamefont
  {Cheng}}, \bibinfo {author} {\bibfnamefont {R.~M.}\ \bibnamefont {Lutchyn}},
  \bibinfo {author} {\bibfnamefont {V.}~\bibnamefont {Galitski}}, \ and\
  \bibinfo {author} {\bibfnamefont {S.}~\bibnamefont {Das~Sarma}},\ }\href
  {\doibase 10.1103/PhysRevB.82.094504} {\bibfield  {journal} {\bibinfo
  {journal} {Phys. Rev. B}\ }\textbf {\bibinfo {volume} {82}},\ \bibinfo
  {pages} {094504} (\bibinfo {year} {2010})}\BibitemShut {NoStop}%
\bibitem [{\citenamefont {Ludwig}\ \emph {et~al.}(2011)\citenamefont {Ludwig},
  \citenamefont {Poilblanc}, \citenamefont {Trebst},\ and\ \citenamefont
  {Troyer}}]{Ludwig2011}%
  \BibitemOpen
  \bibfield  {author} {\bibinfo {author} {\bibfnamefont {A.~W.}\ \bibnamefont
  {Ludwig}}, \bibinfo {author} {\bibfnamefont {D.}~\bibnamefont {Poilblanc}},
  \bibinfo {author} {\bibfnamefont {S.}~\bibnamefont {Trebst}}, \ and\ \bibinfo
  {author} {\bibfnamefont {M.}~\bibnamefont {Troyer}},\ }\href@noop {}
  {\bibfield  {journal} {\bibinfo  {journal} {New Journal of Physics}\ }\textbf
  {\bibinfo {volume} {13}},\ \bibinfo {pages} {045014} (\bibinfo {year}
  {2011})}\BibitemShut {NoStop}%
\bibitem [{\citenamefont {Kraus}\ and\ \citenamefont
  {Stern}(2011)}]{kraus2011majorana}%
  \BibitemOpen
  \bibfield  {author} {\bibinfo {author} {\bibfnamefont {Y.~E.}\ \bibnamefont
  {Kraus}}\ and\ \bibinfo {author} {\bibfnamefont {A.}~\bibnamefont {Stern}},\
  }\href@noop {} {\bibfield  {journal} {\bibinfo  {journal} {New Journal of
  Physics}\ }\textbf {\bibinfo {volume} {13}},\ \bibinfo {pages} {105006}
  (\bibinfo {year} {2011})}\BibitemShut {NoStop}%
\bibitem [{\citenamefont {Lahtinen}\ \emph {et~al.}(2012)\citenamefont
  {Lahtinen}, \citenamefont {Ludwig}, \citenamefont {Pachos},\ and\
  \citenamefont {Trebst}}]{Lahtinen2012}%
  \BibitemOpen
  \bibfield  {author} {\bibinfo {author} {\bibfnamefont {V.}~\bibnamefont
  {Lahtinen}}, \bibinfo {author} {\bibfnamefont {A.~W.}\ \bibnamefont
  {Ludwig}}, \bibinfo {author} {\bibfnamefont {J.~K.}\ \bibnamefont {Pachos}},
  \ and\ \bibinfo {author} {\bibfnamefont {S.}~\bibnamefont {Trebst}},\
  }\href@noop {} {\bibfield  {journal} {\bibinfo  {journal} {Physical Review
  B}\ }\textbf {\bibinfo {volume} {86}},\ \bibinfo {pages} {075115} (\bibinfo
  {year} {2012})}\BibitemShut {NoStop}%
\bibitem [{\citenamefont {Laumann}\ \emph {et~al.}(2012)\citenamefont
  {Laumann}, \citenamefont {Ludwig}, \citenamefont {Huse},\ and\ \citenamefont
  {Trebst}}]{Laumann2012}%
  \BibitemOpen
  \bibfield  {author} {\bibinfo {author} {\bibfnamefont {C.~R.}\ \bibnamefont
  {Laumann}}, \bibinfo {author} {\bibfnamefont {A.~W.}\ \bibnamefont {Ludwig}},
  \bibinfo {author} {\bibfnamefont {D.~A.}\ \bibnamefont {Huse}}, \ and\
  \bibinfo {author} {\bibfnamefont {S.}~\bibnamefont {Trebst}},\ }\href@noop {}
  {\bibfield  {journal} {\bibinfo  {journal} {Physical Review B}\ }\textbf
  {\bibinfo {volume} {85}},\ \bibinfo {pages} {161301} (\bibinfo {year}
  {2012})}\BibitemShut {NoStop}%
\bibitem [{\citenamefont {Silaev}(2013)}]{silaev2013majorana}%
  \BibitemOpen
  \bibfield  {author} {\bibinfo {author} {\bibfnamefont {M.}~\bibnamefont
  {Silaev}},\ }\href@noop {} {\bibfield  {journal} {\bibinfo  {journal}
  {Physical Review B}\ }\textbf {\bibinfo {volume} {88}},\ \bibinfo {pages}
  {064514} (\bibinfo {year} {2013})}\BibitemShut {NoStop}%
\bibitem [{\citenamefont {Ariad}\ \emph {et~al.}(2015)\citenamefont {Ariad},
  \citenamefont {Grosfeld},\ and\ \citenamefont {Seradjeh}}]{Ariad2015}%
  \BibitemOpen
  \bibfield  {author} {\bibinfo {author} {\bibfnamefont {D.}~\bibnamefont
  {Ariad}}, \bibinfo {author} {\bibfnamefont {E.}~\bibnamefont {Grosfeld}}, \
  and\ \bibinfo {author} {\bibfnamefont {B.}~\bibnamefont {Seradjeh}},\
  }\href@noop {} {\bibfield  {journal} {\bibinfo  {journal} {Physical Review
  B}\ }\textbf {\bibinfo {volume} {92}},\ \bibinfo {pages} {035136} (\bibinfo
  {year} {2015})}\BibitemShut {NoStop}%
\bibitem [{\citenamefont {Li}\ and\ \citenamefont {Franz}(2018)}]{Li2018}%
  \BibitemOpen
  \bibfield  {author} {\bibinfo {author} {\bibfnamefont {C.}~\bibnamefont
  {Li}}\ and\ \bibinfo {author} {\bibfnamefont {M.}~\bibnamefont {Franz}},\
  }\href@noop {} {\bibfield  {journal} {\bibinfo  {journal} {Physical Review
  B}\ }\textbf {\bibinfo {volume} {98}},\ \bibinfo {pages} {115123} (\bibinfo
  {year} {2018})}\BibitemShut {NoStop}%
\bibitem [{\citenamefont {Bauer}\ and\ \citenamefont
  {Sigrist}(2012)}]{Bauer2012}%
  \BibitemOpen
  \bibfield  {author} {\bibinfo {author} {\bibfnamefont {E.}~\bibnamefont
  {Bauer}}\ and\ \bibinfo {author} {\bibfnamefont {M.}~\bibnamefont
  {Sigrist}},\ }\href@noop {} {\emph {\bibinfo {title} {{Non-centrosymmetric
  superconductors: Introduction and Overview}}}},\ Vol.\ \bibinfo {volume}
  {847}\ (\bibinfo  {publisher} {Springer Science \& Business Media},\ \bibinfo
  {year} {2012})\BibitemShut {NoStop}%
\bibitem [{\citenamefont {Smidman}\ \emph {et~al.}(2017)\citenamefont
  {Smidman}, \citenamefont {Salamon}, \citenamefont {Yuan},\ and\ \citenamefont
  {Agterberg}}]{Smidman2017}%
  \BibitemOpen
  \bibfield  {author} {\bibinfo {author} {\bibfnamefont {M.}~\bibnamefont
  {Smidman}}, \bibinfo {author} {\bibfnamefont {M.}~\bibnamefont {Salamon}},
  \bibinfo {author} {\bibfnamefont {H.}~\bibnamefont {Yuan}}, \ and\ \bibinfo
  {author} {\bibfnamefont {D.}~\bibnamefont {Agterberg}},\ }\href@noop {}
  {\bibfield  {journal} {\bibinfo  {journal} {Reports on Progress in Physics}\
  }\textbf {\bibinfo {volume} {80}},\ \bibinfo {pages} {036501} (\bibinfo
  {year} {2017})}\BibitemShut {NoStop}%
\bibitem [{\citenamefont {Borisenko}\ \emph {et~al.}(2016)\citenamefont
  {Borisenko}, \citenamefont {Evtushinsky}, \citenamefont {Liu}, \citenamefont
  {Morozov}, \citenamefont {Kappenberger}, \citenamefont {Wurmehl},
  \citenamefont {B{\"u}chner}, \citenamefont {Yaresko}, \citenamefont {Kim},
  \citenamefont {Hoesch} \emph {et~al.}}]{Borisenko2016}%
  \BibitemOpen
  \bibfield  {author} {\bibinfo {author} {\bibfnamefont {S.}~\bibnamefont
  {Borisenko}}, \bibinfo {author} {\bibfnamefont {D.}~\bibnamefont
  {Evtushinsky}}, \bibinfo {author} {\bibfnamefont {Z.-H.}\ \bibnamefont
  {Liu}}, \bibinfo {author} {\bibfnamefont {I.}~\bibnamefont {Morozov}},
  \bibinfo {author} {\bibfnamefont {R.}~\bibnamefont {Kappenberger}}, \bibinfo
  {author} {\bibfnamefont {S.}~\bibnamefont {Wurmehl}}, \bibinfo {author}
  {\bibfnamefont {B.}~\bibnamefont {B{\"u}chner}}, \bibinfo {author}
  {\bibfnamefont {A.}~\bibnamefont {Yaresko}}, \bibinfo {author} {\bibfnamefont
  {T.}~\bibnamefont {Kim}}, \bibinfo {author} {\bibfnamefont {M.}~\bibnamefont
  {Hoesch}},  \emph {et~al.},\ }\href@noop {} {\bibfield  {journal} {\bibinfo
  {journal} {Nature Physics}\ }\textbf {\bibinfo {volume} {12}},\ \bibinfo
  {pages} {311} (\bibinfo {year} {2016})}\BibitemShut {NoStop}%
\bibitem [{\citenamefont {Haindl}(2021)}]{Haindl2021}%
  \BibitemOpen
  \bibfield  {author} {\bibinfo {author} {\bibfnamefont {S.}~\bibnamefont
  {Haindl}},\ }\href@noop {} {\emph {\bibinfo {title} {Iron-Based
  Superconducting Thin Films}}}\ (\bibinfo  {publisher} {Springer},\ \bibinfo
  {year} {2021})\BibitemShut {NoStop}%
\bibitem [{\citenamefont {Caviglia}\ \emph {et~al.}(2010)\citenamefont
  {Caviglia}, \citenamefont {Gabay}, \citenamefont {Gariglio}, \citenamefont
  {Reyren}, \citenamefont {Cancellieri},\ and\ \citenamefont
  {Triscone}}]{Caviglia2010}%
  \BibitemOpen
  \bibfield  {author} {\bibinfo {author} {\bibfnamefont {A.}~\bibnamefont
  {Caviglia}}, \bibinfo {author} {\bibfnamefont {M.}~\bibnamefont {Gabay}},
  \bibinfo {author} {\bibfnamefont {S.}~\bibnamefont {Gariglio}}, \bibinfo
  {author} {\bibfnamefont {N.}~\bibnamefont {Reyren}}, \bibinfo {author}
  {\bibfnamefont {C.}~\bibnamefont {Cancellieri}}, \ and\ \bibinfo {author}
  {\bibfnamefont {J.-M.}\ \bibnamefont {Triscone}},\ }\href@noop {} {\bibfield
  {journal} {\bibinfo  {journal} {Physical Review Letters}\ }\textbf {\bibinfo
  {volume} {104}},\ \bibinfo {pages} {126803} (\bibinfo {year}
  {2010})}\BibitemShut {NoStop}%
\bibitem [{\citenamefont {Shanavas}\ and\ \citenamefont
  {Satpathy}(2014)}]{Shanavas2014}%
  \BibitemOpen
  \bibfield  {author} {\bibinfo {author} {\bibfnamefont {K.}~\bibnamefont
  {Shanavas}}\ and\ \bibinfo {author} {\bibfnamefont {S.}~\bibnamefont
  {Satpathy}},\ }\href@noop {} {\bibfield  {journal} {\bibinfo  {journal}
  {Physical Review Letters}\ }\textbf {\bibinfo {volume} {112}},\ \bibinfo
  {pages} {086802} (\bibinfo {year} {2014})}\BibitemShut {NoStop}%
\bibitem [{\citenamefont {Singh}\ \emph {et~al.}(2022)\citenamefont {Singh},
  \citenamefont {Venditti}, \citenamefont {Saiz}, \citenamefont {Herranz},
  \citenamefont {S\'anchez}, \citenamefont {Jouan}, \citenamefont
  {Feuillet-Palma}, \citenamefont {Lesueur}, \citenamefont {Grilli},
  \citenamefont {Caprara},\ and\ \citenamefont {Bergeal}}]{Singh2022}%
  \BibitemOpen
  \bibfield  {author} {\bibinfo {author} {\bibfnamefont {G.}~\bibnamefont
  {Singh}}, \bibinfo {author} {\bibfnamefont {G.}~\bibnamefont {Venditti}},
  \bibinfo {author} {\bibfnamefont {G.}~\bibnamefont {Saiz}}, \bibinfo {author}
  {\bibfnamefont {G.}~\bibnamefont {Herranz}}, \bibinfo {author} {\bibfnamefont
  {F.}~\bibnamefont {S\'anchez}}, \bibinfo {author} {\bibfnamefont
  {A.}~\bibnamefont {Jouan}}, \bibinfo {author} {\bibfnamefont
  {C.}~\bibnamefont {Feuillet-Palma}}, \bibinfo {author} {\bibfnamefont
  {J.}~\bibnamefont {Lesueur}}, \bibinfo {author} {\bibfnamefont
  {M.}~\bibnamefont {Grilli}}, \bibinfo {author} {\bibfnamefont
  {S.}~\bibnamefont {Caprara}}, \ and\ \bibinfo {author} {\bibfnamefont
  {N.}~\bibnamefont {Bergeal}},\ }\href {\doibase 10.1103/PhysRevB.105.064512}
  {\bibfield  {journal} {\bibinfo  {journal} {Phys. Rev. B}\ }\textbf {\bibinfo
  {volume} {105}},\ \bibinfo {pages} {064512} (\bibinfo {year}
  {2022})}\BibitemShut {NoStop}%
\bibitem [{\citenamefont {Mattheiss}(1972)}]{Mattheiss1972}%
  \BibitemOpen
  \bibfield  {author} {\bibinfo {author} {\bibfnamefont {L.}~\bibnamefont
  {Mattheiss}},\ }\href@noop {} {\bibfield  {journal} {\bibinfo  {journal}
  {Physical Review B}\ }\textbf {\bibinfo {volume} {6}},\ \bibinfo {pages}
  {4740} (\bibinfo {year} {1972})}\BibitemShut {NoStop}%
\bibitem [{\citenamefont {Xi}\ \emph {et~al.}(2015)\citenamefont {Xi},
  \citenamefont {Zhao}, \citenamefont {Wang}, \citenamefont {Berger},
  \citenamefont {Forr{\'o}}, \citenamefont {Shan},\ and\ \citenamefont
  {Mak}}]{Xi2015}%
  \BibitemOpen
  \bibfield  {author} {\bibinfo {author} {\bibfnamefont {X.}~\bibnamefont
  {Xi}}, \bibinfo {author} {\bibfnamefont {L.}~\bibnamefont {Zhao}}, \bibinfo
  {author} {\bibfnamefont {Z.}~\bibnamefont {Wang}}, \bibinfo {author}
  {\bibfnamefont {H.}~\bibnamefont {Berger}}, \bibinfo {author} {\bibfnamefont
  {L.}~\bibnamefont {Forr{\'o}}}, \bibinfo {author} {\bibfnamefont
  {J.}~\bibnamefont {Shan}}, \ and\ \bibinfo {author} {\bibfnamefont {K.~F.}\
  \bibnamefont {Mak}},\ }\href@noop {} {\bibfield  {journal} {\bibinfo
  {journal} {Nature Nanotechnology}\ }\textbf {\bibinfo {volume} {10}},\
  \bibinfo {pages} {765} (\bibinfo {year} {2015})}\BibitemShut {NoStop}%
\bibitem [{\citenamefont {Xi}\ \emph {et~al.}(2016)\citenamefont {Xi},
  \citenamefont {Wang}, \citenamefont {Zhao}, \citenamefont {Park},
  \citenamefont {Law}, \citenamefont {Berger}, \citenamefont {Forr{\'o}},
  \citenamefont {Shan},\ and\ \citenamefont {Mak}}]{Xi2016}%
  \BibitemOpen
  \bibfield  {author} {\bibinfo {author} {\bibfnamefont {X.}~\bibnamefont
  {Xi}}, \bibinfo {author} {\bibfnamefont {Z.}~\bibnamefont {Wang}}, \bibinfo
  {author} {\bibfnamefont {W.}~\bibnamefont {Zhao}}, \bibinfo {author}
  {\bibfnamefont {J.-H.}\ \bibnamefont {Park}}, \bibinfo {author}
  {\bibfnamefont {K.~T.}\ \bibnamefont {Law}}, \bibinfo {author} {\bibfnamefont
  {H.}~\bibnamefont {Berger}}, \bibinfo {author} {\bibfnamefont
  {L.}~\bibnamefont {Forr{\'o}}}, \bibinfo {author} {\bibfnamefont
  {J.}~\bibnamefont {Shan}}, \ and\ \bibinfo {author} {\bibfnamefont {K.~F.}\
  \bibnamefont {Mak}},\ }\href@noop {} {\bibfield  {journal} {\bibinfo
  {journal} {Nature Physics}\ }\textbf {\bibinfo {volume} {12}},\ \bibinfo
  {pages} {139} (\bibinfo {year} {2016})}\BibitemShut {NoStop}%
\bibitem [{\citenamefont {Ugeda}\ \emph {et~al.}(2016)\citenamefont {Ugeda},
  \citenamefont {Bradley}, \citenamefont {Zhang}, \citenamefont {Onishi},
  \citenamefont {Chen}, \citenamefont {Ruan}, \citenamefont
  {Ojeda-Aristizabal}, \citenamefont {Ryu}, \citenamefont {Edmonds},
  \citenamefont {Tsai} \emph {et~al.}}]{Ugeda2016}%
  \BibitemOpen
  \bibfield  {author} {\bibinfo {author} {\bibfnamefont {M.~M.}\ \bibnamefont
  {Ugeda}}, \bibinfo {author} {\bibfnamefont {A.~J.}\ \bibnamefont {Bradley}},
  \bibinfo {author} {\bibfnamefont {Y.}~\bibnamefont {Zhang}}, \bibinfo
  {author} {\bibfnamefont {S.}~\bibnamefont {Onishi}}, \bibinfo {author}
  {\bibfnamefont {Y.}~\bibnamefont {Chen}}, \bibinfo {author} {\bibfnamefont
  {W.}~\bibnamefont {Ruan}}, \bibinfo {author} {\bibfnamefont {C.}~\bibnamefont
  {Ojeda-Aristizabal}}, \bibinfo {author} {\bibfnamefont {H.}~\bibnamefont
  {Ryu}}, \bibinfo {author} {\bibfnamefont {M.~T.}\ \bibnamefont {Edmonds}},
  \bibinfo {author} {\bibfnamefont {H.-Z.}\ \bibnamefont {Tsai}},  \emph
  {et~al.},\ }\href@noop {} {\bibfield  {journal} {\bibinfo  {journal} {Nature
  Physics}\ }\textbf {\bibinfo {volume} {12}},\ \bibinfo {pages} {92} (\bibinfo
  {year} {2016})}\BibitemShut {NoStop}%
\bibitem [{\citenamefont {Wang}\ \emph {et~al.}(2017)\citenamefont {Wang},
  \citenamefont {Huang}, \citenamefont {Lin}, \citenamefont {Cui},
  \citenamefont {Chen}, \citenamefont {Zhu}, \citenamefont {Liu}, \citenamefont
  {Zeng}, \citenamefont {Zhou}, \citenamefont {Yu} \emph {et~al.}}]{Wang2017}%
  \BibitemOpen
  \bibfield  {author} {\bibinfo {author} {\bibfnamefont {H.}~\bibnamefont
  {Wang}}, \bibinfo {author} {\bibfnamefont {X.}~\bibnamefont {Huang}},
  \bibinfo {author} {\bibfnamefont {J.}~\bibnamefont {Lin}}, \bibinfo {author}
  {\bibfnamefont {J.}~\bibnamefont {Cui}}, \bibinfo {author} {\bibfnamefont
  {Y.}~\bibnamefont {Chen}}, \bibinfo {author} {\bibfnamefont {C.}~\bibnamefont
  {Zhu}}, \bibinfo {author} {\bibfnamefont {F.}~\bibnamefont {Liu}}, \bibinfo
  {author} {\bibfnamefont {Q.}~\bibnamefont {Zeng}}, \bibinfo {author}
  {\bibfnamefont {J.}~\bibnamefont {Zhou}}, \bibinfo {author} {\bibfnamefont
  {P.}~\bibnamefont {Yu}},  \emph {et~al.},\ }\href@noop {} {\bibfield
  {journal} {\bibinfo  {journal} {Nature Communications}\ }\textbf {\bibinfo
  {volume} {8}},\ \bibinfo {pages} {1} (\bibinfo {year} {2017})}\BibitemShut
  {NoStop}%
\bibitem [{\citenamefont {de~la Barrera}\ \emph {et~al.}(2018)\citenamefont
  {de~la Barrera}, \citenamefont {Sinko}, \citenamefont {Gopalan},
  \citenamefont {Sivadas}, \citenamefont {Seyler}, \citenamefont {Watanabe},
  \citenamefont {Taniguchi}, \citenamefont {Tsen}, \citenamefont {Xu},
  \citenamefont {Xiao} \emph {et~al.}}]{Barrera2018}%
  \BibitemOpen
  \bibfield  {author} {\bibinfo {author} {\bibfnamefont {S.~C.}\ \bibnamefont
  {de~la Barrera}}, \bibinfo {author} {\bibfnamefont {M.~R.}\ \bibnamefont
  {Sinko}}, \bibinfo {author} {\bibfnamefont {D.~P.}\ \bibnamefont {Gopalan}},
  \bibinfo {author} {\bibfnamefont {N.}~\bibnamefont {Sivadas}}, \bibinfo
  {author} {\bibfnamefont {K.~L.}\ \bibnamefont {Seyler}}, \bibinfo {author}
  {\bibfnamefont {K.}~\bibnamefont {Watanabe}}, \bibinfo {author}
  {\bibfnamefont {T.}~\bibnamefont {Taniguchi}}, \bibinfo {author}
  {\bibfnamefont {A.~W.}\ \bibnamefont {Tsen}}, \bibinfo {author}
  {\bibfnamefont {X.}~\bibnamefont {Xu}}, \bibinfo {author} {\bibfnamefont
  {D.}~\bibnamefont {Xiao}},  \emph {et~al.},\ }\href@noop {} {\bibfield
  {journal} {\bibinfo  {journal} {Nature Communications}\ }\textbf {\bibinfo
  {volume} {9}},\ \bibinfo {pages} {1} (\bibinfo {year} {2018})}\BibitemShut
  {NoStop}%
\bibitem [{\citenamefont {Xiao}\ \emph {et~al.}(2012)\citenamefont {Xiao},
  \citenamefont {Liu}, \citenamefont {Feng}, \citenamefont {Xu},\ and\
  \citenamefont {Yao}}]{Xiao2012}%
  \BibitemOpen
  \bibfield  {author} {\bibinfo {author} {\bibfnamefont {D.}~\bibnamefont
  {Xiao}}, \bibinfo {author} {\bibfnamefont {G.-B.}\ \bibnamefont {Liu}},
  \bibinfo {author} {\bibfnamefont {W.}~\bibnamefont {Feng}}, \bibinfo {author}
  {\bibfnamefont {X.}~\bibnamefont {Xu}}, \ and\ \bibinfo {author}
  {\bibfnamefont {W.}~\bibnamefont {Yao}},\ }\href@noop {} {\bibfield
  {journal} {\bibinfo  {journal} {Physical Review Letters}\ }\textbf {\bibinfo
  {volume} {108}},\ \bibinfo {pages} {196802} (\bibinfo {year}
  {2012})}\BibitemShut {NoStop}%
\bibitem [{\citenamefont {Zhu}\ \emph {et~al.}(2011)\citenamefont {Zhu},
  \citenamefont {Cheng},\ and\ \citenamefont {Schwingenschl{\"o}gl}}]{Zhu2011}%
  \BibitemOpen
  \bibfield  {author} {\bibinfo {author} {\bibfnamefont {Z.~Y.}\ \bibnamefont
  {Zhu}}, \bibinfo {author} {\bibfnamefont {Y.~C.}\ \bibnamefont {Cheng}}, \
  and\ \bibinfo {author} {\bibfnamefont {U.}~\bibnamefont
  {Schwingenschl{\"o}gl}},\ }\href@noop {} {\bibfield  {journal} {\bibinfo
  {journal} {Physical Review B}\ }\textbf {\bibinfo {volume} {84}},\ \bibinfo
  {pages} {153402} (\bibinfo {year} {2011})}\BibitemShut {NoStop}%
\bibitem [{\citenamefont {Ko{\'s}mider}\ \emph {et~al.}(2013)\citenamefont
  {Ko{\'s}mider}, \citenamefont {Gonz{\'a}lez},\ and\ \citenamefont
  {Fern{\'a}ndez-Rossier}}]{Kosmider2013}%
  \BibitemOpen
  \bibfield  {author} {\bibinfo {author} {\bibfnamefont {K.}~\bibnamefont
  {Ko{\'s}mider}}, \bibinfo {author} {\bibfnamefont {J.~W.}\ \bibnamefont
  {Gonz{\'a}lez}}, \ and\ \bibinfo {author} {\bibfnamefont {J.}~\bibnamefont
  {Fern{\'a}ndez-Rossier}},\ }\href@noop {} {\bibfield  {journal} {\bibinfo
  {journal} {Physical Review B}\ }\textbf {\bibinfo {volume} {88}},\ \bibinfo
  {pages} {245436} (\bibinfo {year} {2013})}\BibitemShut {NoStop}%
\bibitem [{\citenamefont {Baglo}\ \emph {et~al.}(2022)\citenamefont {Baglo},
  \citenamefont {Chen}, \citenamefont {Murphy}, \citenamefont {Leenen},
  \citenamefont {McCollam}, \citenamefont {Sutherland},\ and\ \citenamefont
  {Grosche}}]{Baglo2022}%
  \BibitemOpen
  \bibfield  {author} {\bibinfo {author} {\bibfnamefont {J.}~\bibnamefont
  {Baglo}}, \bibinfo {author} {\bibfnamefont {J.}~\bibnamefont {Chen}},
  \bibinfo {author} {\bibfnamefont {K.}~\bibnamefont {Murphy}}, \bibinfo
  {author} {\bibfnamefont {R.}~\bibnamefont {Leenen}}, \bibinfo {author}
  {\bibfnamefont {A.}~\bibnamefont {McCollam}}, \bibinfo {author}
  {\bibfnamefont {M.~L.}\ \bibnamefont {Sutherland}}, \ and\ \bibinfo {author}
  {\bibfnamefont {F.~M.}\ \bibnamefont {Grosche}},\ }\href@noop {} {\bibfield
  {journal} {\bibinfo  {journal} {Physical Review Letters}\ }\textbf {\bibinfo
  {volume} {129}},\ \bibinfo {pages} {046402} (\bibinfo {year}
  {2022})}\BibitemShut {NoStop}%
\bibitem [{\citenamefont {Reyren}\ \emph {et~al.}(2007)\citenamefont {Reyren},
  \citenamefont {Thiel}, \citenamefont {Caviglia}, \citenamefont {Kourkoutis},
  \citenamefont {Hammerl}, \citenamefont {Richter}, \citenamefont {Schneider},
  \citenamefont {Kopp}, \citenamefont {Ruetschi}, \citenamefont {Jaccard} \emph
  {et~al.}}]{Reyren2007}%
  \BibitemOpen
  \bibfield  {author} {\bibinfo {author} {\bibfnamefont {N.}~\bibnamefont
  {Reyren}}, \bibinfo {author} {\bibfnamefont {S.}~\bibnamefont {Thiel}},
  \bibinfo {author} {\bibfnamefont {A.}~\bibnamefont {Caviglia}}, \bibinfo
  {author} {\bibfnamefont {L.~F.}\ \bibnamefont {Kourkoutis}}, \bibinfo
  {author} {\bibfnamefont {G.}~\bibnamefont {Hammerl}}, \bibinfo {author}
  {\bibfnamefont {C.}~\bibnamefont {Richter}}, \bibinfo {author} {\bibfnamefont
  {C.~W.}\ \bibnamefont {Schneider}}, \bibinfo {author} {\bibfnamefont
  {T.}~\bibnamefont {Kopp}}, \bibinfo {author} {\bibfnamefont {A.-S.}\
  \bibnamefont {Ruetschi}}, \bibinfo {author} {\bibfnamefont {D.}~\bibnamefont
  {Jaccard}},  \emph {et~al.},\ }\href@noop {} {\bibfield  {journal} {\bibinfo
  {journal} {Science}\ }\textbf {\bibinfo {volume} {317}},\ \bibinfo {pages}
  {1196} (\bibinfo {year} {2007})}\BibitemShut {NoStop}%
\bibitem [{\citenamefont {Hsu}\ \emph {et~al.}(2017)\citenamefont {Hsu},
  \citenamefont {Vaezi}, \citenamefont {Fischer},\ and\ \citenamefont
  {Kim}}]{Hsu2017}%
  \BibitemOpen
  \bibfield  {author} {\bibinfo {author} {\bibfnamefont {Y.-T.}\ \bibnamefont
  {Hsu}}, \bibinfo {author} {\bibfnamefont {A.}~\bibnamefont {Vaezi}}, \bibinfo
  {author} {\bibfnamefont {M.~H.}\ \bibnamefont {Fischer}}, \ and\ \bibinfo
  {author} {\bibfnamefont {E.-A.}\ \bibnamefont {Kim}},\ }\href@noop {}
  {\bibfield  {journal} {\bibinfo  {journal} {Nature Communications}\ }\textbf
  {\bibinfo {volume} {8}},\ \bibinfo {pages} {14985} (\bibinfo {year}
  {2017})}\BibitemShut {NoStop}%
\bibitem [{\citenamefont {Yi}\ \emph {et~al.}(2022)\citenamefont {Yi},
  \citenamefont {Hu}, \citenamefont {Wang}, \citenamefont {Xiao}, \citenamefont
  {Cai}, \citenamefont {Hickey}, \citenamefont {Dong}, \citenamefont {Zhao},
  \citenamefont {Zhou}, \citenamefont {Zhang} \emph {et~al.}}]{Yi2022}%
  \BibitemOpen
  \bibfield  {author} {\bibinfo {author} {\bibfnamefont {H.}~\bibnamefont
  {Yi}}, \bibinfo {author} {\bibfnamefont {L.-H.}\ \bibnamefont {Hu}}, \bibinfo
  {author} {\bibfnamefont {Y.}~\bibnamefont {Wang}}, \bibinfo {author}
  {\bibfnamefont {R.}~\bibnamefont {Xiao}}, \bibinfo {author} {\bibfnamefont
  {J.}~\bibnamefont {Cai}}, \bibinfo {author} {\bibfnamefont {D.~R.}\
  \bibnamefont {Hickey}}, \bibinfo {author} {\bibfnamefont {C.}~\bibnamefont
  {Dong}}, \bibinfo {author} {\bibfnamefont {Y.-F.}\ \bibnamefont {Zhao}},
  \bibinfo {author} {\bibfnamefont {L.-J.}\ \bibnamefont {Zhou}}, \bibinfo
  {author} {\bibfnamefont {R.}~\bibnamefont {Zhang}},  \emph {et~al.},\
  }\href@noop {} {\bibfield  {journal} {\bibinfo  {journal} {Nature Materials}\
  }\textbf {\bibinfo {volume} {21}},\ \bibinfo {pages} {1366} (\bibinfo {year}
  {2022})}\BibitemShut {NoStop}%
\bibitem [{\citenamefont {Forsythe}\ \emph {et~al.}(2018)\citenamefont
  {Forsythe}, \citenamefont {Zhou}, \citenamefont {Watanabe}, \citenamefont
  {Taniguchi}, \citenamefont {Pasupathy}, \citenamefont {Moon}, \citenamefont
  {Koshino}, \citenamefont {Kim},\ and\ \citenamefont {Dean}}]{Forsythe2018}%
  \BibitemOpen
  \bibfield  {author} {\bibinfo {author} {\bibfnamefont {C.}~\bibnamefont
  {Forsythe}}, \bibinfo {author} {\bibfnamefont {X.}~\bibnamefont {Zhou}},
  \bibinfo {author} {\bibfnamefont {K.}~\bibnamefont {Watanabe}}, \bibinfo
  {author} {\bibfnamefont {T.}~\bibnamefont {Taniguchi}}, \bibinfo {author}
  {\bibfnamefont {A.}~\bibnamefont {Pasupathy}}, \bibinfo {author}
  {\bibfnamefont {P.}~\bibnamefont {Moon}}, \bibinfo {author} {\bibfnamefont
  {M.}~\bibnamefont {Koshino}}, \bibinfo {author} {\bibfnamefont
  {P.}~\bibnamefont {Kim}}, \ and\ \bibinfo {author} {\bibfnamefont {C.~R.}\
  \bibnamefont {Dean}},\ }\href@noop {} {\bibfield  {journal} {\bibinfo
  {journal} {Nature Nanotechnology}\ }\textbf {\bibinfo {volume} {13}},\
  \bibinfo {pages} {566} (\bibinfo {year} {2018})}\BibitemShut {NoStop}%
\bibitem [{\citenamefont {Yankowitz}\ \emph {et~al.}(2018)\citenamefont
  {Yankowitz}, \citenamefont {Jung}, \citenamefont {Laksono}, \citenamefont
  {Leconte}, \citenamefont {Chittari}, \citenamefont {Watanabe}, \citenamefont
  {Taniguchi}, \citenamefont {Adam}, \citenamefont {Graf},\ and\ \citenamefont
  {Dean}}]{Yankowitz2018}%
  \BibitemOpen
  \bibfield  {author} {\bibinfo {author} {\bibfnamefont {M.}~\bibnamefont
  {Yankowitz}}, \bibinfo {author} {\bibfnamefont {J.}~\bibnamefont {Jung}},
  \bibinfo {author} {\bibfnamefont {E.}~\bibnamefont {Laksono}}, \bibinfo
  {author} {\bibfnamefont {N.}~\bibnamefont {Leconte}}, \bibinfo {author}
  {\bibfnamefont {B.~L.}\ \bibnamefont {Chittari}}, \bibinfo {author}
  {\bibfnamefont {K.}~\bibnamefont {Watanabe}}, \bibinfo {author}
  {\bibfnamefont {T.}~\bibnamefont {Taniguchi}}, \bibinfo {author}
  {\bibfnamefont {S.}~\bibnamefont {Adam}}, \bibinfo {author} {\bibfnamefont
  {D.}~\bibnamefont {Graf}}, \ and\ \bibinfo {author} {\bibfnamefont {C.~R.}\
  \bibnamefont {Dean}},\ }\href@noop {} {\bibfield  {journal} {\bibinfo
  {journal} {Nature}\ }\textbf {\bibinfo {volume} {557}},\ \bibinfo {pages}
  {404} (\bibinfo {year} {2018})}\BibitemShut {NoStop}%
\bibitem [{\citenamefont {Shi}\ \emph {et~al.}(2019)\citenamefont {Shi},
  \citenamefont {Ma},\ and\ \citenamefont {Song}}]{Shi2019}%
  \BibitemOpen
  \bibfield  {author} {\bibinfo {author} {\bibfnamefont {L.-k.}\ \bibnamefont
  {Shi}}, \bibinfo {author} {\bibfnamefont {J.}~\bibnamefont {Ma}}, \ and\
  \bibinfo {author} {\bibfnamefont {J.~C.}\ \bibnamefont {Song}},\ }\href@noop
  {} {\bibfield  {journal} {\bibinfo  {journal} {2D Materials}\ }\textbf
  {\bibinfo {volume} {7}},\ \bibinfo {pages} {015028} (\bibinfo {year}
  {2019})}\BibitemShut {NoStop}%
\bibitem [{\citenamefont {Xu}\ \emph {et~al.}(2021)\citenamefont {Xu},
  \citenamefont {Horn}, \citenamefont {Zhu}, \citenamefont {Tang},
  \citenamefont {Ma}, \citenamefont {Li}, \citenamefont {Liu}, \citenamefont
  {Watanabe}, \citenamefont {Taniguchi}, \citenamefont {Hone} \emph
  {et~al.}}]{Xu2021}%
  \BibitemOpen
  \bibfield  {author} {\bibinfo {author} {\bibfnamefont {Y.}~\bibnamefont
  {Xu}}, \bibinfo {author} {\bibfnamefont {C.}~\bibnamefont {Horn}}, \bibinfo
  {author} {\bibfnamefont {J.}~\bibnamefont {Zhu}}, \bibinfo {author}
  {\bibfnamefont {Y.}~\bibnamefont {Tang}}, \bibinfo {author} {\bibfnamefont
  {L.}~\bibnamefont {Ma}}, \bibinfo {author} {\bibfnamefont {L.}~\bibnamefont
  {Li}}, \bibinfo {author} {\bibfnamefont {S.}~\bibnamefont {Liu}}, \bibinfo
  {author} {\bibfnamefont {K.}~\bibnamefont {Watanabe}}, \bibinfo {author}
  {\bibfnamefont {T.}~\bibnamefont {Taniguchi}}, \bibinfo {author}
  {\bibfnamefont {J.~C.}\ \bibnamefont {Hone}},  \emph {et~al.},\ }\href@noop
  {} {\bibfield  {journal} {\bibinfo  {journal} {Nature Materials}\ }\textbf
  {\bibinfo {volume} {20}},\ \bibinfo {pages} {645} (\bibinfo {year}
  {2021})}\BibitemShut {NoStop}%
\bibitem [{\citenamefont {Li}\ \emph {et~al.}(2021)\citenamefont {Li},
  \citenamefont {Dietrich}, \citenamefont {Forsythe}, \citenamefont
  {Taniguchi}, \citenamefont {Watanabe}, \citenamefont {Moon},\ and\
  \citenamefont {Dean}}]{Li2021a}%
  \BibitemOpen
  \bibfield  {author} {\bibinfo {author} {\bibfnamefont {Y.}~\bibnamefont
  {Li}}, \bibinfo {author} {\bibfnamefont {S.}~\bibnamefont {Dietrich}},
  \bibinfo {author} {\bibfnamefont {C.}~\bibnamefont {Forsythe}}, \bibinfo
  {author} {\bibfnamefont {T.}~\bibnamefont {Taniguchi}}, \bibinfo {author}
  {\bibfnamefont {K.}~\bibnamefont {Watanabe}}, \bibinfo {author}
  {\bibfnamefont {P.}~\bibnamefont {Moon}}, \ and\ \bibinfo {author}
  {\bibfnamefont {C.~R.}\ \bibnamefont {Dean}},\ }\href@noop {} {\bibfield
  {journal} {\bibinfo  {journal} {Nature Nanotechnology}\ }\textbf {\bibinfo
  {volume} {16}},\ \bibinfo {pages} {525} (\bibinfo {year} {2021})}\BibitemShut
  {NoStop}%
\bibitem [{\citenamefont {Tilak}\ \emph {et~al.}(2022)\citenamefont {Tilak},
  \citenamefont {Li}, \citenamefont {Taniguchi}, \citenamefont {Watanabe},\
  and\ \citenamefont {Andrei}}]{Tilak2022}%
  \BibitemOpen
  \bibfield  {author} {\bibinfo {author} {\bibfnamefont {N.}~\bibnamefont
  {Tilak}}, \bibinfo {author} {\bibfnamefont {G.}~\bibnamefont {Li}}, \bibinfo
  {author} {\bibfnamefont {T.}~\bibnamefont {Taniguchi}}, \bibinfo {author}
  {\bibfnamefont {K.}~\bibnamefont {Watanabe}}, \ and\ \bibinfo {author}
  {\bibfnamefont {E.~Y.}\ \bibnamefont {Andrei}},\ }\href@noop {} {\bibfield
  {journal} {\bibinfo  {journal} {Nano Letters}\ }\textbf {\bibinfo {volume}
  {23}},\ \bibinfo {pages} {73} (\bibinfo {year} {2022})}\BibitemShut {NoStop}%
\bibitem [{\citenamefont {Zhou}\ \emph {et~al.}(2023)\citenamefont {Zhou},
  \citenamefont {Hou}, \citenamefont {Huang}, \citenamefont {Wang},
  \citenamefont {Fu}, \citenamefont {Liu}, \citenamefont {Yuan}, \citenamefont
  {Xi}, \citenamefont {Xu}, \citenamefont {Lin},\ and\ \citenamefont
  {Gao}}]{zhou_stack_2023}%
  \BibitemOpen
  \bibfield  {author} {\bibinfo {author} {\bibfnamefont {Z.}~\bibnamefont
  {Zhou}}, \bibinfo {author} {\bibfnamefont {F.}~\bibnamefont {Hou}}, \bibinfo
  {author} {\bibfnamefont {X.}~\bibnamefont {Huang}}, \bibinfo {author}
  {\bibfnamefont {G.}~\bibnamefont {Wang}}, \bibinfo {author} {\bibfnamefont
  {Z.}~\bibnamefont {Fu}}, \bibinfo {author} {\bibfnamefont {W.}~\bibnamefont
  {Liu}}, \bibinfo {author} {\bibfnamefont {G.}~\bibnamefont {Yuan}}, \bibinfo
  {author} {\bibfnamefont {X.}~\bibnamefont {Xi}}, \bibinfo {author}
  {\bibfnamefont {J.}~\bibnamefont {Xu}}, \bibinfo {author} {\bibfnamefont
  {J.}~\bibnamefont {Lin}}, \ and\ \bibinfo {author} {\bibfnamefont
  {L.}~\bibnamefont {Gao}},\ }\href {\doibase 10.1038/s41586-023-06404-x}
  {\bibfield  {journal} {\bibinfo  {journal} {Nature}\ } (\bibinfo {year}
  {2023}),\ 10.1038/s41586-023-06404-x}\BibitemShut {NoStop}%
\bibitem [{\citenamefont {Agterberg}(2003)}]{Agterberg2003}%
  \BibitemOpen
  \bibfield  {author} {\bibinfo {author} {\bibfnamefont {D.}~\bibnamefont
  {Agterberg}},\ }\href@noop {} {\bibfield  {journal} {\bibinfo  {journal}
  {Physica C: Superconductivity}\ }\textbf {\bibinfo {volume} {387}},\ \bibinfo
  {pages} {13} (\bibinfo {year} {2003})}\BibitemShut {NoStop}%
\bibitem [{\citenamefont {Rajagopal}\ and\ \citenamefont
  {Vasudevan}(1966)}]{Rajagopal1966}%
  \BibitemOpen
  \bibfield  {author} {\bibinfo {author} {\bibfnamefont {A.}~\bibnamefont
  {Rajagopal}}\ and\ \bibinfo {author} {\bibfnamefont {R.}~\bibnamefont
  {Vasudevan}},\ }\href@noop {} {\bibfield  {journal} {\bibinfo  {journal}
  {Physics Letters}\ }\textbf {\bibinfo {volume} {20}},\ \bibinfo {pages} {585}
  (\bibinfo {year} {1966})}\BibitemShut {NoStop}%
\end{thebibliography}%


\begin{thebibliography}{1}
\expandafter\ifx\csname natexlab\endcsname\relax\def\natexlab#1{#1}\fi
\expandafter\ifx\csname bibnamefont\endcsname\relax
  \def\bibnamefont#1{#1}\fi
\expandafter\ifx\csname bibfnamefont\endcsname\relax
  \def\bibfnamefont#1{#1}\fi
\expandafter\ifx\csname citenamefont\endcsname\relax
  \def\citenamefont#1{#1}\fi
\expandafter\ifx\csname url\endcsname\relax
  \def\url#1{\texttt{#1}}\fi
\expandafter\ifx\csname urlprefix\endcsname\relax\def\urlprefix{URL }\fi
\providecommand{\bibinfo}[2]{#2}
\providecommand{\eprint}[2][]{\url{#2}}

\bibitem[{\citenamefont{Ezawa et~al.}(2013)\citenamefont{Ezawa, Tanaka, and
  Nagaosa}}]{Ezawa2013}
\bibinfo{author}{\bibfnamefont{M.}~\bibnamefont{Ezawa}},
  \bibinfo{author}{\bibfnamefont{Y.}~\bibnamefont{Tanaka}}, \bibnamefont{and}
  \bibinfo{author}{\bibfnamefont{N.}~\bibnamefont{Nagaosa}},
  \bibinfo{journal}{Scientific reports} \textbf{\bibinfo{volume}{3}},
  \bibinfo{pages}{1} (\bibinfo{year}{2013}).

\end{thebibliography}

\end{document}

\title{Supplementary Materials: {Topological superconductivity induced by spin-orbit coupling, perpendicular magnetic field and superlattice potential}}
\author{Jonathan Schirmer$^1$, J. K. Jain$^1$, and C. -X. Liu$^1$}
\affiliation{$^1$Department of Physics, 104 Davey Lab, The Pennsylvania State University, University Park, Pennsylvania 16802}
\maketitle

The Supplementary Materials section contains the following. Section \ref{sec2} contains details on particle-hole symmetry and how to take full advantage of it when solving the mean-field problem. In Section \ref{sec3}, we construct the mean-field groundstate while utilizing particle-hole symmetry. In Section \ref{sec4}, we construct the Gor'kov Green's function and write correlation functions, including those relevant for the self-consistency equations. In Section \ref{sec6} we derive an expression for the current operator, and in Section \ref{sec7}, the spin-orbit coupling Hamiltonian. Section \ref{tsc_sec} closely examines a case where the spectral gap closes and re-opens across a Chern number-changing transitions (topological phase transition).

\section{Particle-Hole Symmetry}\label{sec2}
The BdG Hamiltonian has a particle-hole symmetry. Define $\mathcal{P}=\mathcal{K}\tau_x$ where $\tau_x$ is a Pauli matrix acting on the particle and hole subspaces and $\mathcal{K}$ is complex conjugation. Then
\be
\mathcal{P}^{-1} H_{\rm BdG}(\vec{k}) \mathcal{P} = -H_{\rm BdG}(-\vec{k})
\ee
The implications of this symmetry for our problem are crucial, so it is worth unpacking them more carefully. Consider the subspace of equal and opposite fixed momenta $(\vec{k},-\vec{k})$, along with the BdG Hamiltonian matrices $H_{\rm BdG}(\vec{k})$ and $H_{\rm BdG}(-\vec{k})$ at those momenta. Suppose that $\mathcal{U}(\vec{k})$ and $\mathcal{U}(-\vec{k})$ are unitary matrices which diagonalize each:
\be
\begin{split}
\mathcal{U}^\dagger(\vec{k})H_{\rm BdG}(\vec{k})\mathcal{U}(\vec{k}) &= {\rm diag}\Big(E_1(\vec{k}),\ldots,E_{4MN}(\vec{k}) \Big)\\
\mathcal{U}^\dagger(-\vec{k})H_{\rm BdG}(-\vec{k})\mathcal{U}(-\vec{k}) &=  {\rm diag}\Big(E_1(-\vec{k}),\ldots,E_{4MN}(-\vec{k}) \Big)
\end{split}
\ee
where the energies are listed in non-decreasing order $E_1(\pm \vec{k}) \leq \cdots \leq E_{4MN}(\pm \vec{k})$ and $MN$ is the number of sites in the magnetic unit cell. Note that there are many unitary matrices which yield the energies in non-decreasing order if some of the energies are degenerate. Using particle-hole symmetry, we have
\be \label{eq15}
\begin{split}
&\mathcal{P}\mathcal{U}^\dagger(\vec{k})\mathcal{P}^{-1}\mathcal{P}H_{\rm BdG}(\vec{k})\mathcal{P}^{-1}\mathcal{P}\mathcal{U}(\vec{k})\mathcal{P}^{-1} \\ &= \mathcal{P}{\rm diag}\Big(E_1(\vec{k}),\ldots,E_{4MN}(\vec{k}) \Big)\mathcal{P}^{-1}\\
&\Rightarrow\big(\mathcal{P}\mathcal{U}^\dagger(\vec{k})\mathcal{P}^{-1} \big)H_{\rm BdG}(-\vec{k})\big(\mathcal{P}\mathcal{U}(\vec{k})\mathcal{P}^{-1} \big)\\& ={\rm diag}\Big(-E_{2MN+1}(\vec{k}),\ldots,-E_{4MN}(\vec{k}),\\&~~~~~~~-E_1(\vec{k}),\ldots,-E_{2MN}(\vec{k})) \Big)
\end{split}
\ee 
 So $\mathcal{P}\mathcal{U}(\vec{k})\mathcal{P}^{-1}$ is a unitary matrix which diagonalizes $H_{\rm BdG}(-\vec{k})$ but $\mathcal{P}\mathcal{U}(\vec{k})\mathcal{P}^{-1} \neq \mathcal{U}(-\vec{k})$ in general, since its action on $H_{\rm BdG}(-\vec{k})$ does not necessarily yield a diagonal matrix with non-decreasing elements. However, they can be related to one another by a permutation matrix $\mathcal{R}$ which reorders the basis. By looking at \eqref{eq15}, and keeping in mind that $E_1(\vec{k}),\ldots,E_{4MN}(\vec{k})$ is non-decreasing, we can see that 
 \be
 \mathcal{U}(-\vec{k})=\mathcal{P}\mathcal{U}(\vec{k})\mathcal{P}^{-1}\mathcal{R}
 \ee
 where
 \be
 \mathcal{R}=\begin{pmatrix}
 A_{2MN \times 2MN} & 0 \\
 0 & A_{2MN \times 2MN}
 \end{pmatrix}
 \ee
 and $A_{Q \times Q}$ is a $Q \times Q$ anti-diagonal matrix with unit entries.
  So if we know the eigenvectors of $H_{\rm BdG}(\vec{k})$ (that is, we diagonalize and find $\mathcal{U}(\vec{k})$), we know the eigenvectors of $H_{\rm BdG}(-\vec{k})$, since we simply can perform a particle hole transformation on $\mathcal{U}(\vec{k})$. This means that we only need to diagonalize the Hamiltonian for half of the MBZ, say on the domain ${\rm HBZ}=\{(k_x,k_y)|k_x \geq 0, k_y\geq0 \} \cup \{(k_x,k_y)|k_x > 0, k_y<0 \}$. Note that for a finite system where only a discrete set of momenta are allowed, one should tune the boundary conditions such that if $\vec{k}$ is an allowed momentum, then $-\vec{k}$ is also an allowed momentum. For $L$ odd (see the definition of $L$ in the main text just above Eq. 10), this entails periodic boundary conditions, while for $L$ even, anti-periodic boundary conditions. Furthermore, particle-hole symmetry implies that we must have the following equality of sets
 \be \label{sets}
 \begin{split}
 &\big\{ E_1(-\vec{k}),\ldots,E_{4MN}(-\vec{k}) \big\} \\ =&\big\{-E_{2MN+1}(\vec{k}),\ldots,-E_{4MN}(\vec{k}),-E_1(\vec{k}), \\&\ldots,-E_{2MN}(\vec{k}) \big\}
 \end{split}
 \ee
The left hand side of \eqref{sets} is in non-decreasing order, by assumption -- if we reorder the right hand side to be non-decreasing, we must have that
\be
E_q(-\vec{k})=-E_{4MN-q+1}(\vec{k})~~~~~ q=1,\dots,4MN
\ee
And, correspondingly, since $\mathcal{P}\mathcal{U}(\vec{k})\mathcal{P}^{-1}\mathcal{R}$ diagonalizes $H_{\rm BdG}(-\vec{k})$ with the energies ordered non-decreasingly, we have
\be \label{eq20}
\begin{split}
\gamma_q^\dagger(-\vec{k})&=\sum_{p=1}^{4MN} \psi^\dagger_p(-\vec{k}) \big(\mathcal{P}\mathcal{U}(\vec{k})\mathcal{P}^{-1}\mathcal{R}\big)_{pq}\\&=\sum_{r,s,t=1}^{4MN} {\psi_r(\vec{k})} \mathcal{U}_{rs}^*(\vec{k})(\tau_x)_{st}\mathcal{R}_{tq}\\&=\sum_{r,s=1}^{4MN} {\psi_r(\vec{k})} \mathcal{U}_{rs}^*(\vec{k})\delta_{4MN-s+1,q}\\
&=\sum_{r=1}^{4MN} {\psi_r(\vec{k})} \mathcal{U}_{r,4MN-q+1}^*(\vec{k})=\gamma_{4MN-q+1}(\vec{k})
\end{split}
\ee
where we have used that $(\tau_x \mathcal{R})_{sq}=(A_{4MN\times 4MN})_{sq} =\delta_{4MN-s+1,q} $. Consequently, the creation operators at $-\vec{k}$ are annihilation operators at $\vec{k}$.

\section{Mean-field groundstate} \label{sec3}
Writing the mean-field Hamiltonian in terms of the BdG eigenoperators, we have
\be \label{eq21}
\begin{split}
\cH_{\rm MF}&=\frac{1}{2}\sum_{\vec{k}}\sum_{q=1}^{4MN} E_q(\vec{k})\gamma^\dagger_q(\vec{k})\gamma_q(\vec{k}) +\frac{1}{2}\sum_{\vec{k}}\Tr \big[h(\vec{k}) \big]+\sum_j \frac{|\Delta_j|^2}{U} \\
&=\frac{1}{2}\sum_{\vec{k} \in \rm HBZ}\Bigg( \sum_{q=1}^{2MN}E_q(\vec{k})\gamma^\dagger_q(\vec{k})\gamma_q(\vec{k})+\sum_{q=2MN+1}^{4MN}E_q(\vec{k})\gamma^\dagger_q(\vec{k})\gamma_q(\vec{k})+\sum_{q=1}^{2MN} E_q(-\vec{k})\gamma^\dagger_q(-\vec{k})\gamma_q(-\vec{k})\\&+\sum_{q=2MN+1}^{4MN} E_q(-\vec{k})\gamma^\dagger_q(-\vec{k})\gamma_q(-\vec{k})\Bigg) +\frac{1}{2}\sum_{\vec{k}}\Tr \big[h(\vec{k}) \big]+\sum_j \frac{|\Delta_j|^2}{U}\\
&=\sum_{\vec{k} \in \rm HBZ}\Bigg(\sum_{q=2MN+1}^{4MN}E_q(\vec{k})\gamma^\dagger_q(\vec{k})\gamma_q(\vec{k})+\sum_{q=2MN+1}^{4MN} E_q(-\vec{k})\gamma^\dagger_q(-\vec{k})\gamma_q(-\vec{k}) - \frac{1}{2}\sum_{q=2MN+1}^{4MN}E_q(\vec{k})- \frac{1}{2}\sum_{q=2MN+1}^{4MN}E_q(-\vec{k})\Bigg) \\&+\frac{1}{2}\sum_{\vec{k}}\Tr \big[h(\vec{k}) \big]+\sum_j \frac{|\Delta_j|^2}{U} \\
&=\sum_{\vec{k}}\Bigg(\sum_{q=2MN+1}^{4MN}E_q(\vec{k})\gamma^\dagger_q(\vec{k})\gamma_q(\vec{k}) {-\frac{1}{2} \sum_{q=2MN+1}^{4MN}E_q(\vec{k})}\Bigg) +\frac{1}{2}\sum_{\vec{k}}\Tr \big[h(\vec{k}) \big]+\sum_j \frac{|\Delta_j|^2}{U}
\end{split}
\ee

That is, only the operators $\gamma_q(\vec{k})$ which correspond to the ``top half'' of the BdG spectrum appear in the Hamiltonian and, consequently, we can work exclusively with those to construct the mean-field groundstate. The others can be obtained by conjugation using \eqref{eq20}, if needed.

The mean-field groundstate is defined as the eigenstate of $\cH_{\rm MF}$ with minimum energy. From \eqref{eq21}, we see that such a state can be constructed by filling the Dirac sea of BdG quasiparticles
\be \label{mfgs}
\begin{split}
|\Gamma\rangle &=\bigg( \prod_{\vec{k}} \prod^{4MN}_{\substack{q~{\rm s.t.}\\E_q(\vec{k}) < 0}} \gamma^\dagger_q(\vec{k}) \bigg) |0\rangle\\&=  \prod_{\vec{k}} \Bigg(\prod^{4MN}_{\substack{q=2MN+1\\~{\rm s.t.}\\E_q(\vec{k}) < 0}} \gamma^\dagger_q(\vec{k}) \prod^{4MN}_{\substack{q=2MN+1\\{\rm s.t.}\\E_q(-\vec{k}) > 0}} \gamma_q(-\vec{k}) \Bigg) |0\rangle\\
&=\prod_{\vec{k}} \Bigg(\prod^{4MN}_{\substack{q=2MN+1\\~{\rm s.t.}\\E_q(\vec{k}) < 0}} \gamma^\dagger_q(\vec{k}) \prod^{4MN}_{\substack{q=2MN+1\\{\rm s.t.}\\E_q(\vec{k}) > 0}} \gamma_q(\vec{k}) \Bigg) |0\rangle,
\end{split}
\ee
where we have used the particle-hole redundancy of the $\gamma$'s to write the groundstate in terms of the ``top half'' of the BdG quasiparticle operators. 
If there is an $E_q(\vec{k})=0$, then there is a groundstate degeneracy and we may use the corresponding $\gamma$ and $\gamma^\dagger$ acting on the expression above to construct the groundstate subspace.

Alternatively, if we wish to work only in the HBZ, we may write \eqref{mfgs} as
\be
|\Gamma\rangle=\prod_{\vec{k} \in \rm HBZ} \Bigg(\prod^{4MN}_{\substack{q=1\\~{\rm s.t.}\\E_q(\vec{k}) < 0}} \gamma^\dagger_q(\vec{k}) \prod^{4MN}_{\substack{q=1\\{\rm s.t.}\\E_q(\vec{k}) > 0}} \gamma_q(\vec{k}) \Bigg) |0\rangle
\ee
\vspace{3mm}
\section{Gor'kov Green's Function} \label{sec4}
We define the Gor'kov Green's function at momentum $\vec{k}$ as
\be
\begin{split}
\mathcal{G}_{pq}(\vec{k}) &= \langle \psi^\dagger_p(\vec{k})\psi_q(\vec{k}) \rangle=\sum_{m,n=1}^{4MN}\mathcal{U}^\dagger_{mp}(\vec{k})\langle\gamma_m^\dagger(\vec{k}) \gamma_n(\vec{k})\rangle\mathcal{U}_{qn}(\vec{k})\\&=\sum_{m=1}^{4MN}\mathcal{U}^\dagger_{mp}(\vec{k})\mathcal{U}_{qm}(\vec{k}) f(\beta E_m(\vec{k})) \\
&=\bigg( \mathcal{U}(\vec{k}) F\big(E(\vec{k}) \big) \mathcal{U}^\dagger(\vec{k}) \bigg)^{\rm T}_{pq}
\end{split}
\ee
where $ F\big(E(\vec{k})\big)_{mn}=f(\beta E_m(\vec{k})) \delta_{mn}$, the expectation value is with respect to $|\Gamma\rangle$ and $f(\beta E)=\frac{1}{\exp(\beta E)+1}$ is the Fermi-Dirac distribution function. The indices $p$ and $q$ run over both spin and particle-hole labels. We have written $\psi^\dagger_p(\vec{k})=\sum_m \gamma_m^\dagger(\vec{k})\mathcal{U}^\dagger_{mp}(\vec{k})$ and $\psi_q(\vec{k})=\sum_{n} \gamma_{n}(\vec{k})\mathcal{U}_{qn}(\vec{k})$

The matrix $\mathcal{G}_{pq}(\vec{k})$ contains all the correlation functions of interest:
\be \label{gdmatrix}
\mathcal{G}(\vec{k})=
\begin{pmatrix}
\langle c^\dagger_\up(\vec{k})c_\up(\vec{k}) \rangle & \langle c^\dagger_\up(\vec{k})c_\down(\vec{k}) \rangle & \langle c^\dagger_\up(\vec{k})c^\dagger_\up(-\vec{k}) \rangle & \langle c^\dagger_\up(\vec{k})c^\dagger_\down(-\vec{k}) \rangle \\
\langle c^\dagger_\down(\vec{k})c_\up(\vec{k}) \rangle & \langle c^\dagger_\down(\vec{k})c_\down(\vec{k}) \rangle & \langle c^\dagger_\down(\vec{k})c^\dagger_\up(-\vec{k}) \rangle & \langle c^\dagger_\down(\vec{k})c^\dagger_\down(-\vec{k}) \rangle \\
\langle c_\up(-\vec{k})c_\up(\vec{k}) \rangle & \langle c_\up(-\vec{k})c_\down(\vec{k}) \rangle & \langle c_\up(-\vec{k})c^\dagger_\up(-\vec{k}) \rangle & \langle c_\up(-\vec{k})c^\dagger_\down(-\vec{k}) \rangle \\
\langle c_\down(-\vec{k})c_\up(\vec{k}) \rangle & \langle c_\down(-\vec{k})c_\down(\vec{k}) \rangle & \langle c_\down(-\vec{k})c^\dagger_\up(-\vec{k}) \rangle & \langle c_\down(-\vec{k})c^\dagger_\down(-\vec{k}) \rangle 
\end{pmatrix}
\ee
The correlation function which is relevant for the self-consistency equation is
\be \label{eq26}
\Delta_j=U\langle c_{j,\downarrow}c_{j,\uparrow} \rangle = \frac{U}{L^2} \sum_{\vec{k},\vec{k}'} e^{i (\vec{k}+\vec{k}')\cdot \vec{R}_j} \underbrace{\langle c_{\tilde{j},\downarrow}(\vec{k}')c_{\tilde{j},\uparrow}(\vec{k}) \rangle}_{\delta_{\vec{k}',-\vec{k}} \mathcal{G}_{\tilde{j}+3MN,\tilde{j}}(\vec{k})}=\frac{U}{L^2} \sum_{\vec{k}} \mathcal{G}_{\tilde{j}+3MN,\tilde{j}}(\vec{k})
\ee
It may also be useful to examine other angular momentum components of the order parameter, for example $p$-wave. This is obtained in a similar fashion as above
\be
\langle c_{j,\up}c_{j+\delta,\up} \rangle= \frac{1}{L^2} \sum_{\vec{k},\vec{k}'} e^{i (\vec{k}\cdot \vec{R}_j+\vec{k}'\cdot \vec{R}_{j+\delta})} \underbrace{\langle c_{\tilde{j},\up}(\vec{k}')c_{\tilde{j}+\delta,\uparrow}(\vec{k}) \rangle}_{\delta_{\vec{k}',-\vec{k}} \mathcal{G}_{\tilde{j}+2MN,\tilde{j}+\delta}(\vec{k})}= \frac{1}{L^2} \sum_{\vec{k}} e^{i \vec{k}\cdot (\vec{R}_j-\vec{R}_{j+\delta})} \mathcal{G}_{\tilde{j}+2MN,\tilde{j}+\delta}(\vec{k})
\ee

 \section{Current Operator} \label{sec6}
We give a derivation of the current operator by varying the Hamiltonian with respect to the hopping phases. To motivate this procedure, consider the Hamiltonian of $N$ interacting particles
\be
\begin{split}
\cH &= \frac{1}{2m} \sum_{i=1}^N (\vec{p}_i - q\vec{A}(\vec{r}_i))^2 + V( \vec{r}_1, \ldots, \vec{r}_N) \\ 
&\Rightarrow - \frac{\delta \cH}{\delta \big(q\vec{A}(\vec{r})\big)} \\&= \frac{1}{m}\sum_{i=1}^N\delta(\vec{r}_i - \vec{r}) \big(\vec{p}_i - q\vec{A}(\vec{r})\big) = \vec{J}(\vec{r}),
\end{split}
\ee
which is the velocity flow operator. We then have
\be
\int d^D\vec{r}~\vec{J}(\vec{r}) =  \frac{1}{m}\sum_{i=1}^N (\vec{p}_i - q\vec{A}(\vec{r}_i)),
\ee
which is evidently the net particle current of the system.

 We will use this result to find the current in our lattice model, where the hopping phases $A_\delta(\vec{r}_i) \leftrightarrow q\vec{A}(\vec{r})$. From the hopping part of the lattice Hamiltonian, we obtain the usual expression
 \be
 -\frac{\partial \cH_0}{\partial A_\epsilon(\vec{r}_i)} = i \sum_\sigma \Big(e^{iA_\epsilon (\vec{r}_i)}c^\dagger_{i+\epsilon,\sigma} c_{i,\sigma}-e^{-iA_\epsilon(\vec{r}_i)}c^{\dagger}_{i,\sigma} c_{i+\epsilon,\sigma}\Big).
 \ee
 \begin{widetext}
 The spin-orbit coupling also contributes, however:
 \be
 \begin{split}
 -\frac{\partial \cH_{\rm SO}}{\partial A_x(\vec{r}_i)} &= i V_{\rm SO} \bigg(e^{iA_x(\vec{r}_i)} \big(c^\dagger_{i+\unit{x},\down}c_{i,\up} -c^\dagger_{i+\unit{x},\up} c_{i,\down}\big) - e^{-iA_x(\vec{r}_i)} \big(c^\dagger_{i,\up}c_{i+\unit{x},\down} -c^\dagger_{i,\down} c_{i+\unit{x},\up}\big)\bigg) \\
  -\frac{\partial \cH_{\rm SO}}{\partial A_y(\vec{r}_i)} &= - V_{\rm SO} \bigg(e^{iA_y(\vec{r}_i)} \big(c^\dagger_{i+\unit{y},\down}c_{i,\up} + c^\dagger_{i+\unit{y},\up}c_{i,\down}\big) + e^{-iA_y(\vec{r}_i)} \big(c^\dagger_{i,\up}c_{i+\unit{y},\down} +c^\dagger_{i,\down} c_{i+\unit{y},\up}\big)\bigg)
 \end{split}
 \ee
The interaction term and pairing terms contribute nothing to the current operator, since they do not involve the hopping phases $A_\delta(\vec{r}_i)$. So the average current is 
\be
\begin{split}
\langle J_x(\vec{r}_i) \rangle = & i \sum_\sigma \Big(e^{iA_x (\vec{r}_i)}\langle c^\dagger_{i+\unit{x},\sigma} c_{i,\sigma}\rangle-e^{-iA_x(\vec{r}_i)}\langle c^{\dagger}_{i,\sigma} c_{i+\unit{x},\sigma}\rangle\Big) \\ &+ i V_{\rm SO} \bigg(e^{iA_x(\vec{r}_i)} \big(\langle c^\dagger_{i+\unit{x},\down}c_{i,\up} \rangle - \langle c^\dagger_{i+\unit{x},\up} c_{i,\down}\rangle \big) - e^{-iA_x(\vec{r}_i)} \big( \langle c^\dagger_{i,\up}c_{i+\unit{x},\down}\rangle - \langle c^\dagger_{i,\down} c_{i+\unit{x},\up}\rangle \big)\bigg)\\
\langle J_y(\vec{r}_i) \rangle = &i \sum_\sigma \Big(e^{iA_y (\vec{r}_i)}\langle c^\dagger_{i+\unit{y},\sigma} c_{i,\sigma}\rangle-e^{-iA_y(\vec{r}_i)}\langle c^{\dagger}_{i,\sigma} c_{i+\unit{y},\sigma}\rangle\Big) \\ & - V_{\rm SO} \bigg(e^{iA_y(\vec{r}_i)} \big(\langle c^\dagger_{i+\unit{y},\down}c_{i,\up}\rangle + \langle c^\dagger_{i+\unit{y},\up}c_{i,\down}\rangle \big) + e^{-iA_y(\vec{r}_i)} \big(\langle c^\dagger_{i,\up}c_{i+\unit{y},\down}\rangle + \langle c^\dagger_{i,\down} c_{i+\unit{y},\up}\rangle \big)\bigg).
\end{split}
\ee

The expectation values in the above expression are components of the Gor'kov Green's function $\mathcal{G}(\vec{k})$ (cf. Eq. \eqref{gdmatrix}):
\be
\begin{split}
\langle c^\dagger_{i+\epsilon,\sigma} c_{i,\sigma'}\rangle = \frac{1}{L^2}\sum_{\vec{k}} e^{i \vec{k}\cdot (\vec{R}_i - \vec{R}_{i+\epsilon})} \tilde{\mathcal{G}}_{i+\epsilon,\sigma; i, \sigma'}(\vec{k}),
\end{split}
\ee
where 
\be
\tilde{\mathcal{G}}_{i+\epsilon,\sigma; i, \sigma'}(\vec{k}) = 
\begin{pmatrix}
\mathcal{G}_{i+\epsilon; i}(\vec{k}) & \mathcal{G}_{i+\epsilon; i+MN}(\vec{k}) \\
\mathcal{G}_{i+\epsilon+MN; i}(\vec{k}) & \mathcal{G}_{i+\epsilon+MN; i+MN}(\vec{k}).
\end{pmatrix}
\ee
It will also be useful to write the expression for the boosted current $\bar{\vec{J}}(\vec{r}_i)= U^\dagger_B(\vec{Q})\vec{J}(\vec{r}_i)U_B(\vec{Q})$. We have
\be
\begin{split}
\bar{J}_x(\vec{r}_i) = & i \sum_\sigma \Big(e^{iA_x (\vec{r}_i)-i\vec{Q}\cdot\unit{x}} c^\dagger_{i+\unit{x},\sigma} c_{i,\sigma}-e^{-iA_x(\vec{r}_i)+i\vec{Q}\cdot\unit{x}} c^{\dagger}_{i,\sigma} c_{i+\unit{x},\sigma}\Big) \\ &+ i V_{\rm SO} \bigg(e^{iA_x(\vec{r}_i)-i\vec{Q}\cdot\unit{x}} \big( c^\dagger_{i+\unit{x},\down}c_{i,\up}  -  c^\dagger_{i+\unit{x},\up} c_{i,\down} \big) - e^{-iA_x(\vec{r}_i)+i\vec{Q}\cdot\unit{x}} \big(  c^\dagger_{i,\up}c_{i+\unit{x},\down} -  c^\dagger_{i,\down} c_{i+\unit{x},\up} \big)\bigg)\\
 \bar{J}_y(\vec{r}_i)  = &i \sum_\sigma \Big(e^{iA_y (\vec{r}_i)-i\vec{Q}\cdot\unit{y}} c^\dagger_{i+\unit{y},\sigma} c_{i,\sigma}-e^{-iA_y(\vec{r}_i)+i\vec{Q}\cdot\unit{y}} c^{\dagger}_{i,\sigma} c_{i+\unit{y},\sigma}\Big) \\ & - V_{\rm SO} \bigg(e^{iA_y(\vec{r}_i)-i\vec{Q}\cdot\unit{y}} \big( c^\dagger_{i+\unit{y},\down}c_{i,\up} +  c^\dagger_{i+\unit{y},\up}c_{i,\down} \big) + e^{-iA_y(\vec{r}_i)+i\vec{Q}\cdot\unit{y}} \big( c^\dagger_{i,\up}c_{i+\unit{y},\down} +  c^\dagger_{i,\down} c_{i+\unit{y},\up} \big)\bigg)
\end{split}
\ee
\section{Derivation of the lattice spin-orbit coupling Hamiltonian} \label{sec7}
In this section, we will derive the form of $\cH_{SO}$ used in this work, namely
\be \label{HSO_explain}
\begin{split}
 \cH_{\rm SO}&= V_{\rm SO} \sum_{j} \bigg( e^{iA_{\unit{x}} (\vec{r}_j)}\Big( c^\dagger_{j+\unit{x},\down}c_{j,\up} -c^\dagger_{j+\unit{x},\up} c_{j,\down} \Big)  \\ &+i e^{iA_{\unit{y}} (\vec{r}_j)} \Big(c^\dagger_{j+\unit{y},\down}c_{j,\up} + c^\dagger_{j+\unit{y},\up}c_{j,\down} \Big)  \bigg)  + {\rm h.c.}
 \end{split}
\ee
from the continuum form for Rashba spin-orbit coupling $H_R=\alpha\hat{z}\cdot(\vec{\sigma\times\vec{\pi}})$. To do this, we will first derive it without the presence of a gauge field $\vec{A}$, then we will include the gauge field at the last step by inserting Peierls phases to achieve a gauge invariant Hamiltonian. Thus we start with $H_R=\alpha\hat{z}\cdot(\vec{\sigma\times\vec{p}}) = \alpha(\sigma_x p_y - \sigma_y p_x)$. To go from the continuum to the lattice, we make the replacement $p_{\alpha}/\hbar \rightarrow \sin(p_{\alpha}/\hbar)$ ($\alpha = x,y$). Note that we have set the lattice constant equal to unity (and therefore $p_\alpha$ and $\hbar$ have the same units). The action of $\sin(p_{\alpha}/\hbar)$ on states in the position basis is determined by
\be
\hbar\sin(p_{\alpha}/\hbar) |\vec{i}, s\rangle = \frac{\hbar}{2i}\big( |\vec{i}+\unit{\alpha},s\rangle - |\vec{i}-\unit{\alpha},s \rangle \big)
\ee
were $s = \up, \down$. Therefore
\begin{equation}
\begin{split}
\langle \vec{i},s |H_R|\vec{j}, s'\rangle &=\hbar \alpha \langle \vec{i},s |(\sigma_y \sin(p_x/\hbar) - \sigma_x \sin(p_y/\hbar)|\vec{j}, s'\rangle \\ &=  \frac{\hbar\alpha}{2} \Big[\delta_{s, -s'} \zeta(s) \big(\delta_{i+\unit{x}, j} - \delta_{i-\unit{x}, j} \big) \\& + i\delta_{s, -s'} \big( \delta_{i+\unit{y}, j} - \delta_{i-\unit{y}, j} \big)  \Big]
\end{split}
\end{equation}
where $-s'$ is the opposite spin of $s'$, $\zeta(\up)=-1$ and $\zeta(\down)=1$. Using these matrix elements, we construct the second quantized Hamiltonian
\begin{widetext}
\begin{equation}
\begin{split}
 \cH_{\rm SO} &= \sum_{s,s'} \sum_{i,j} \Big(\langle \vec{i},s |H_R|\vec{j}, s'\rangle c^\dagger_{i, s} c_{j, s'}  \Big) \\  &=\frac{\hbar\alpha}{2}\sum_{j} \bigg(\Big( c^\dagger_{j+\unit{x},\down}c_{j,\up} -c^\dagger_{j+\unit{x},\up} c_{j,\down} \Big) +i  \Big(c^\dagger_{j+\unit{y},\down}c_{j,\up} + c^\dagger_{j+\unit{y},\up}c_{j,\down} \Big)-\Big( c^\dagger_{j,\down}c_{j+\unit{x},\up} -c^\dagger_{j,\up} c_{j+\unit{x},\down} \Big) -i  \Big(c^\dagger_{j,\down}c_{j+\unit{y},\up} + c^\dagger_{j,\up}c_{j+\unit{y},\down} \Big)   \bigg)
\end{split}
\end{equation}
\end{widetext}
\end{widetext}
Finally, to include the effect of the magnetic field, we multiply operators of the form $c^\dagger_{i+\unit{\delta}, s} c_{i, s'}$ by the phase factors $e^{iA_{\unit{\delta}}(\vec{r}_i)}$, which yields Eq. \eqref{HSO_explain} with $V_{\rm SO} =\hbar \alpha / 2$.

\section{{Topological Phase Transitions}} \label{tsc_sec}

{As seen in Figs. 4 (c) of the main text, there is an abundance of topological phase transitions, i.e. points where the Chern number changes. In this section, we give details about the nature of these transitions. We find two categories of topological phase transitions: (1) gap-closing transitions, where the BdG spectral gap closes across the transitions and (2) first-order transitions, where the BdG gap remains open, and the vortex lattice undergoes a structural reconfiguration. Due to the first-order structural transition of the second type, the electron gap can be jump from the positive value to negative value and thus we will not see the BdG spectral gap to close during the transition, which has been discussed in the early literature \cite{Ezawa2013}. 
In this section, we will examine an example of both categories from the data in the right panel of Fig. 4 of the main manuscript. The green arrows in the figure indicate points where a gap-closing transition occurs and the black arrows indicate points where a first-order transition occurs. We discuss the transitions at these points in detail below.}

\subsubsection{{Gap-Closing Transition}}\label{ex1}
\begin{figure}
\includegraphics[width=.6\linewidth]{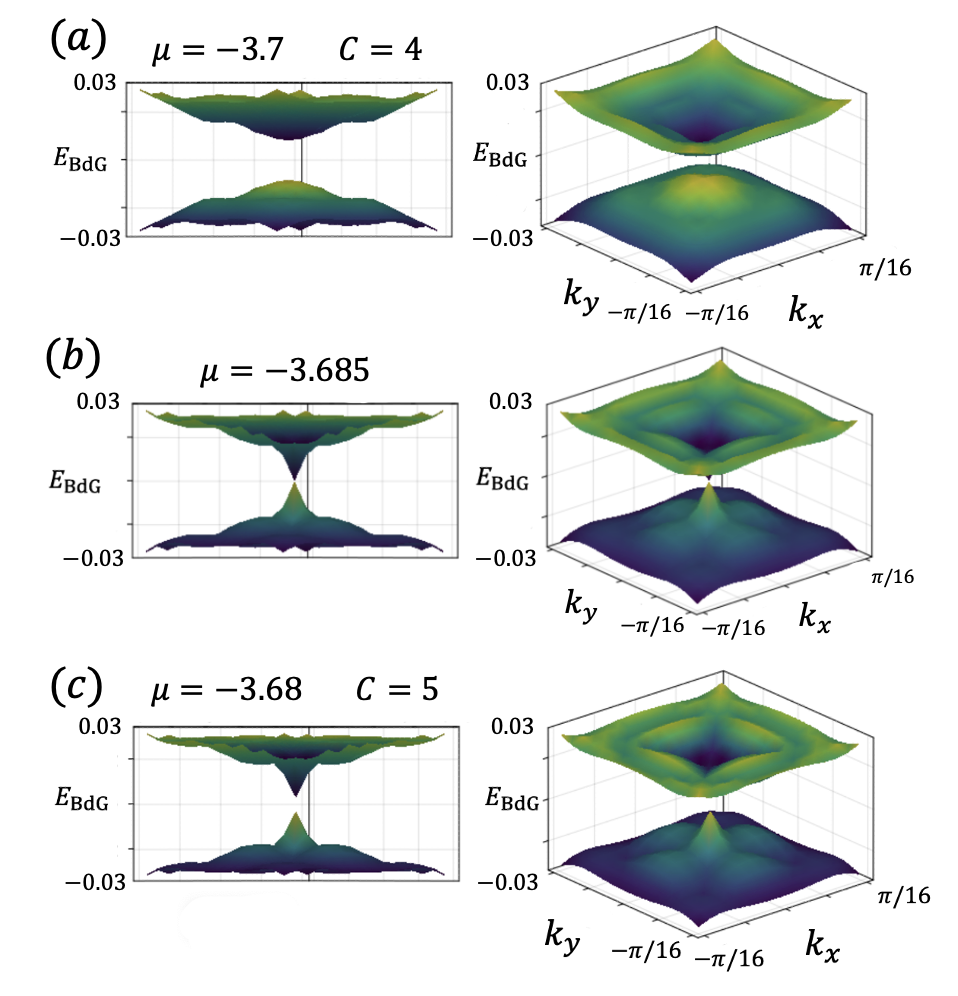}
\caption{{(a) The BdG spectrum (two panels, showing different perspectives) for $U=3.34$, $\mu=-3.7$, $V_{\text{SO}} = 0.1$, and $V_{\text{sp}} = 0.2$. The Chern number is $C =4$. (b) The BdG spectrum (two panels, showing different perspectives) for $\mu=-3.685$, with all other parameters the same, at the gap-closing (and thus the phase transition) point. The Chern number is undefined here since the BdG spectral gap is zero. (c) The BdG spectrum (two panels, showing different perspectives) for $\mu=-3.68$, with all other parameters the same. The Chern number has changed to $C =5$.}} \label{fig:gap-closing-example}
\end{figure}

{It is usually expected that a change in Chern number is associated with a closing of the spectral gap. 
Here we discuss an example where the gap closes and reopens, and the Chern number changes. In Fig. \ref{fig:gap-closing-example}, we show the BdG spectrum for a topological phase transition from $C = 4$ to $C = 5$ upon a small change in the chemical potential from $\mu=-3.7$ to $\mu=-3.68$ (other parameters are $U=3.34$,  $V_{\text{SO}} = 0.1$, and $V_{\text{sp}} = 0.2$). The points are indicated with green arrows in the right panel of Fig. 4 of the main text. Fig. \ref{fig:gap-closing-example}(a) shows the spectrum at $\mu=-3.7$. There is clearly a spectral gap. As the chemical potential is increased, we find that the spectral gap closes at $\mu = -3.685$. This is shown in Fig. \ref{fig:gap-closing-example}(b). Note that the Chern number is not well-defined at this point since the spectral gap is zero. Finally, the gap reopens at $\mu=-3.8$, and the Chern number changes from $C=4$ to $C=5$. Meanwhile, the vortex lattice structure remains the same throughout this process. 
%The vortex lattice structure on the two sides of the transition is shown in Fig. \ref{fig:vortex-example}. 
In both cases, the structure is a lattice of giant vortices.}

\subsubsection{{First-Order Transition}}\label{ex2}
\begin{figure*}
    \centering
    \includegraphics[width=.8\linewidth]{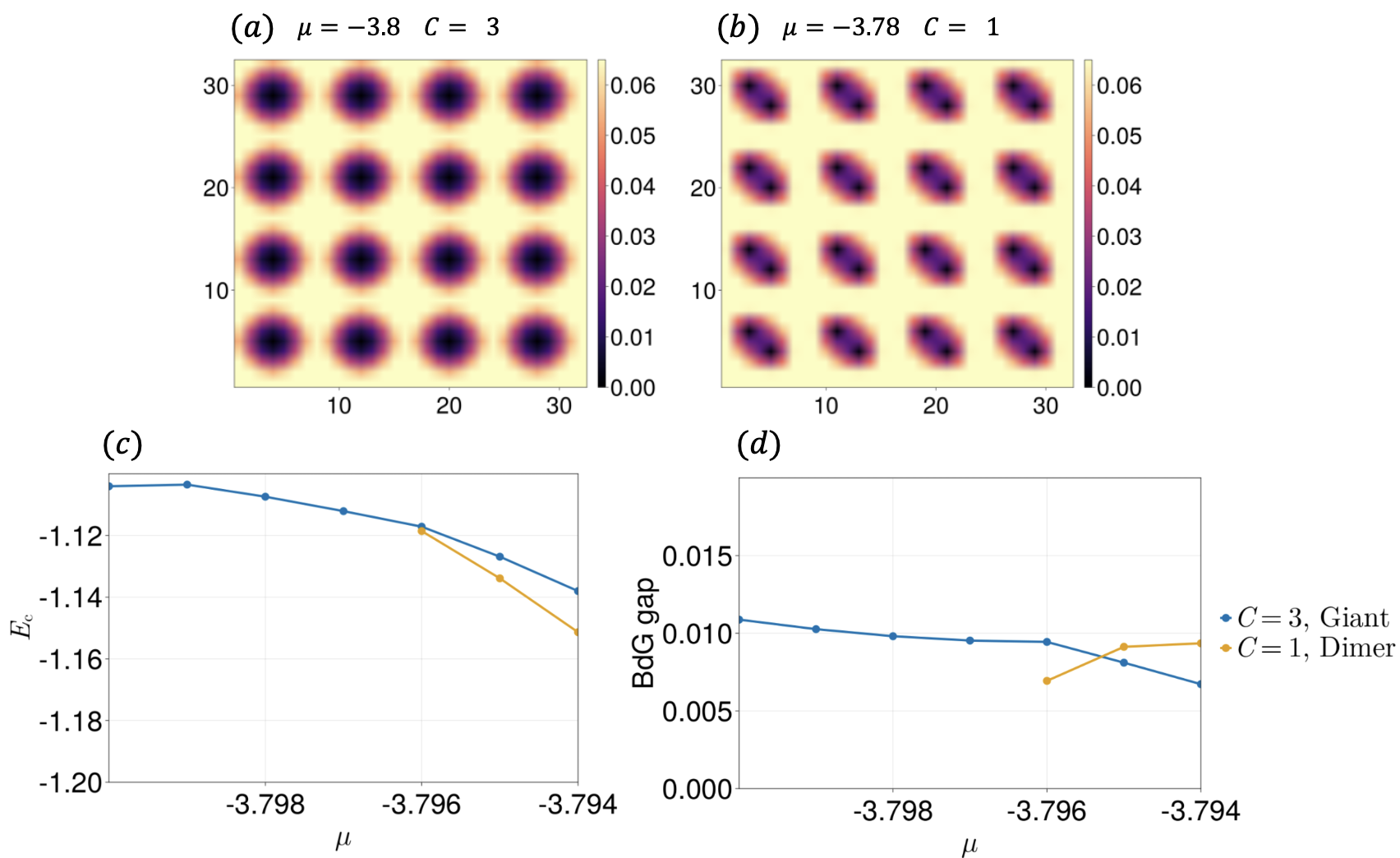}
    \caption{{(a) The vortex lattice configuration is a lattice of giant vortices at $\mu=-3.8$ with Chern number $C=3$. (b) The vortex lattice configuration is a lattice of vortex dimers at $\mu=-3.78$ with Chern number $C=1$. (c) As the chemical potential is tuned from $\mu=-3.8$ to $\mu=-3.78$, there is a first-order transition from the giant vortex lattice with $C=3$ to the dimer vortex lattice with $C=1$. The condensation energies $E_c$ for the giant vortex lattice (blue points) and the dimer vortex lattice (orange points) are shown in the range $-3.8 \leq \mu \leq -3.794$. The lines are guides to the eye. The absence of orange points for $-3.8 \leq \mu \leq -3.797$ indicates that the dimer vortex lattice does not solve the self-consistency equations in that range. (d) The BdG spectral gap in the range $-3.8 \leq \mu \leq -3.794$.}}
    \label{fig:first-order}
\end{figure*}

{We also find that the Chern number may change without a BdG spectral gap closing \cite{Ezawa2013}. This happens when there is a structural transition of the vortex lattice. We give an example of a giant vortex lattice at $\mu=-3.8$ with $C=3$ transitioning to a dimer vortex lattice at $\mu=-3.78$ with $C=1$. The corresponding points are indicated with black arrows in the right panel of Fig. 4 in the main text and the vortex configurations are shown in Figs. \ref{fig:first-order}(a) and \ref{fig:first-order}(b). Fig. \ref{fig:first-order}(c) shows the condensation energy $E_c$ as a function of $\mu$ for giant vortex lattice and dimer lattice configurations. For $-3.8 \leq \mu \leq -3.797$, only the giant vortex lattice is found to be a solution of the self-consistency equations -- the dimer vortex configuration is not a solution in this range. At $\mu = -3.796$, however, the dimer vortex configuration appears as a solution, and it has slightly lower energy than the giant vortex lattice. The giant vortex lattice and the dimer vortex lattice configurations coexist as solutions in the range $-3.796 \leq \mu \leq -3.794$, with the dimer vortex lattice having lower energy. The coexistence of these configurations suggests that this is a first-order phase transition. For $-3.793 \leq \mu \leq -3.8$ the giant vortex lattice disappears as a solution to the self-consistency equations, and only the dimer vortex lattice remains. Fig. \ref{fig:first-order}(d) shows the BdG spectral gap across the transition, which occurs at $\mu \approx -3.796$. The gap remains open across the transition, despite the Chern number changing.}

\bibliography{my2}